\documentclass[a4paper,11pt]{article}
\pdfoutput=1 

\usepackage{jheppub} 

\usepackage[T1]{fontenc} 

\usepackage{dsfont}
\usepackage{feynmp}
\usepackage{slashed}

\usepackage{multirow}

\allowdisplaybreaks

\def\beq{\begin{equation}}
\def\eeq#1{\label{#1}\end{equation}}
\def\eeqn{\end{equation}}

\def\beqa{\begin{eqnarray}}
\def\eeqa#1{\label{#1}\end{eqnarray}}
\def\eeqan{\end{eqnarray}}
\def\CR{\nonumber \\}

\def\bseq{\begin{subequations}}
\def\eseq#1{\label{#1}\end{subequations}}
\def\eseqn{\end{subequations}}

\def\L{{\cal L}}
\def\O{{\cal O}}
\def\I{{\cal I}}
\def\Q{{\cal Q}}
\def\b{\text{b}}
\def\c{\text{c}}
\def\identity{\mathds{1}}
\def\Tr{\mathrm{Tr}}
\def\tr{\mathrm{tr}}
\def\Psl{\slashed{P}}
\def\qsl{\slashed{q}}

\def\lf{\frac{i}{16\pi^2}}

\def\pref{\frac{c_s}{16\pi^2}}
\def\logm#1{\log\frac{#1}{\mu^2}}
\def\dsq#1{\Delta_{#1}^2}

\preprint{MCTP-16-23
\vspace{-8pt}
\begin{flushright}DESY-16-188\;\end{flushright}
}

\title{Covariant diagrams for one-loop matching}

 \author[a,b]{Zhengkang Zhang}
 \affiliation[a]{Michigan Center for Theoretical Physics (MCTP), University of Michigan, \\450 Church Street, Ann Arbor, MI 48109, USA}
 \affiliation[b]{Deutsches Elektronen-Synchrotron (DESY), \\Notkestra\ss e 85, 22607 Hamburg, Germany}

\emailAdd{zzkevin@umich.edu}

\abstract{
We present a diagrammatic formulation of recently-revived covariant functional approaches to one-loop matching from an ultraviolet (UV) theory to a low-energy effective field theory. 
Various terms following from a covariant derivative expansion (CDE) are represented by diagrams which, unlike conventional Feynman diagrams, involve gauge-covariant quantities and are thus dubbed ``covariant diagrams.'' The use of covariant diagrams helps organize and simplify one-loop matching calculations, which we illustrate with examples. Of particular interest is the derivation of UV model-independent universal results, which reduce matching calculations of specific UV models to applications of master formulas. We show how such derivation can be done in a more concise manner than the previous literature, and discuss how additional structures that are not directly captured by existing universal results, including mixed heavy-light loops, open covariant derivatives, and mixed statistics, can be easily accounted for.
}

\begin{document} 
\maketitle
\flushbottom

\section{Introduction}
\label{sec:intro}

Matching from an ultraviolet (UV) theory to a low-energy effective field theory (EFT) beyond tree level has gained renewed interest in recent years. On the phenomenological side, 
one-loop matching is in many cases necessary for accurate translation of experimental constraints on the Standard Model (SM) EFT parameter space into those on specific new physics models. 
On the theoretical side, it is interesting to realize that matching calculations can be accomplished in more elegant and oftentimes simpler ways than using Feynman diagrams. For the latter aspect, the idea is to directly tackle the path integral, and identify and expand heavy fields' contributions to the functional determinant arising at one-loop level to obtain effective operators involving the light fields. Such functional approaches to matching have at least two important virtues:
\begin{itemize}
\item By performing a {\it covariant} derivative expansion (CDE), one can work with gauge-covariant quantities in all steps of the calculation, and thus automatically arrive at gauge-invariant effective operators in the end. This is unlike conventional Feynman diagram methods, where gauge-invariant final results are obtained only after putting together individual pieces which may not be separately gauge-invariant. 
\item The generality of such approaches has brought up the possibility of obtaining {\it universal} results. With general assumptions on the form of the UV theory, evaluation of the functional determinants involved proceeds in a model-independent way, which can thus be done once and for all. The result will be widely-applicable master formulas, from which matching calculations for specific models are reduced to matrix algebra.
\end{itemize}

The development and use of covariant functional approaches to matching dates back to the 1980s; see e.g.~\cite{Gaillard85,Chan86,Cheyette87}. The subject was revived recently, thanks to the work~\cite{HLM14} by Henning, Lu and Murayama (HLM). Following the CDE approach of Gaillard~\cite{Gaillard85} and Cheyette~\cite{Cheyette87}, HLM presented in~\cite{HLM14} a universal master formula for one-loop matching, assuming degenerate mass spectrum for the heavy particles. Applications of this master formula to various examples have been illustrated by HLM in~\cite{HLM14}, and also by others in~\cite{Huo:scalar,Huo:fermion,Huo:MSSM}. The HLM master formula was generalized by Drozd, Ellis, Quevillon and You~\cite{DEQY} to the case of nondegenerate heavy particle masses. The same Gaillard-Cheyette CDE approach is followed in~\cite{DEQY}, and the resulting master formula is dubbed the ``Universal One-Loop Effective Action'' (UOLEA), to emphasize the universality of the approach, as discussed in the second bullet point above. The UOLEA was applied to the example of integrating out nondegenerate stops in~\cite{DEQY}.

It was later pointed out, however, that the HLM/UOLEA master formulas, in their original forms at least, do not capture possible contributions from mixed heavy-light loops~\cite{DKS} (see also~\cite{BGP}). The reason can be most easily understood by noting that light fields are treated as background fields in~\cite{HLM14,DEQY} and are thus not allowed to run in loops. Fixes to this problem were soon proposed, following different CDEs~\cite{HLM16}, or alternatively still within the UOLEA framework~\cite{EQYZ}. Although technically quite different, both approaches in~\cite{HLM16} and~\cite{EQYZ} share a similar spirit, namely to include quantum fluctuations of light fields also, and then identify and subtract off nonlocal pieces from the functional determinant to obtain local effective operators. These studies provide, at the very least, a proof of principle that mixed heavy-light loops can be accounted for in covariant functional approaches to matching. This latter point was further corroborated recently in~\cite{FPR}, following an alternative CDE approach that builds upon~\cite{DittmaierG9501,DittmaierG9505}. Compared with~\cite{HLM16,EQYZ}, matching calculations are simplified in~\cite{FPR} partly due to the use of expansion by regions techniques~\cite{EBR-BenekeS,EBR-Smirnov,EBR-Jantzen}, which allow local pieces of the functional determinant to be directly identified, so that no subtraction procedure is needed.

These recent developments of functional matching techniques are, unfortunately, also accompanied by different levels of technical complication compared with~\cite{HLM14,DEQY}. 
It should be emphasized that the motivation for studying functional matching lies not only in theoretical curiosity, but also, at least equally importantly, in practical usefulness. 
In this latter respect, the goal is to develop a set of tools for matching that is easy to use even for those not necessarily familiar with all the technical subtleties of functional methods. There are at least two possibilities for achieving this goal:
\begin{itemize}
\item Ideally we wish to obtain a truly {\it universal} master formula, as an extension of the results presented in~\cite{HLM14,DEQY}. Such an extension requires incorporation of not only mixed heavy-light contributions mentioned above, but also e.g.\ open covariant derivatives (covariant derivatives acting openly to the right as opposed to appearing in commutators) and mixed statistics (both bosonic and fermionic fields in the loop).
\item Even if deriving such extended universal results turns out to be too involved to be completed very soon, we may still take advantage of the {\it covariant} feature of functional approaches, and consider alternatives to Feynman diagram methods that simplify calculations and offer useful intuition, even though on a case-by-case basis. This will also bring new options for more efficient automation of matching calculations\footnote{See e.g.~\cite{MatchMaker} for recent progress on automation of Feynman diagrammatic matching.}.
\end{itemize}

It is the purpose of this paper to present a tool that will be useful for making progress along both these lines. The idea is to have a diagrammatic formulation of one-loop functional matching which is as systematic as the conventional Feynman diagram approach, but differs crucially from the latter by preserving gauge covariance in intermediate steps. It is perhaps not surprising that this is possible, since recent studies of functional matching~\cite{HLM16,EQYZ,FPR} all follow diagrammatic intuitions to some extent. We will show explicitly how to establish such a gauge-covariant diagrammatic formulation, building upon the approach of~\cite{FPR} (which we provide a more rigorous derivation of)\footnote{The approach of~\cite{HLM14,DEQY} also allows for a diagrammatic formulation, which is however more complicated technically and will not be discussed further.}, and how to use it in one-loop matching calculations. The diagrams introduced are dubbed ``covariant diagrams'' --- they are in a sense gauge-covariant versions of Feynman diagrams. Just like Feynman diagrams, which keep track of terms in an expansion of correlation functions, covariant diagrams keep track of terms in a CDE in functional matching. Let us clarify that enumerating and computing covariant diagrams is equivalent to selecting and evaluating various terms of interest that result from a CDE. But as we will see, it is both technically simpler and conceptually more intuitive than the latter, and meanwhile preserves the universality feature of functional matching procedures.

\subsection{Outline of the paper}

For the sake of pedagogy, we will present many details of derivations and computations. The hope is that readers can easily reproduce all the intermediate steps as well as final results in this paper, and readily apply the techniques to other examples of interest. Given the considerable length that results, we provide an extended outline of the paper to guide the reading.
\begin{itemize}
\item In Section~\ref{sec:functionalmatching}, we reproduce the functional matching procedure of~\cite{FPR}, expanding the latter with a more formal and rigorous functional derivation. 
Readers familiar with the general idea and procedure of functional matching may wish to briefly look at the following key equations and skip the technical details in a first reading:
\begin{itemize}
\item Eq.~\eqref{matchingcondition}; 
\item Eqs.~\eqref{GammaLUVtree} and~\eqref{GammaLUVloop}, with $\Phi_\c$ defined in Eq.~\eqref{Phicdef} and $\Q_\text{UV}$ defined in Eq.~\eqref{QUV}; 
\item Eqs.~\eqref{GammaEFTtree} and~\eqref{GammaEFTloop}, with $\Q_\text{EFT}$ defined in Eq.~\eqref{QEFT}; 
\item Eqs.~\eqref{LEFTtree} and~\eqref{SEFTloopfinal}, with $\Delta_{H,L}$ and $X_{HL,LH}$ defined in~Eq.~\eqref{QUVelements}; 
\item Eqs.~\eqref{logdetsoft} and~\eqref{SEFTloopTr}, with the meaning of ``hard'' and ``soft'' explained below Eq.~\eqref{EBR}; 
\item Eq.~\eqref{LEFTloop};
\item Eq.~\eqref{LEFTloopfinal}, with $X_H$ defined in Eq.~\eqref{DeltaH}. 
\end{itemize}
\item In Section~\ref{sec:covariantdiagrams}, we derive a diagrammatic formulation of covariant functional matching. The basic ingredients are obtained in Section~\ref{sec:cd-heavy}, and additional structures are gradually added in the subsequent subsections. {\it Section~\ref{sec:cd-sum} is the core of the paper,} where we summarize the derivation and present a step-by-step recipe for using covariant diagrams in one-loop matching calculations.
\item In Section~\ref{sec:example}, we work out several examples to demonstrate the use of the covariant diagrams and the simplification that results.
\begin{itemize}
\item The UOLEA master formula is rederived in Section~\ref{sec:UOLEA}, with a much simpler procedure compared with~\cite{DEQY}. Also, fewer master integrals are involved in the final results, and explicit expressions in terms of heavy particle masses are simplified in many cases.
\item The rest of Section~\ref{sec:example} contains examples of matching specific UV models to EFTs, with the aim to show that covariant diagrams are capable of dealing with several additional structures not captured by previous universal results in a straightforward manner. 
\end{itemize}
With these examples, it is reasonable to expect that the use of covariant diagrams will be helpful for organizing and simplifying derivations of extended universal master formulas --- such calculations are underway and will be presented elsewhere. On the other hand, our examples show that even before extended universal results are available, one can already use covariant diagrams to easily perform matching calculations for specific models in a gauge-covariant manner, as an alternative to Feynman diagrammatic matching.
\item We conclude in Section~\ref{sec:conclusions}, and tabulate some useful master integrals and explicit expressions of the UOLEA operator coefficients in the appendices.
\end{itemize}

\section{Gauge-covariant functional matching}
\label{sec:functionalmatching}

The problem of matching can be formulated as follows: given an UV Lagrangian $\L_\text{UV}[\Phi, \phi]$ for a set of heavy fields $\Phi$ of masses $\{M_i\}$ and a set of light fields $\phi$ of masses $\{m_{i'}\}\ll\{M_i\}$, 
\beq
\L_\text{EFT} [\phi] = \;? \quad \text{s.t.} \quad \Gamma_\text{L,UV}[\phi_\b] = \Gamma_\text{EFT}[\phi_\b] \,.
\eeq{matchingcondition}
Here $\Gamma_\text{L,UV}$ is the one-light-particle-irreducible (1LPI) effective action calculated in the UV theory, while $\Gamma_\text{EFT}$ is the one-particle-irreducible (1PI) effective action (a.k.a.\ quantum action) calculated in the low-energy EFT. They will be computed as functionals of background fields $\phi_\b$ by the standard procedures of the background field method (see e.g.~\cite{PeskinSchroeder,Abbott}). Eq.~\eqref{matchingcondition} ensures that the UV theory and the EFT give identical physical predictions regarding the light fields. 

In this section, we shall focus on the simplest case of real scalar fields for illustration. The results derived below can be easily generalized to other types of fields.

\subsection{Calculating $\Gamma_\text{L,UV}[\phi_\b]$}

To compute $\Gamma_\text{L,UV}[\phi_\b]$, we start from the path integral,
\beq
Z_\text{UV} [J_\Phi, J_\phi] = \int [D\Phi] [D\phi]\, e^{\,i\int d^dx (\L_\text{UV}[\Phi, \phi] +J_\Phi \Phi +J_\phi \phi)} \,,
\eeqn
and separate all fields contained in the heavy and light field multiplets into classical backgrounds (labeled by subscripts ``b'') and quantum fluctuations (labeled by primes),
\beq
\Phi = \Phi_\b +\Phi' 
\,, \qquad 
\phi = \phi_\b +\phi' 
\,.
\eeq{separation}
The background fields and sources are related by
\beq
0 
\,=\, \frac{\delta\L_\text{UV}}{\delta\Phi} [\Phi_\b, \phi_\b] +J_\Phi
\,=\, \frac{\delta\L_\text{UV}}{\delta\phi} [\Phi_\b, \phi_\b] +J_\phi
\,.
\eeq{bkgfield-source}
The 1LPI effective action $\Gamma_\text{L,UV}[\phi_\b]$ is obtained as the Legendre transform of the path integral with respect to the light fields,
\beq
\Gamma_\text{L,UV} [\phi_\b] = -i \log Z_\text{UV} [ J_\Phi=0, J_\phi] -\int d^dx\, J_\phi \phi_\b \,.
\eeqn
Note that $J_\Phi$ is set to zero because we are interested in correlation functions with no external sources of the heavy fields.

With the separation in Eq.~\eqref{separation}, the UV theory Lagrangian plus source terms can be written as
\beq
\L_\text{UV} [\Phi, \phi] +J_\Phi \Phi +J_\phi \phi 
= \L_\text{UV} [\Phi_\b, \phi_\b] +J_\Phi \Phi_\b +J_\phi \phi_\b 
-\frac{1}{2} \left(\Phi'^T \; \phi'^T\right) \Q_\text{UV} [\Phi_\b, \phi_\b]
\left(
\begin{matrix}
\Phi' \\
\phi'
\end{matrix}
\right)
+\,\dots
\eeq{LUVexpand}
where the quadratic operator
\beq
\Q _\text{UV}[\Phi_\b, \phi_\b]\equiv \left(
\begin{matrix}
-\frac{\delta^2\L_\text{UV}}{\delta\Phi^2} [\Phi_\b, \phi_\b] &\,\, -\frac{\delta^2\L_\text{UV}}{\delta\Phi\delta\phi} [\Phi_\b, \phi_\b] \\
-\frac{\delta^2\L_\text{UV}}{\delta\phi\delta\Phi} [\Phi_\b, \phi_\b] &\,\, -\frac{\delta^2\L_\text{UV}}{\delta\phi^2} [\Phi_\b, \phi_\b]
\end{matrix}
\right).
\eeq{QUV}
Note that in Eq.~\eqref{LUVexpand}, terms linear in $\phi'$ or $\Phi'$ vanish due to Eq.~\eqref{bkgfield-source}. We therefore obtain the tree-level result as the stationary point approximation,
\beqa
Z_\text{UV}^\text{tree} [J_\Phi, J_\phi] 
&=& \int [D\Phi] [D\phi] e^{\,i\int d^dx (\L_\text{UV}[\Phi_\b, \phi_\b] +J_\Phi \Phi_\b +J_\phi \phi_\b)} 
\propto e^{\,i\int d^dx (\L_\text{UV}[\Phi_\b, \phi_\b] +J_\Phi \Phi_\b +J_\phi \phi_\b}) \CR
\Rightarrow \Gamma_\text{L,UV}^\text{tree} [\phi_\b] &=& \int d^dx \,\L_\text{UV}\bigl[\Phi_\c[\phi_\b], \phi_\b\bigr] ,
\eeqa{GammaLUVtree}
up to an irrelevant constant term, where $\Phi_\c[\phi_\b]$ (subscript ``c'' for ``classical'') is defined by
\beq
\Phi_\c[\phi_\b] \equiv \Phi_\b [J_\Phi=0] \quad 
\text{i.e.} \quad 
\frac{\delta\L_\text{UV}}{\delta\Phi} \bigl[\Phi_\c[\phi_\b], \phi_\b\bigr] 
\equiv \left. \frac{\delta\L_\text{UV}[\Phi, \phi]}{\delta\Phi} \right|_{\Phi = \Phi_\c[\phi_\b], \,\phi = \phi_\b} 
= 0.
\eeq{Phicdef}
In other words, $\Phi_\c[\phi_\b]$ solves the classical equations of motion for the heavy fields when the light fields are treated as backgrounds.

Up to one-loop level, we have
\beqa
Z_\text{UV} [J_\Phi, J_\phi] &\simeq& Z_\text{UV}^\text{tree} \int [D\Phi'] [D\phi'] \exp\left\{-\frac{i}{2}\int d^dx 
\left(\Phi'^T \; \phi'^T\right) \Q_\text{UV} [\Phi_\b, \phi_\b]
\left(
\begin{matrix}
\Phi' \\
\phi'
\end{matrix}
\right)
\right\} \CR
&\propto& Z_\text{UV}^\text{tree} \, \bigl(\det\Q_\text{UV}[\Phi_\b, \phi_\b] \bigr)^{-\frac{1}{2}} \CR[1ex]
\Rightarrow \Gamma_\text{L,UV}^\text{1-loop} [\phi_\b] &=& \frac{i}{2} \log\det\Q_\text{UV} \bigl[\Phi_\c[\phi_\b], \phi_\b\bigr] \,,
\eeqa{GammaLUVloop}
which is familiar from standard calculations of 1PI effective actions.

\subsection{Calculating $\Gamma_\text{EFT}[\phi_\b]$}

On the EFT side, suppose
\beq
\L_\text{EFT} [\phi] = \L_\text{EFT}^\text{tree} [\phi] +\L_\text{EFT}^\text{1-loop} [\phi] +\dots
\eeqn
where $\L_\text{EFT}^\text{tree}$ and $\L_\text{EFT}^\text{1-loop}$ contain effective operators generated at tree and one-loop level, respectively. The path integral can be evaluated up to one-loop level,
\beqa
Z_\text{EFT} [J_\phi] 
&=& \int [D\phi]\, e^{\,i\int d^dx (\L_\text{EFT}[\phi] +J_\phi \phi)} \CR
&\simeq&  e^{\,i\int d^dx (\L_\text{EFT}[\phi_\b] +J_\phi \phi_\b)} \int [D\phi']\, e^{-\frac{i}{2}\int d^dx\, \phi'^T \Q_\text{EFT}^\text{tree}[\phi_\b]\, \phi'} \CR
&\propto& e^{\,i\int d^dx (\L_\text{EFT}^\text{tree}[\phi_\b] +\L_\text{EFT}^\text{1-loop}[\phi_\b] +J_\phi \phi_\b)} \bigl(\det\Q_\text{EFT}^\text{tree}[\phi_\b] \bigr)^{-\frac{1}{2}} ,
\eeqan
where the quadratic operator
\beq
\Q_\text{EFT} [\phi_\b] \equiv -\frac{\delta^2\L_\text{EFT}}{\delta\phi^2} [\phi_\b] \,.
\eeq{QEFT}
Again, in the exponent, terms linear in $\phi'$ vanish due to the relation
\beq
\frac{\delta\L_\text{EFT}}{\delta\phi} [\phi_\b] +J_\phi \,=\, 0 \,.
\eeqn
We therefore obtain the 1PI effective action in the EFT up to one-loop level,
\beqa
\Gamma_\text{EFT} [\phi_\b] &=& -i \log Z_\text{EFT} [J_\phi] -\int d^dx\, J_\phi \phi_\b \CR
&\simeq& \int d^dx \left(\L_\text{EFT}^\text{tree}[\phi_\b] +\L_\text{EFT}^\text{1-loop}[\phi_\b] \right) +\frac{i}{2} \log\det\Q_\text{EFT}^\text{tree} [\phi_\b] \,, \\
\Rightarrow\; \Gamma_\text{EFT}^\text{tree} [\phi_\b] &=& \int d^dx \L_\text{EFT}^\text{tree}[\phi_\b] \,, \label{GammaEFTtree}\\
\Gamma_\text{EFT}^\text{1-loop} [\phi_\b] &=& \int d^dx \L_\text{EFT}^\text{1-loop}[\phi_\b] +\frac{i}{2} \log\det\Q_\text{EFT}^\text{tree} [\phi_\b] \,. \label{GammaEFTloop}
\eeqan
The meaning of the above equations is clear. The tree-level quantum action is given by the tree-level terms in the classical action, while at one-loop level, the quantum action contains two pieces --- one-loop-size effective operators used at tree level, and tree-level-size effective operators used at one-loop level.

\subsection{Matching $\Gamma_\text{L,UV}[\phi_\b]$ and $\Gamma_\text{EFT}[\phi_\b]$}

Equating Eqs.~\eqref{GammaLUVtree}, \eqref{GammaLUVloop} and Eqs.~\eqref{GammaEFTtree}, \eqref{GammaEFTloop}, we obtain the EFT Lagrangian that satisfies the matching condition~\eqref{matchingcondition}. At tree level,
\beqa
\L_\text{EFT}^\text{tree}[\phi] = \L_\text{UV}\bigl[\Phi_\c[\phi], \phi\bigr] \to \L_\text{UV}\bigl[\hat\Phi_\c[\phi], \phi\bigr] \,,
\eeqa{LEFTtree}
where $\hat\Phi_\c[\phi]$ is the local operator expansion of the nonlocal object $\Phi_\c[\phi]$. The extra step from $\Phi_\c[\phi]$ to $\hat\Phi_\c[\phi]$ is necessary so that $\L_\text{EFT}[\phi]$ consists of local operators. As a trivial example, suppose
\beq
\L_\text{UV} [\Phi, \phi] = \L_0[\phi] +\Phi^T F[\phi] -\frac{1}{2}\Phi^T \bigl( -P^2 +M^2 \bigr)\, \Phi \,,
\eeqn
where $P_\mu\equiv iD_\mu$. The advantage of introducing this notation is that $P_\mu$ is a hermitian operator. $\Phi_\c[\phi]$ is obtained by solving the classical equation of motion [see Eq.~\eqref{Phicdef}],
\beq
\frac{\delta\L_\text{UV}}{\delta\Phi} = F[\phi] -\bigl( -P^2 +M^2 \bigr)\, \Phi = 0 \quad
\Rightarrow\quad \Phi_\c[\phi] = \frac{1}{-P^2 +M^2} F[\phi] \,.
\eeqn
This is a nonlocal quantity due to the appearance of $P^2$ in the denominator. The corresponding local operator expansion, which should appear in the EFT, reads
\beq
\hat\Phi_\c [\phi] =\frac{1}{M^2} F[\phi] +\frac{1}{M^2} P^2 \frac{1}{M^2} F[\phi] +\dots
\eeqn

Moving on to one-loop level, we have
\beq
\int d^dx\, \L_\text{EFT}^\text{1-loop}[\phi] 
= \frac{i}{2} \log\det\Q_\text{UV} \bigl[\Phi_\c[\phi], \phi\bigr] -\frac{i}{2} \log\det\Q_\text{EFT}^\text{tree} [\phi] \,.
\eeq{SEFTloop}
To proceed, we follow~\cite{FPR} and block-diagonalize $\Q_\text{UV}$. With the following short-hand notation for the elements of $\Q_\text{UV}$,
\beq
\Q_\text{UV}[\Phi, \phi] = 
\left(
\begin{matrix}
-\frac{\delta^2\L_\text{UV}}{\delta\Phi^2} [\Phi, \phi] &\,\, -\frac{\delta^2\L_\text{UV}}{\delta\Phi\delta\phi} [\Phi, \phi] \\
-\frac{\delta^2\L_\text{UV}}{\delta\phi\delta\Phi} [\Phi, \phi] &\,\, -\frac{\delta^2\L_\text{UV}}{\delta\phi^2} [\Phi, \phi]
\end{matrix}
\right) 
\equiv 
\left(
\begin{matrix}
\Delta_H [\Phi, \phi] &\,\, X_{HL} [\Phi, \phi] \\
X_{LH} [\Phi, \phi] &\,\, \Delta_L [\Phi, \phi]
\end{matrix}
\right) ,
\eeq{QUVelements}
it is easy to show that
\beq
V^\dagger \Q_\text{UV} V =
\left(
\begin{matrix}
\Delta_H -X_{HL} \Delta_L^{-1} X_{LH} & 0 \\
0 & \Delta_L
\end{matrix}
\right)
\quad\text{with}\quad
V =
\left(
\begin{matrix}
\identity & \; 0 \; \\
-\Delta_L^{-1} X_{LH} & \; \identity \;
\end{matrix}
\right).
\eeq{QUVdiagonal}
Note that for real scalar fields, $X_{HL}=X_{LH}$ and both are hermitian. When generalized to complex fields, $X_{HL} = X_{LH}^\dagger$. With Eq.~\eqref{QUVdiagonal}, the first term on the RHS of Eq.~\eqref{SEFTloop} becomes
\beq
\frac{i}{2} \log\det\Q_\text{UV} \bigl[\Phi_\c[\phi], \phi\bigr]  = \frac{i}{2} \log\det \left( \Delta_H -X_{HL} \Delta_L^{-1} X_{LH} \right) +\frac{i}{2} \log\det\Delta_L \,,
\eeq{logdetQUV}
where the arguments $\bigl[\Phi_\c[\phi], \phi\bigr] \to \bigl[\hat\Phi_\c[\phi], \phi\bigr]$ have been dropped on the RHS for simplicity. Note that $\Phi_\c[\phi]$ should be replaced by $\hat\Phi_\c[\phi]$ to form local operators of the EFT.

Let us now look at the second term on the RHS of Eq.~\eqref{SEFTloop}. With Eqs.~\eqref{QEFT} and~\eqref{LEFTtree}, we have
\beqa
\Q_\text{EFT}^\text{tree} [\phi] &=& -\frac{\delta^2 \L_\text{UV}\bigl[\hat\Phi_\c[\phi], \phi\bigr]}{\delta\phi^2}
= -\frac{\delta}{\delta\phi} \left(\frac{\delta \L_\text{UV}}{\delta\phi}\bigl[\hat\Phi_\c[\phi], \phi\bigr] +\frac{\delta\hat\Phi_\c[\phi]}{\delta\phi} \frac{\delta\L_\text{UV}}{\delta\Phi}\bigl[\hat\Phi_\c[\phi], \phi\bigr] \right) \CR
&=& -\frac{\delta}{\delta\phi} \left(\frac{\delta \L_\text{UV}}{\delta\phi}\bigl[\hat\Phi_\c[\phi], \phi\bigr] \right)
= -\frac{\delta^2 \L_\text{UV}}{\delta\phi^2}\bigl[\hat\Phi_\c[\phi], \phi\bigr] -\frac{\delta\hat\Phi_\c[\phi]}{\delta\phi}\frac{\delta^2 \L_\text{UV}}{\delta\Phi\delta\phi}\bigl[\hat\Phi_\c[\phi], \phi\bigr] \CR
&=& \Delta_L \bigl[\hat\Phi_\c[\phi], \phi\bigr] -X_{LH} \hat\Delta_H^{-1} X_{HL}\bigl[\hat\Phi_\c[\phi], \phi\bigr] \,.
\eeqa{QEFTtree}
When going from the first line to the second, we have used $\frac{\delta\L_\text{UV}}{\delta\Phi}\bigl[\hat\Phi_\c[\phi], \phi\bigr] = \frac{\delta\L_\text{UV}}{\delta\Phi}\bigl[\Phi_\c[\phi], \phi\bigr] =0$ --- this is true because the EoM can be solved order by order in $\frac{1}{M}$ to obtain a local operator expansion $\hat\Phi_\c[\phi]$. To arrive at the last line of Eq.~\eqref{QEFTtree}, note that
\beqa
0 &=& \frac{\delta}{\delta\phi}\left( \frac{\delta\L_\text{UV}}{\delta\Phi}\bigl[\hat\Phi_\c[\phi], \phi\bigr] \right)
= \frac{\delta^2\L_\text{UV}}{\delta\phi\delta\Phi}\bigl[\hat\Phi_\c[\phi], \phi\bigr] +\frac{\delta\hat\Phi_\c[\phi]}{\delta\phi} \frac{\delta^2\L_\text{UV}}{\delta\Phi^2}\bigl[\hat\Phi_\c[\phi], \phi\bigr] \CR
&=& -X_{LH}\bigl[\hat\Phi_\c[\phi], \phi\bigr] -\frac{\delta\hat\Phi_\c[\phi]}{\delta\phi} \Delta_H\bigl[\hat\Phi_\c[\phi], \phi\bigr] \CR
&\Rightarrow& \; \frac{\delta\hat\Phi_\c[\phi]}{\delta\phi} = -X_{LH} \Delta_H^{-1} \bigl[\hat\Phi_\c[\phi], \phi\bigr] 
\to -X_{LH} \hat\Delta_H^{-1} \bigl[\hat\Phi_\c[\phi], \phi\bigr] \,,
\eeqan
where $\hat\Delta_H^{-1}$ is the local operator expansion of $\Delta_H^{-1}$. We therefore obtain
\beqa
-\frac{i}{2} \log\det\Q_\text{EFT}^\text{tree} [\phi] &=& -\frac{i}{2} \log\det \left( \Delta_L  -X_{LH} \hat\Delta_H^{-1} X_{HL} \right) \CR
&=& -\frac{i}{2} \log\det\Delta_L -\frac{i}{2} \log\det \left( \identity -\Delta_L^{-1}X_{LH} \hat\Delta_H^{-1} X_{HL} \right) \CR
&=& -\frac{i}{2} \log\det\Delta_L -\frac{i}{2} \log\det \left( \identity -\hat\Delta_H^{-1} X_{HL} \Delta_L^{-1}X_{LH} \right) \CR
&=& -\frac{i}{2} \log\det\Delta_L +\frac{i}{2} \log\det\hat\Delta_H -\frac{i}{2} \log\det \left( \hat\Delta_H -X_{HL} \Delta_L^{-1}X_{LH} \right) \,, \CR
\eeqan
with the arguments $\bigl[\hat\Phi_\c[\phi], \phi\bigr]$ implicit.
Adding this equation to Eq.~\eqref{logdetQUV}, we finally obtain, according to Eq.~\eqref{SEFTloop},
\beqa
\int d^dx\, \L_\text{EFT}^\text{1-loop}[\phi] &=& \frac{i}{2} \Bigl(\, \log\det \bigl( \Delta_H -X_{HL} \Delta_L^{-1}X_{LH} \bigr) -\log\det \bigl( \hat\Delta_H -X_{HL} \Delta_L^{-1}X_{LH} \bigr) \Bigr) \CR
&& +\frac{i}{2} \log\det\hat\Delta_H \,,
\eeqa{SEFTloopfinal}
where again, the arguments $\bigl[\hat\Phi_\c[\phi], \phi\bigr]$ are implicit. As expected, $\log\det\Delta_L$ which comes from pure light loops cancels between the two terms.

\subsection{Hard vs.\ soft}

The formula obtained above for one-loop matching using functional methods, Eq.~\eqref{SEFTloopfinal}, is quite abstract. To make use of it, a key observation, as emphasized in~\cite{FPR}, is that with dimensional regularization (which we adopt, together with the $\overline{\text{MS}}$ scheme, throughout this paper), each ``$\log\det$'' can be separated into ``hard'' and ``soft'' region contributions, namely
\beq
\log\det X = \left. \log\det X \right|_\text{hard} +\left. \log\det X \right|_\text{soft}.
\eeq{EBR}
What ``hard'' and ``soft'' mean is the following: for the ``loop integrals'' that appear in the computation of $\log\det X$, which involve heavy and light particle masses $\{M_i\}$, $\{m_{i'}\}$, and a ``loop momentum'' (i.e.\ integration variable) $q$,
\begin{itemize}
\item the hard region contribution is obtained by first expanding the integrand for $|q^2|\sim M_i^2 \gg |m_{i'}^2|$, 
and then performing the integration over the {\it full} momentum space;
\item the soft region contribution is obtained by first expanding the integrand for $|q^2|\sim |m_{i'}^2| \ll M_i^2$, and then performing the integration over the {\it full} momentum space.
\end{itemize}
The nontrivial identity~\eqref{EBR} is known as the method of expansion by regions, which has been well-known in Feynman diagrammatic multi-loop calculations; see e.g.~\cite{EBR-BenekeS,EBR-Smirnov,EBR-Jantzen}. As a simple one-loop example, consider the following {\it IR- and UV-finite} integral (in $d=4-\epsilon$ dimensions):
\beqa
\int\frac{d^dq}{(2\pi)^d}\, \frac{1}{(q^2-M^2)(q^2-m^2)^2} &=& \lf \left[ \frac{1}{M^2-m^2} \left( 1 -\log\frac{M^2}{m^2} \right) -\frac{m^2}{(M^2-m^2)^2} \log\frac{M^2}{m^2} \right] \CR
&=& \lf\frac{1}{M^2} \left( 1 -\log\frac{M^2}{m^2} \right) +\O (M^{-4}) \,.
\eeqan
The hard and soft regions yield {\it IR- and UV-divergent} integrals, respectively:
\newpage
\bseq
\beqa
\left. \int\frac{d^dq}{(2\pi)^d}\, \frac{1}{(q^2-M^2)(q^2-m^2)^2} \right|_\text{hard} &=& \int\frac{d^dq}{(2\pi)^d}\, \frac{1}{(q^2-M^2)q^4} (1+\dots) \CR
&=& \lf \frac{1}{M^2} \left( \frac{2}{\bar\epsilon} +1 -\log\frac{M^2}{\mu^2} \right) +\O (M^{-4}) , \CR\\
\left. \int\frac{d^dq}{(2\pi)^d}\, \frac{1}{(q^2-M^2)(q^2-m^2)^2} \right|_\text{soft} &=& \int\frac{d^dq}{(2\pi)^d} \left[ -\frac{1}{M^2} \frac{1}{(q^2-m^2)^2} +\dots \right] \CR
&=& \lf \frac{1}{M^2} \left( -\frac{2}{\bar\epsilon} +\log\frac{m^2}{\mu^2} \right) +\O (M^{-4}) ,
\eeqan
\eseqn
where $\frac{2}{\bar\epsilon} \equiv \frac{2}{\epsilon} -\gamma +\log 4\pi$ with $\epsilon = 4-d$. However, the $\frac{1}{\epsilon}$ singularities cancel when the two equations are added, and the finite result of the original integral is reproduced.

Now we can simplify Eq.~\eqref{SEFTloopfinal}. The crucial statements are
\bseq
\beqa
&& \log\det \bigl( \hat\Delta_H -X_{HL} \Delta_L^{-1}X_{LH} \bigr) = \left. \log\det \bigl( \Delta_H -X_{HL} \Delta_L^{-1}X_{LH} \bigr) \right|_\text{soft}, \\
&& \log\det\hat\Delta_H = \left. \log\det\Delta_H \right|_\text{soft} = 0 \,.
\eeqan
\eseq{logdetsoft}
It is not hard to understand that replacing $\Delta_H$ by $\hat\Delta_H$ singles out the soft part, because $M_i$ dependence comes only from $\Delta_H$, and a local operator expansion corresponds to the limit $M_i\to\infty$. On the other hand, $\left. \log\det\Delta_H \right|_\text{soft}$ vanishes because for pure heavy loops, expanding in the soft region gives rise to scaleless integrals. Combining Eqs.~\eqref{SEFTloopfinal}, \eqref{EBR} and~\eqref{logdetsoft}, we finally arrive at the following formula,
\beqa
\int d^dx\, \L_\text{EFT}^\text{1-loop}[\phi] &=& \frac{i}{2} \left. \log\det \bigl( \Delta_H -X_{HL} \Delta_L^{-1}X_{LH} \bigr) \right|_\text{hard} \CR
&=& \frac{i}{2} \left. \Tr\,\log \bigl( \Delta_H -X_{HL} \Delta_L^{-1}X_{LH} \bigr) \right|_\text{hard} .
\eeqa{SEFTloopTr}

\subsection{Evaluating the functional trace}
\label{sec:evaluateTr}

The initial steps of evaluating the functional trace \eqref{SEFTloopTr} are standard, which we reproduce here for the sake of completeness. Recall that entries of the infinite-dimensional matrix $\Delta_H -X_{HL} \Delta_L^{-1}X_{LH}$, which we shall call $\Delta$ to simplify notation, are labeled by spacetime indices $x$ (momentum indices $q$) when the UV theory Lagrangian is written in position (momentum) space, i.e.\ in terms of $\Phi(x)$, $\phi(x)$ ($\tilde\Phi(q)$, $\tilde\phi(q)$), plus possible internal indices. $\Delta$ contains $x$ and $i\partial_x$ in position space, which become operators $\hat x$ and $\hat p$ in general. We evaluate its trace using the momentum eigenstate basis, and follow standard manipulations familiar from quantum mechanics,
\beqa
\Tr\, \Delta(\hat x, \hat p) &=& \int\frac{d^dq}{(2\pi)^d}\, \langle q |\, \tr \,\Delta(\hat x, \hat p) \,| q \rangle 
= \int d^dx \int\frac{d^dq}{(2\pi)^d}\, \langle q |\, x\rangle \langle x | \, \tr \,\Delta(\hat x, \hat p)  \,| q \rangle \CR
&=& \int d^dx \int\frac{d^dq}{(2\pi)^d}\, e^{iq\cdot x} \,\tr \,\Delta(x, i\partial_x) \, e^{-iq\cdot x} 
= \int d^dx \int\frac{d^dq}{(2\pi)^d}\, \tr \,\Delta(x, i\partial_x+q) \CR
&=& \int d^dx \int\frac{d^dq}{(2\pi)^d}\, \tr \,\Delta(x, i\partial_x-q) \,,
\eeqan
where ``$\tr$'' is over internal indices only, and we have used $\langle x | q \rangle = e^{-iq\cdot x}$. The last equality follows from a conventional change of integration variable $q\to -q$. As a result, 
\beq
\L_\text{EFT}^\text{1-loop}[\phi] = \frac{i}{2} \left. \int\frac{d^dq}{(2\pi)^d}\, \tr\log \bigl( \Delta_H -X_{HL} \Delta_L^{-1}X_{LH} \bigr)_{P \to P-q} \right|_\text{hard} .
\eeq{LEFTloop}

At this point, there is one additional transformation that can be made~\cite{Gaillard85,Cheyette87,HLM14,DEQY}, but is optional. The idea is to put all covariant derivatives $P_\mu$ into commutators, e.g.\ $[P_\mu, P_\nu]$, $[P_\mu, X(x)]$, by sandwiching the $\tr\log$ between $e^{P\cdot\partial_q}$ and $e^{-P\cdot\partial_q}$ (which trivially become 1's when acting on identities on both sides) and using the Baker-Campbell-Hausdorff (BCH) formula. This transformation is convenient in the sense that all intermediate steps from here on will involve $P_\mu$'s only through commutators, as does the final result\footnote{Recall that $P_\mu$ as a operator acts on everything to its right, so e.g.\ $iD_\mu\phi$\,'s in the final result for $\L_\text{EFT}$ really mean $[P_\mu, \phi]$. On the other hand, gauge field strengths can be written as $[P_\mu, P_\nu]$ up to normalization.}. But meanwhile, it makes the computation more tedious because of a plethora of terms resulting from applying the BCH formula. This is especially true when the quadratic operator $\Q_\text{UV}$ contains open covariant derivatives, namely $P_\mu$'s acting openly to the right as opposed to appearing in commutators, in addition to those from kinetic terms. Another disadvantage is that with the introduction of $\partial_q$ which does not commute with $q$, the logarithm cannot be expanded in a simple way due to the fact that $\log(AB) \ne \log A +\log B$ when $[A,B]\ne0$\;\footnote{Recall that ``$\tr$'' is over internal indices only, so $\tr\, [\partial_q, f(q)] \ne 0$. Also, $\int\frac{d^dq}{(2\pi)^d}\, [\partial_q, f(q)] = \int\frac{d^dq}{(2\pi)^d}\, f'(q)$ may not vanish due to UV divergences.}. As a way out, an auxiliary integral is introduced in~\cite{HLM14,DEQY}, which nevertheless complicates the integrations to be done. Therefore, we choose to follow~\cite{HLM16,FPR} and proceed without making this additional transformation.

\subsection{Covariant derivative expansion (CDE)}
\label{sec:CDE}

The next step is to perform a CDE, i.e.\ to make an expansion in power series of $P_\mu$ while keeping $P_\mu$ intact (as opposed to separating it into $i\partial_\mu$ and $gA_\mu$). Suppose, quite generally,
\beq
\Delta_H = -P^2 +M^2 +X_H \,,
\eeq{DeltaH}
where
\beq
M = \text{diag}\,(M_1, M_2,\dots)
\eeqn
is the mass matrix of the heavy field multiplet $\Phi$\,\footnote{It is always possible to simultaneously diagonalize the $P_\mu$ and $M$ matrices, since mass mixing can only happen among fields with identical gauge quantum numbers, as far as unbroken gauge symmetries are concerned. On the other hand, if the UV theory is written in the broken phase of a spontaneously broken gauge symmetry, there could also be mass mixing induced by spontaneous symmetry breaking. In that case, gauge fields associated with the broken symmetries would not appear in $P_\mu$ in the first place, so the diagonalization is still possible.}. In general, $X_H$ may take the form
\beq
X_H [\Phi, \phi, P_\mu] = U_H[\Phi, \phi] + P_\mu Z_H^\mu[\Phi, \phi] + Z_H^{\dagger\mu}[\Phi, \phi] P_\mu + P_\mu P_\nu Z_H^{\mu\nu}[\Phi, \phi] + Z_H^{\dagger\mu\nu}[\Phi, \phi] P_\nu P^\mu +\dots
\eeq{XHgeneral}
In the hard region, the logarithm in Eq.~\eqref{LEFTloop} can be expanded as follows:
\beqa
&& \log \bigl( \Delta_H -X_{HL} \Delta_L^{-1}X_{LH} \bigr)_{P \to P-q} 
= \log \bigl( -q^2 +M^2 +2q\cdot P -P^2 +X_H -X_{HL} \Delta_L^{-1}X_{LH} \bigr) \CR
&&\quad = \log (-q^2 +M^2) +\log \Bigl[ 1 -\bigl(q^2-M^2\bigr)^{-1} \bigl(2q\cdot P -P^2 +X_H -X_{HL} \Delta_L^{-1}X_{LH} \bigr) \Bigr] \CR
&&\quad = \log (-q^2 +M^2) -\sum_{n=1}^{\infty} \frac{1}{n} \Bigl[ \bigl(q^2-M^2\bigr)^{-1} \bigl(2q\cdot P -P^2 +X_H -X_{HL} \Delta_L^{-1}X_{LH} \bigr) \Bigr]^n ,
\eeqan
where the substitution $P \to P-q$ is assumed in $X_H$ and $X_{HL} \Delta_L^{-1}X_{LH}$. Therefore, up to an  additive constant,
\beqa
\L_\text{EFT}^\text{1-loop}[\phi] &=& -\frac{i}{2} \,\tr \sum_{n=1}^{\infty} \frac{1}{n} \int\frac{d^dq}{(2\pi)^d} \Bigl[ \bigl(q^2-M^2\bigr)^{-1} \CR
&&\qquad \left.  \bigl(2q\cdot P -P^2 +\left. X_H\right|_{P\to P-q} -\left. X_{HL} \Delta_L^{-1}X_{LH}\right|_{P\to P-q} \bigr) \Bigr]^n \right|_\text{hard} .
\eeqa{LEFTloopfinal}
As before, $X_{H,HL,LH}$ and $\Delta_L$ have arguments $\bigl[\hat\Phi_\c[\phi], \phi\bigr]$. Eq.~\eqref{LEFTloopfinal} holds for the special case of real scalars but can be straightforwardly generalized. It will be our starting point for deriving a covariant diagrammatic formulation of one-loop matching in the next section.

\section{Covariant diagrams}
\label{sec:covariantdiagrams}

\subsection{Pure heavy loops}
\label{sec:cd-heavy}

We first look at the simplest case, where the following three restrictions are satisfied:
\begin{itemize}
\item $X_{HL} = X_{LH} = 0$, i.e.\ no mixed heavy-light contributions to one-loop matching. This already covers a broad class of UV models where heavy fields do not couple linearly to light degrees of freedom and $\Phi_\c=0$.
\item $X_H$ does not contain open covariant derivatives, i.e.\ $X_H=U_H$; see Eq.~\eqref{XHgeneral}.
\item The field multiplet $\Phi$ contains only bosonic fields.
\end{itemize}
After developing the basics of covariant diagrams for this simplest case, we will lift the above restrictions one by one in the next three subsections.

For real scalars, we can directly use Eq.~\eqref{LEFTloopfinal}, which becomes, under the above restrictions,
\beq
\L_\text{EFT}^\text{1-loop}[\phi] = -\frac{i}{2} \,\tr \sum_{n=1}^{\infty} \frac{1}{n} \int\frac{d^dq}{(2\pi)^d} \Bigl[ \bigl(q^2-M^2\bigr)^{-1}  \bigl(2q\cdot P -P^2 +U_H \bigr) \Bigr]^n .
\eeq{LEFTloopUH}
Note that, with no light masses involved, the hard part of the integral is trivially equal to the original integral. A key observation is that each term in the sum in Eq.~\eqref{LEFTloopUH} factorizes into a loop integral over $q$ and a trace involving $P_\mu$ and $U_H$ that gives rise to effective operators. The nonvanishing loop integrals involved have the generic form
\beq
\int\frac{d^dq}{(2\pi)^d} \frac{q^{\mu_1}\cdots q^{\mu_{2n_c}}}{(q^2-M_i^2)^{n_i}(q^2-M_j^2)^{n_j}\cdots}
\,\equiv\, g^{\mu_1\dots\mu_{2n_c}} \,\I[q^{2n_c}]_{ij\dots}^{n_i n_j\dots} ,
\eeq{MIdefH}
where $g^{\mu_1\dots\mu_{2n_c}}$ is the completely symmetric tensor, e.g.\ $g^{\mu\nu\rho\sigma}=g^{\mu\nu}g^{\rho\sigma} +g^{\mu\rho}g^{\nu\sigma} +g^{\mu\sigma}g^{\nu\rho}$. Eq.~\eqref{MIdefH} defines the {\it master integrals} $\,\I[q^{2n_c}]_{ij\dots}^{n_i n_j\dots}$. We use the symbol ``$\,\I\,$'' to distinguish from the master integrals in~\cite{DEQY} which are denoted by ``$\,\mathrm{I}\,$'' and involve an extra auxiliary integral. Some useful master integrals are summarized in Appendix~\ref{app:MI}.

Eq.~\eqref{LEFTloopUH} has a straightforward diagrammatic representation as a sum of one-loop diagrams with propagators $\frac{1}{q^2-M^2}$ and vertex insertions $2q\cdot P$, $-P^2$ and $U_H$. The loop integral can be read off from a diagram simply by counting the numbers of propagators (for each species) and $2q\cdot P$ vertices. As a result of evaluating the loop integral as in Eq.~\eqref{MIdefH}, various terms in $g^{\mu_1\dots\mu_{2n_c}}$ Lorentz-contract the $P_\mu$'s from $2q\cdot P$ insertions in different ways, and all possibilities are summed over. We can keep track of such contractions by connecting two $2q\cdot P$ vertices by a dotted line. The above procedure can be easily understood with an example,
\vspace{4pt}
\beq
\begin{gathered}
\begin{fmffile}{ex-heavy}
\begin{fmfgraph*}(40,40)
\fmfsurround{vP2,vU1,vP1,vU2}
\fmf{plain,left=0.4,label=$i$,l.d=3pt}{vU2,vP1,vU1}
\fmf{plain,left=0.4,label=$j$,l.d=3pt}{vU1,vP2,vU2}
\fmfv{decor.shape=circle,decor.filled=full,decor.size=3thick}{vP1,vP2}
\fmfv{decor.shape=circle,decor.filled=empty,decor.size=3thick}{vU1,vU2}
\fmf{dots,width=thick}{vP1,vP2}
\end{fmfgraph*}
\end{fmffile}
\end{gathered}
\;\; \,=\, -\frac{i}{2}\, \frac{1}{2} \, \I[q^2]_{ij}^{22} \,\tr\bigl( (2P^\mu) \,U_{H\,ij} \,(2P_\mu) \,U_{H\,ji} \bigr),
\eeq{HloopEx}
where the diagram is read clockwise, and filled and empty circles represent $2q\cdot P$ and $U_H$ insertions, respectively (recall that $P_\mu$ is diagonal in the field multiplet space and hence does not change the propagator label). Eq.~\eqref{HloopEx} represents a term in the expansion~\eqref{LEFTloopUH}. The only element in Eq.~\eqref{HloopEx} which we have not discussed is the symmetry factor $\frac{1}{2}$, coming from $\frac{1}{n}=\frac{1}{4}$ (four propagators) multiplied by 2 (two identical contributions $\tr( (2P^\mu) \,U_{H\,ij} \,(2P_\mu) \,U_{H\,ji})$ and $\tr(U_{H\,ij} \,(2P_\mu) \,U_{H\,ji}\, (2P^\mu) )$). An easy way to find this symmetry factor is to note the $\mathbb{Z}_2$ symmetry of the diagram under rotation. It is not hard to show that in general, the presence of a $\mathbb{Z}_S$ symmetry of a diagram under rotation indicates a symmetry factor $\frac{1}{S}$. We see that our diagrammatic formulation automatically collects terms from the CDE containing equivalent operator traces, and thus makes finding such factors a trivial task.

One can draw all possible diagrams like the one in Eq.~\eqref{HloopEx} to keep track of all terms in the expansion~\eqref{LEFTloopUH} up to a certain order. These terms, which contain operator structures with open covariant derivatives, would eventually organize into independent operators with covariant derivatives appearing only in commutators (recall that the final result can always be written in a form that involves $P_\mu$'s only via commutators). For example, we could enumerate all diagrams containing two $P_\mu$'s and two $U_H$'s, which include the one in Eq.~\eqref{HloopEx}, a second diagram with adjacent $P_\mu$ contractions, and a third diagram with a $-P^2$ insertion. The latter two diagrams represent
\beq
-\frac{i}{2}\, \I[q^2]_{ij}^{31} \,\tr\bigl( (2P^\mu)  \,(2P_\mu) \,U_{H\,ij}\,U_{H\,ji} \bigr)
-\frac{i}{2}\, \I_{ij}^{21} \,\tr\bigl( (-P^2)  \,U_{H\,ij}\,U_{H\,ji} \bigr) \,,
\eeq{HloopExR}
with no symmetry factors. Here and in the following, we abbreviate $\I[q^0]_{ij\dots}^{n_i n_j\dots}$ as $\I_{ij\dots}^{n_i n_j\dots}$. Adding up the three terms in Eqs.~\eqref{HloopEx} and~\eqref{HloopExR}, and making use of the identity $\I_{ij}^{21} = 2\,\I[q^2]_{ij}^{22} +4\,\I[q^2]_{ij}^{31}$,\footnote{This identity can be easily proved by writing $\I[q^2]_{ij}^{22} = \frac{1}{4} (\I_{ij}^{21} +M_j^2\,\I_{ij}^{22})$, $\I[q^2]_{ij}^{31} = \frac{1}{4} (\I_{ij}^{21} +M_i^2\,\I_{ij}^{31})$ and using the formulas in Appendix~\ref{app:MI}.} we arrive at one single operator of the desired form (without open covariant derivatives),
\beqa
&& -i\,\Bigl\{\I[q^2]_{ij}^{22}\,\tr\bigl( P^\mu \,U_{H\,ij} \,P_\mu \,U_{H\,ji} \bigr) + \Bigl(2\, \I[q^2]_{ij}^{31} -\frac{1}{2}\,\I_{ij}^{21} \Bigr) \tr\bigl( P^2 \,U_{H\,ij}\,U_{H\,ji} \bigr)\Bigr\} \CR
&& = -i\,\I[q^2]_{ij}^{22}\,\tr\bigl( P^\mu \,U_{H\,ij} \,P_\mu \,U_{H\,ji} -P^2 \,U_{H\,ij}\,U_{H\,ji} \bigr) \CR
&& = -\frac{i}{2}\,\I[q^2]_{ij}^{22}\,\tr\bigl( 2\,P^\mu \,U_{H\,ij} \,P_\mu \,U_{H\,ji} -P^2 \,U_{H\,ij}\,U_{H\,ji} -P^2 \,U_{H\,ji}\,U_{H\,ij} \bigr) \CR
&& = -\frac{i}{2}\,\I[q^2]_{ij}^{22}\,\tr\bigl([P^\mu , U_{H\,ij}] [P_\mu , U_{H\,ji}]\bigr) \,.
\eeqa{HloopExSum}

Alternatively, we could have anticipated the form of the final result before enumerating the diagrams --- there is only one independent operator involving two $P_\mu$'s and two $U_H$'s, namely $\tr([P^\mu, U_H][P_\mu, U_H])$, so we know all relevant terms in the CDE must add up to
\beqa
&& c_{ij}\, \tr\bigl([P^\mu , U_{H\,ij}] [P_\mu , U_{H\,ji}]\bigr) \CR
&& = 2 c_{ij}\,\tr\bigl(P^\mu \, U_{H\,ij}\, P_\mu \, U_{H\,ji}\bigr)
-(c_{ij}+c_{ji})\,\tr\bigl( P^2 \,U_{H\,ij}\,U_{H\,ji} \bigr)\,.
\eeqa{PUPUexpand}
To determine the coefficient $c_{ij}$, it is actually not necessary to compute all three diagrams as we did above. Since the last two diagrams only contribute to the second term of Eq.~\eqref{PUPUexpand}, we could have obtained $c_{ij}$ without computing them, simply by comparing Eq.~\eqref{HloopEx} to the first term of Eq.~\eqref{PUPUexpand}. The result would be $c_{ij} = -\frac{i}{2}\,\I[q^2]_{ij}^{22}$, in agreement with Eq.~\eqref{HloopExSum}.

In fact, it is generally true that to determine the coefficients of all independent effective operators in the final result, it is sufficient to compute just a subset of all possible diagrams. This is simply because when $P_\mu$'s are involved, the number of independent structures one can write down with open covariant derivatives (two for the example above, $\tr(P^\mu \, U_{H\,ij}\, P_\mu \, U_{H\,ji})$ and $\tr( P^2 \,U_{H\,ij}\,U_{H\,ji})$) is greater than the number of independent operators with $P_\mu$'s appearing only in commutators (only one, $\tr([P^\mu , U_{H\,ij}] [P_\mu , U_{H\,ji}])$). While we do not have an algorithm to determine, in full generality, the minimal set of diagrams to be computed, we have discovered a useful prescription that greatly reduces the workload: all diagrams with either $-P^2$ insertions or {\it adjacent} $P_\mu$ contractions, namely those that yield $\tr(\dots P^2 \dots)$, can be dropped. In the example above, this prescription corresponds to not explicitly writing down and computing Eq.~\eqref{HloopExR} which, as we have seen, only provides redundant information on $c_{ij}$. In fact, in many of the examples in Section~\ref{sec:example}, this prescription will reduce the diagrams to be computed to a minimal set, in the sense that we will have just enough information to determine all the operator coefficients in the final results.

The above discussion can also be applied to other types of bosonic fields. A complex scalar is equivalent to a multiplet of two real scalars, e.g.\ its real and imaginary parts. In practice it is often more convenient to use a multiplet consisting of the complex scalar itself and an appropriately-defined complex conjugate field. We will see explicitly how this is done in the next section. For vector bosons, with the addition of the $R_\xi$ gauge fixing term, the UV Lagrangian contains the following terms quadratic in the quantum fluctuations,
\beq
-\frac{1}{2}\, V_\alpha^{\prime a}\, \Bigl\{ \bigl(-g^{\alpha\beta}\bigr) \bigl(-(P^2)_{ab} +M_V^2\delta_{ab}\bigr) -\Bigl(1-\frac{1}{\xi}\Bigr) \bigl(P^\alpha P^\beta\bigr)_{ab} +U_{H\,ab}^{\alpha\beta}
\Bigr\}\, V_\beta^{\prime b}
\eeqn
It is convenient to use the Feynman gauge $\xi=1$, where $\Delta_H$ takes the form of Eq.~\eqref{DeltaH} as in the scalar case, so that the same procedure of using covariant diagrams can be followed\,\footnote{The associated Goldstone boson and ghost fields can also be treated in the same way as scalars, except that ghost loops come with a factor of $(-1)$ due to the Grassmannian Gaussian integral.}. The only nontrivial extension is that vector boson fields carry Lorentz indices, which are regarded as additional internal indices and should be contracted with $-g_{\alpha\beta}$ (note minus sign!) when computing traces. 
This can be seen as follows,
\beqa
&&
\log\bigl\{ (-g^{\alpha\beta}) (-P^2 +M_V^2) +U_H^{\alpha\beta}\bigr\}_{P\to P-q}
= \log\bigl\{ (-g^{\alpha\beta}) (-q^2 +M_V^2 +2q\cdot P -P^2) +U_H^{\alpha\beta}\bigr\} 
\CR &&
= \log\bigl\{ (-g^{\alpha\gamma})(-q^2 +M_V^2)\bigr\} 
\CR &&\quad
+\log\bigl\{\delta_\gamma^{\;\;\beta} - (-g_{\gamma\delta})(q^2-M_V^2)^{-1} \bigl( (-g^{\delta\beta}) (2q\cdot P-P^2) +U_H^{\delta\beta} \bigr) \bigr\},
\eeqan
with internal indices $a,b$ dropped for simplicity. As an example, when only vector fields are considered, the trace in Eq.~\eqref{HloopEx} should be understood as
\beqa
&& \tr\, (P^\mu \,U_{H\,ij} \,P_\mu \,U_{H\,ji}) = \CR
&&\qquad (-g_{\alpha_1\beta_1}) (-g_{\alpha_2\beta_2}) (-g_{\alpha_3\beta_3}) (-g_{\alpha_4\beta_4}) \,\tr \bigl( (-g^{\beta_4\alpha_1} P^\mu) (U_{H\,ij}^{\beta_1\alpha_2}) (-g^{\beta_2\alpha_3} P_\mu) (U_{H\,ji}^{\beta_3\alpha_4}) \bigr)\,,\qquad
\eeqa{VtrEx}
with all Lorentz indices written out explicitly. The ``tr'' in the second line of Eq.~\eqref{VtrEx} then indicates a trace over the remaining internal indices.

A summary of the building blocks of covariant diagrams and the operator structures they represent in the restricted case discussed in this subsection can be found in Table~\ref{tab:heavy} of Section~\ref{sec:cd-sum}.

\subsection{Mixed heavy-light loops}
\label{sec:cd-mixed}

Next, we allow $X_{HL,LH}$ to be nonzero, while still assuming the absence of open covariant derivatives. Specifically, we consider
\bseq
\beqa
&& X_{HL} = U_{HL}\,,\quad X_{LH} = U_{LH} \,, \\
&& \Delta_L = -P^2 +m^2 +X_L = -P^2 +m^2 +U_L \,.
\eeqan
\eseqn
where
\beq
m = \text{diag}\,(m_1, m_2, \dots)
\eeqn
is the mass matrix of the light field multiplet $\phi$. The additional piece in Eq.~\eqref{LEFTloopfinal} becomes
%
\beqa
-\left. X_{HL} \Delta_L^{-1}X_{LH}\right|_{P\to P-q}
&=& U_{HL} (q^2 -m^2 -2q\cdot P +P^2 -U_L)^{-1} U_{LH} \CR
&\to& U_{HL} \sum_{n=0}^{\infty}\Bigl[ \frac{1}{q^2} (m^2 +2q\cdot P -P^2 +U_L) \Bigr]^n \frac{1}{q^2} U_{LH} \,.
\eeqa{MHLexpand}
The expansion above is suitable in the hard region where $|q^2|\gg |m^2|$. Eq.~\eqref{MHLexpand} as a whole can be thought of as a new type of insertion in the heavy loop, in addition to $2q\cdot P$, $-P^2$, $U_H$ considered in the previous subsection. Equivalently, the expansion of Eq.~\eqref{MHLexpand} instructs us to draw one-loop diagrams involving both heavy and light propagators which represent $\frac{1}{q^2-M^2}$ and $\frac{1}{q^2}$, respectively. $2q\cdot P$, $-P^2$ and $U_H$ can be inserted in heavy propagators as before, while $2q\cdot P$, $-P^2$, $U_L$ and $m^2$ can be inserted in light propagators. $U_{HL}$ ($U_{LH}$) connects an incoming heavy (light) propagator and an outgoing light (heavy) propagator, when the diagrams are read clockwise. Loop integrals now have the form
\beq
\int\frac{d^dq}{(2\pi)^d} \frac{q^{\mu_1}\cdots q^{\mu_{2n_c}}}{(q^2-M_i^2)^{n_i}(q^2-M_j^2)^{n_j}\cdots (q^2)^{n_L}}
\,\equiv\, g^{\mu_1\dots\mu_{2n_c}} \,\I[q^{2n_c}]_{ij\dots 0}^{n_i n_j\dots n_L} .
\eeq{MIdefMHL}
Eq.~\eqref{MIdefMHL} defines an extended set of master integrals $\,\I[q^{2n_c}]_{ij\dots 0}^{n_i n_j\dots n_L}$, some of which are summarized in Appendix~\ref{app:MI}. Note that these loop integrals do not depend on light particle masses because the latter are treated as vertex insertions. This implies, in particular, that in the case of massless particles, there is no need to keep $m^2$ nonzero as an IR regulator.

As a simple example, we show a mixed heavy-light version of Eq.~\eqref{HloopEx},
\vspace{4pt}
\beq
\begin{gathered}
\begin{fmffile}{ex-mixed}
\begin{fmfgraph*}(40,40)
\fmfsurround{vP2,vUHL,vP1,vULH}
\fmf{plain,left=0.4,label=$i$,l.d=3pt}{vULH,vP1,vUHL}
\fmf{dashes,left=0.4,label=$i'$,l.d=3pt}{vUHL,vP2,vULH}
\fmfv{decor.shape=circle,decor.filled=full,decor.size=3thick}{vP1,vP2}
\fmfv{decor.shape=circle,decor.filled=empty,decor.size=3thick}{vUHL,vULH}
\fmf{dots,width=thick}{vP1,vP2}
\end{fmfgraph*}
\end{fmffile}
\end{gathered}
\;\; \,=\, -\frac{i}{2} \, \I[q^2]_{i0}^{22} \,\tr\bigl( (2P^\mu) \,U_{HL\,ii'} \,(2P_\mu) \,U_{LH\,i'i}\bigr) \,,
\eeqn
where light propagators are represented by dashed lines, and labeled by primed indices. Note the absence of a nontrivial symmetry factor in this case.
The additional building blocks of covariant diagrams discussed in this subsection are summarized in Table~\ref{tab:mixed} of Section~\ref{sec:cd-sum}. 

\subsection{Open covariant derivatives}
\label{sec:cd-ocd}

In addition to $U_{H,HL,LH,L}$ considered above, the $X_{H, HL, LH, L}$ matrices may also contain terms involving open covariant derivatives; see Eq.~\eqref{XHgeneral}. These terms are slightly different from the $U$ terms because they are modified by the substitution $P\to P-q$. For example, terms in Eq.~\eqref{XHgeneral} with one open covariant derivative become
\beq
P_\mu Z_H^\mu + Z_H^{\dagger\mu} P_\mu \to P_\mu Z_H^\mu + Z_H^{\dagger\mu} P_\mu -q_\mu Z_H^\mu - Z_H^{\dagger\mu} q_\mu \,,
\eeq{PZshift}
resulting in two types of vertex insertions: $P_\mu Z_H^\mu$ and $Z_H^{\dagger\mu} P_\mu$ are just like $U$ insertions, while $-q_\mu Z_H^\mu$ and $-Z_H^{\dagger\mu} q_\mu$ are similar to $2q\cdot P$ insertions. In the latter case, the $q_\mu$'s involved are part of the loop integral, which gives rise to $g^{\mu_1\dots\mu_{2n_c}}$. Lorentz contractions are thus possible not only between $P_\mu$'s from $2q\cdot P$ insertions, but also $Z_H^{(\dagger)\mu}$'s from $-q_\mu Z_H^\mu$, $-Z_H^{\dagger\mu} q_\mu$ insertions. We shall use the same symbol for the two types of $Z^{(\dagger)}$ insertions --- they are distinguished by whether or not a contraction is indicated (by a dotted line as before). As a simple example,
\bseq
\beqa
\begin{gathered}
\begin{fmffile}{ex-ocd-a}
\begin{fmfgraph*}(40,40)
\fmfsurround{vZHd,vZH}
\fmf{plain,left=1,label=$i$,l.d=3pt}{vZHd,vZH}
\fmf{plain,left=1,label=$j$,l.d=3pt}{vZH,vZHd}
\fmfv{decor.shape=square,decor.filled=20,decor.size=3thick}{vZH}
\fmfv{decor.shape=square,decor.filled=60,decor.size=3thick}{vZHd}
\end{fmfgraph*}
\end{fmffile}
\end{gathered}
\;\; &=& -\frac{i}{2} \,\I_{ij}^{11}\,\tr (P_\mu\, Z_{H\,ij}^\mu\, Z_{H\,ji}^{\dagger\nu}\, P_\nu)\,, \\[24pt]
\begin{gathered}
\begin{fmffile}{ex-ocd-b}
\begin{fmfgraph*}(40,40)
\fmfsurround{vZHd,vZH}
\fmf{plain,left=1,label=$i$,l.d=3pt}{vZHd,vZH}
\fmf{plain,left=1,label=$j$,l.d=3pt}{vZH,vZHd}
\fmfv{decor.shape=square,decor.filled=20,decor.size=3thick}{vZH}
\fmfv{decor.shape=square,decor.filled=60,decor.size=3thick}{vZHd}
\fmf{dots,width=thick}{vZH,vZHd}
\end{fmfgraph*}
\end{fmffile}
\end{gathered}
\;\; &=& -\frac{i}{2} \,\I[q^2]_{ij}^{11}\,\tr ( Z_{H\,ij}^\mu\, Z_{H\,\mu\,ji}^{\dagger})\,,
\eeqan
\eseqn
where light and dark gray squares represent $(P_\mu) Z_H^\mu$ and $Z_H^{\dagger\mu} (P_\mu)$ insertions, respectively. Here and in the following, ``$[q^{2n_c}]$'' is dropped when writing master integrals with $n_c=0$.

We have focused on pure heavy loops in the discussion above for concreteness, but there is no essential difference for mixed heavy-light loops, which may involve $Z_{HL,LH,L}^{(\dagger)}$. A summary of possible $Z^{(\dagger)}$ insertions (up to one-open-covariant-derivative terms) can be found in Table~\ref{tab:ocd} of Section~\ref{sec:cd-sum}. Also, it is straightforward to extend the procedure to terms in the $X$ matrices with more than one open covariant derivatives, though more complex notation may be needed to keep track of Lorentz contractions.

\subsection{Loops with fermions}
\label{sec:cd-ferm}

Up to now we have considered loops with bosonic fields only. Fermionic fields have a different form of quadratic operator $\Q_\text{UV}$, with e.g.\ $-\Psl +M$ in the case of Dirac fermions in place of $-P^2+M^2$. There are at least two approaches one can follow. One is to square the quadratic operator to match the general form in the bosonic case. To give an example for illustration, suppose
\beq
\L_\text{UV} [\Psi, \phi] = \L_0 [\phi] + \bar\Psi (\Psl-M-X_H[\phi])\Psi \,,
\eeqn
where $\phi$ denotes collectively light fields, and $\Psi$ is a heavy Dirac fermion. We assume $X_H=X_{H,e}+X_{H,o}$ with $X_{H,e}$ ($X_{H,o}$) containing terms with even (odd) numbers of gamma matrices. There is no mixed heavy-light contribution to matching in this case, so
\beq
S_\text{EFT}^\text{1-loop} = -i\,\Tr\log (\Psl-M-X_H).
\eeqn
Note the different overall sign compared with bosonic case, due to the Grassmannian nature of the $\Psi$ field. Using the fact that traces of gamma matrices are invariant under changing signs of all gamma matrices, we have
\beqa
S_\text{EFT}^\text{1-loop} 
&=& -\frac{i}{2} \,\bigl[\Tr\log (\Psl-M-X_{H,e}-X_{H,o}) +\Tr\log (-\Psl-M-X_{H,e}+X_{H,o}) \bigr] \CR
&=&  -\frac{i}{2} \,\Tr\log \bigl(-\Psl^2+M^2+2MX_{H,e}+X_H(X_{H,e}-X_{H,o}) -[\Psl,X_{H,e}] +\{\Psl,X_{H,o}\}\bigr) \CR
&=&  -\frac{i}{2} \,\Tr\log \bigl(-P^2+M^2 -\frac{i}{2}\sigma^{\mu\nu}G'_{\mu\nu}+2MX_{H,e} \CR
&&\qquad\qquad\quad +X_H(X_{H,e}-X_{H,o}) -[\Psl,X_{H,e}] +\{\Psl,X_{H,o}\} \bigr)\,,\quad
\eeqa{SEFTloopferm}
where $G'_{\mu\nu} = [D_\mu, D_\nu] =-igG_{\mu\nu}$ and $\sigma^{\mu\nu} = \frac{i}{2}[\gamma^\mu, \gamma^\nu]$. The calculation then proceeds as in the bosonic case, with $-\frac{i}{2}\sigma^{\mu\nu}G'_{\mu\nu}+2MX_{H,e}+X_H(X_{H,e}-X_{H,o}) -[\Psl,X_{H,e}] +\{\Psl,X_{H,o}\}$ playing the role of $X_H$.

In this paper, however, we follow an alternative strategy so as to derive a more straightforward diagrammatic formulation of one-loop functional matching. Still using the example above and, for the moment, further assuming $X_H=U_H$ does not contain open covariant derivatives for simplicity, we repeat the steps in Sections~\ref{sec:evaluateTr} and~\ref{sec:CDE} {\it without} squaring the quadratic operator,
\beqa
\L_\text{EFT}^\text{1-loop} &=& -i\,\int\frac{d^dq}{(2\pi)^d}\, \tr\log (\Psl -\qsl -M -U_H ) \CR
&=&  -i\,\int\frac{d^dq}{(2\pi)^d}\, \tr\log (-\qsl -M) -i\,\int\frac{d^dq}{(2\pi)^d}\, \tr\log \bigl[1-(-\qsl -M)^{-1}(-\Psl +U_H ) \bigr] \CR
&=& \text{const.} +i\, \tr \sum_{n=1}^\infty \frac{1}{n} \int\frac{d^dq}{(2\pi)^d}\, \bigl[(-\qsl -M)^{-1}(-\Psl +U_H ) \bigr]^n .
\eeqa{LEFTloopferm}
This is a fermionic version of Eq.~\eqref{LEFTloopUH}, after the irrelevant constant term is dropped. The diagrammatic representation in this case involves fermionic propagators $(-\qsl -M)^{-1}$ and vertex insertions $-\Psl$ and $U_H$. The rules of drawing covariant diagrams and reading off their expressions are similar to the bosonic case, but we note the following three major differences:
\begin{itemize}
\item The prefactor has a different sign due to the fermionic Gaussian integral. It is convenient to denote the prefactor by $-ic_s$, as is common in the literature. We have seen that for real bosonic degrees of freedom, $c_s=\frac{1}{2}$, while for Dirac fermions, $c_s=-1$. In any case, $c_s$ can be easily seen from the Gaussian integral involved. For example, $c_s=-1$ for ghost fields, and $c_s=-\frac{1}{2}$ for Weyl fermions.
\item Each fermionic propagator contains two terms,
\beq
(-\qsl -M)^{-1} = \frac{-\qsl+M}{q^2-M^2} = \frac{M}{q^2-M^2} +\frac{-q_\mu\gamma^\mu}{q^2-M^2} \,.
\eeq{fermprop}
The first term is just the bosonic propagator multiplied by $M$, while the second term involves $q_\mu$ in the numerator which modifies the loop integral compared with the bosonic case. The situation is the same as that of Eq.~\eqref{PZshift} in the previous subsection. We shall continue to use dotted lines to indicate contractions among Lorentz vectors associated with $q_\mu$ (in this case $\gamma^\mu$). Our rule is to take the first or second term in Eq.~\eqref{fermprop} depending on whether the fermionic propagator is connected to a dotted line. To give an example,
\bseq
\beqa
\begin{gathered}
\begin{fmffile}{ex-ferm-a}
\begin{fmfgraph}(40,40)
\fmfsurround{fp1,vU1,fp2,vU2}
\fmf{plain,left=0.4}{vU1,fp1,vU2,fp2,vU1}
\fmfv{decor.shape=circle,decor.filled=empty,decor.size=3thick}{vU1,vU2}
\end{fmfgraph}
\end{fmffile}
\end{gathered}
\;\; &=& i\, \frac{1}{2} \, \I_i^2 \,M^2 \,\tr\,U_H^2 \,,\\[6pt]
\begin{gathered}
\begin{fmffile}{ex-ferm-b}
\begin{fmfgraph}(40,40)
\fmfsurround{fp1,vU1,fp2,vU2}
\fmf{plain,left=0.4}{vU1,fp1,vU2,fp2,vU1}
\fmfv{decor.shape=circle,decor.filled=empty,decor.size=3thick}{vU1,vU2}
\fmf{dots,width=thick}{fp1,fp2}
\end{fmfgraph}
\end{fmffile}
\end{gathered}
\;\; &=& i\, \frac{1}{2} \, \I[q^2]_i^2 \,\tr\bigl( (-\gamma^\mu) U_H (-\gamma_\mu) U_H \bigr),
\eeqan
\eseqn
where $\frac{1}{2}$ is a symmetry factor, and it is understood that $M_i=M$ in the master integrals. As before, we have used empty circles for $U_H$ insertions.
\item Covariant derivative insertions are in the form of $-\Psl$ which, unlike $2q\cdot P$, is $q$-independent and thus decouples from the loop integral. We shall continue to use filled circles to denote covariant derivative insertions in fermion propagators, but they should not be contracted (i.e.\ connected by dotted lines) with each other in this case.
\end{itemize}
With the new features discussed above taken into account, it is straightforward to generalize the procedures of the previous two subsections to incorporate mixed heavy-light loops and additional structures in the $X$ matrices in the fermionic case. Mixed bosonic-fermionic loops can also be handled --- the derivation in this case is actually very similar to that of mixed heavy-light loops. The sign of $c_s$ is determined by the propagator from which one starts reading a diagram, with no ambiguity. For example, one may have $\tr(\dots U_{BF}\dots U_{FB}\dots)$ or $\tr(\dots U_{FB}\dots U_{BF}\dots)$, depending on whether one starts reading the diagram from a bosonic (B) or fermionic (F) propagator. The values of the two traces are opposite to each other, since $U_{BF}$ and $U_{FB}$ are fermionic and anticommuting (while all $\dots$'s are bosonic), so they give the same result when multiplied by opposite spin factors.

The new ingredients for building covariant diagrams involving Dirac fermions are summarized in Table~\ref{tab:ferm} of Section~\ref{sec:cd-sum}. We further note that, as in the bosonic case discussed in Section~\ref{sec:cd-heavy}, the prescription of dropping terms involving $\tr(\dots P^\mu P_\mu\dots)$ can be adopted. These terms can arise, for example, when two fermionic propagators are contracted which are separated by two $\Psl$ insertions and one uncontracted fermionic propagator, provided that the loop integral is convergent --- this is because $\gamma^\mu\Psl\Psl\gamma_\mu = 4 P^2 +\O(\epsilon)$ where $\epsilon=4-d$.

\subsection{Summary: recipe for one-loop matching}
\label{sec:cd-sum}

\begin{table}[tbp]
\centering
\begin{tabular}{|l|c|c|}
\hline
Element of diagram & Symbol & Expression \\
\hline
heavy propagator (bosonic) &  
\begin{fmffile}{bb-pH-b}
\begin{fmfgraph*}(40,10)
\fmfleft{i1,i2}
\fmfright{o1,o2}
\fmf{plain,label=$i$,l.s=left,l.d=2pt}{i1,o1}
\end{fmfgraph*}
\end{fmffile}
& $1$ \\
$P$ insertion (bosonic, heavy) &
\begin{fmffile}{bb-vP-bh}
\begin{fmfgraph*}(32,20)
\fmfleft{i1,i2,i3}
\fmfright{o1,o2,o3}
\fmf{plain,label=$i$,l.s=left,l.d=2pt}{i2,vP}
\fmf{plain,label=$j$,l.s=left,l.d=1pt}{vP,o2}
\fmf{phantom}{i1,b,o1}
\fmf{dots,width=thick,tension=0}{b,vP}
\fmfv{decor.shape=circle,decor.filled=full,decor.size=3thick}{vP}
\end{fmfgraph*}
\end{fmffile}
& $2P_\mu \delta_{ij}$ \\
$U$ insertion (heavy-heavy) &
\begin{fmffile}{bb-vU-hh}
\begin{fmfgraph*}(40,10)
\fmfleft{i1,i2}
\fmfright{o1,o2}
\fmf{plain,label=$i$,l.s=left,l.d=2pt}{i1,vU}
\fmf{plain,label=$j$,l.s=left,l.d=1pt}{vU,o1}
\fmfv{decor.shape=circle,decor.filled=empty,decor.size=3thick}{vU}
\end{fmfgraph*}
\end{fmffile}
& $U_{H\,ij}$ \\
\hline
\end{tabular}
\caption{Building blocks of covariant diagrams for integrating out heavy bosonic fields (and fermionic fields as well if one follows the approach of Eq.~\eqref{SEFTloopferm} to square their quadratic operator), in the absence of mixed heavy-light contributions and open covariant derivatives in the $\bf X$ matrix, as derived in Section~\ref{sec:cd-heavy}. All previous universal results in the literature~\cite{HLM14,DEQY} can be easily reproduced by computing one-loop covariant diagrams built from these elements; see Section~\ref{sec:UOLEA}.
\label{tab:heavy}}
\end{table}


\begin{table}[tbp]
\centering
\begin{tabular}{|l|c|c|}
\hline
Element of diagram & Symbol & Expression \\
\hline
light propagator (bosonic) &  
\begin{fmffile}{bb-pL-b}
\begin{fmfgraph*}(40,10)
\fmfleft{i1,i2}
\fmfright{o1,o2}
\fmf{dashes,label=$i'$,l.s=left,l.d=2pt}{i1,o1}
\end{fmfgraph*}
\end{fmffile}
& $1$ \\
light mass insertion (bosonic) &
\begin{fmffile}{bb-vm-b}
\begin{fmfgraph*}(40,10)
\fmfleft{i1,i2}
\fmfright{o1,o2}
\fmf{dashes,label=$i'$,l.s=left,l.d=2pt}{i1,vm}
\fmf{dashes,label=$j'$,l.s=right,l.d=1pt}{o1,vm}
\fmfv{decor.shape=cross,decor.size=3thick}{vm}
\end{fmfgraph*}
\end{fmffile}
& $ m_{i'}^2\,\delta_{i'j'}$ \\
$P$ insertion (bosonic, light) &
\begin{fmffile}{bb-vP-bl}
\begin{fmfgraph*}(32,20)
\fmfleft{i1,i2,i3}
\fmfright{o1,o2,o3}
\fmf{dashes,label=$i'$,l.s=left,l.d=2pt}{i2,vP}
\fmf{dashes,label=$j'$,l.s=right,l.d=1pt}{o2,vP}
\fmf{phantom}{i1,b,o1}
\fmf{dots,width=thick,tension=0}{b,vP}
\fmfv{decor.shape=circle,decor.filled=full,decor.size=3thick}{vP}
\end{fmfgraph*}
\end{fmffile}
& $2P_\mu \,\delta_{i'j'}$ \\
$U$ insertion (heavy-light) &
\begin{fmffile}{bb-vU-hl}
\begin{fmfgraph*}(40,10)
\fmfleft{i1,i2}
\fmfright{o1,o2}
\fmf{plain,label=$i$,l.s=left,l.d=2pt}{i1,vU}
\fmf{dashes,label=$j'$,l.s=right,l.d=1pt}{o1,vU}
\fmfv{decor.shape=circle,decor.filled=empty,decor.size=3thick}{vU}
\end{fmfgraph*}
\end{fmffile}
& $U_{HL\,ij'}$ \\
$U$ insertion (light-heavy) &
\begin{fmffile}{bb-vU-lh}
\begin{fmfgraph*}(40,10)
\fmfleft{i1,i2}
\fmfright{o1,o2}
\fmf{dashes,label=$i'$,l.s=left,l.d=2pt}{i1,vU}
\fmf{plain,label=$j$,l.s=right,l.d=1pt}{o1,vU}
\fmfv{decor.shape=circle,decor.filled=empty,decor.size=3thick}{vU}
\end{fmfgraph*}
\end{fmffile}
& $U_{LH\,i'j}$ \\
$U$ insertion (light-light) &
\begin{fmffile}{bb-vU-ll}
\begin{fmfgraph*}(40,10)
\fmfleft{i1,i2}
\fmfright{o1,o2}
\fmf{dashes,label=$i'$,l.s=left,l.d=2pt}{i1,vU}
\fmf{dashes,label=$j'$,l.s=right,l.d=1pt}{o1,vU}
\fmfv{decor.shape=circle,decor.filled=empty,decor.size=3thick}{vU}
\end{fmfgraph*}
\end{fmffile}
& $U_{L\,i'j'}$ \\
\hline
\end{tabular}
\caption{Additional building blocks of covariant diagrams in the presence of mixed heavy-light contributions to matching, as derived in Section~\ref{sec:cd-mixed}. Example applications can be found in Sections~\ref{sec:triplet-scalar}, \ref{sec:triplet-gauge} and~\ref{sec:singlet}.
\label{tab:mixed}}
\end{table}


\begin{table}[tbp]
\centering
\begin{tabular}{|l|c|c|}
\hline
Element of diagram & Symbol & Expression \\
\hline
$Z$ insertion (uncontracted, heavy-heavy) &
\begin{fmffile}{bb-vZ-uhh}
\begin{fmfgraph*}(40,10)
\fmfleft{i1,i2}
\fmfright{o1,o2}
\fmf{plain,label=$i$,l.s=left,l.d=2pt}{i1,vZ}
\fmf{plain,label=$j$,l.s=left,l.d=1pt}{vZ,o1}
\fmfv{decor.shape=square,decor.filled=20,decor.size=3thick}{vZ}
\end{fmfgraph*}
\end{fmffile}
& $P_\mu Z_{H\,ij}^\mu$ \\
$Z$ insertion (uncontracted, heavy-light) &
\begin{fmffile}{bb-vZ-uhl}
\begin{fmfgraph*}(40,10)
\fmfleft{i1,i2}
\fmfright{o1,o2}
\fmf{plain,label=$i$,l.s=left,l.d=2pt}{i1,vZ}
\fmf{dashes,label=$j'$,l.s=right,l.d=1pt}{o1,vZ}
\fmfv{decor.shape=square,decor.filled=20,decor.size=3thick}{vZ}
\end{fmfgraph*}
\end{fmffile}
& $P_\mu Z_{HL\,ij'}^\mu$ \\
$Z$ insertion (uncontracted, light-heavy) &
\begin{fmffile}{bb-vZ-ulh}
\begin{fmfgraph*}(40,10)
\fmfleft{i1,i2}
\fmfright{o1,o2}
\fmf{dashes,label=$i'$,l.s=left,l.d=2pt}{i1,vZ}
\fmf{plain,label=$j$,l.s=right,l.d=1pt}{o1,vZ}
\fmfv{decor.shape=square,decor.filled=20,decor.size=3thick}{vZ}
\end{fmfgraph*}
\end{fmffile}
& $P_\mu Z_{LH\,i'j}^\mu$ \\
$Z$ insertion (uncontracted, light-light) &
\begin{fmffile}{bb-vZ-ull}
\begin{fmfgraph*}(40,10)
\fmfleft{i1,i2}
\fmfright{o1,o2}
\fmf{dashes,label=$i'$,l.s=left,l.d=2pt}{i1,vZ}
\fmf{dashes,label=$j'$,l.s=right,l.d=1pt}{o1,vZ}
\fmfv{decor.shape=square,decor.filled=20,decor.size=3thick}{vZ}
\end{fmfgraph*}
\end{fmffile}
& $P_\mu Z_{L\,i'j'}^\mu$ \\
$Z$ insertion (contracted, heavy-heavy) &
\begin{fmffile}{bb-vZ-chh}
\begin{fmfgraph*}(32,20)
\fmfleft{i1,i2,i3}
\fmfright{o1,o2,o3}
\fmf{plain,label=$i$,l.s=left,l.d=2pt}{i2,vZ}
\fmf{plain,label=$j$,l.s=right,l.d=1pt}{o2,vZ}
\fmf{phantom}{i1,b,o1}
\fmf{dots,width=thick,tension=0}{b,vZ}
\fmfv{decor.shape=square,decor.filled=20,decor.size=3thick}{vZ}
\end{fmfgraph*}
\end{fmffile}
& $-Z_{H\,ij}^\mu$ \\
$Z$ insertion (contracted, heavy-light) &
\begin{fmffile}{bb-vZ-chl}
\begin{fmfgraph*}(32,20)
\fmfleft{i1,i2,i3}
\fmfright{o1,o2,o3}
\fmf{plain,label=$i$,l.s=left,l.d=2pt}{i2,vZ}
\fmf{dashes,label=$j'$,l.s=right,l.d=1pt}{o2,vZ}
\fmf{phantom}{i1,b,o1}
\fmf{dots,width=thick,tension=0}{b,vZ}
\fmfv{decor.shape=square,decor.filled=20,decor.size=3thick}{vZ}
\end{fmfgraph*}
\end{fmffile}
& $-Z_{HL\,ij'}^\mu$ \\
$Z$ insertion (contracted, light-heavy) &
\begin{fmffile}{bb-vZ-clh}
\begin{fmfgraph*}(32,20)
\fmfleft{i1,i2,i3}
\fmfright{o1,o2,o3}
\fmf{dashes,label=$i'$,l.s=left,l.d=2pt}{i2,vZ}
\fmf{plain,label=$j$,l.s=right,l.d=1pt}{o2,vZ}
\fmf{phantom}{i1,b,o1}
\fmf{dots,width=thick,tension=0}{b,vZ}
\fmfv{decor.shape=square,decor.filled=20,decor.size=3thick}{vZ}
\end{fmfgraph*}
\end{fmffile}
& $-Z_{LH\,i'j}^\mu$ \\
$Z$ insertion (contracted, light-light) &
\begin{fmffile}{bb-vZ-cll}
\begin{fmfgraph*}(32,20)
\fmfleft{i1,i2,i3}
\fmfright{o1,o2,o3}
\fmf{dashes,label=$i'$,l.s=left,l.d=2pt}{i2,vZ}
\fmf{dashes,label=$j'$,l.s=right,l.d=1pt}{o2,vZ}
\fmf{phantom}{i1,b,o1}
\fmf{dots,width=thick,tension=0}{b,vZ}
\fmfv{decor.shape=square,decor.filled=20,decor.size=3thick}{vZ}
\end{fmfgraph*}
\end{fmffile}
& $-Z_{L\,i'j'}^\mu$ \\
\hline
$Z^\dagger$ insertion (uncontracted, heavy-heavy) &
\begin{fmffile}{bb-vZd-uhh}
\begin{fmfgraph*}(40,10)
\fmfleft{i1,i2}
\fmfright{o1,o2}
\fmf{plain,label=$i$,l.s=left,l.d=2pt}{i1,vZ}
\fmf{plain,label=$j$,l.s=left,l.d=1pt}{vZ,o1}
\fmfv{decor.shape=square,decor.filled=60,decor.size=3thick}{vZ}
\end{fmfgraph*}
\end{fmffile}
& $Z_{H\,ij}^{\dagger\mu} P_\mu$ \\
$Z^\dagger$ insertion (uncontracted, heavy-light) &
\begin{fmffile}{bb-vZd-uhl}
\begin{fmfgraph*}(40,10)
\fmfleft{i1,i2}
\fmfright{o1,o2}
\fmf{plain,label=$i$,l.s=left,l.d=2pt}{i1,vZ}
\fmf{dashes,label=$j'$,l.s=right,l.d=1pt}{o1,vZ}
\fmfv{decor.shape=square,decor.filled=60,decor.size=3thick}{vZ}
\end{fmfgraph*}
\end{fmffile}
& $Z_{LH\,ij'}^{\dagger\mu} P_\mu$ \\
$Z^\dagger$ insertion (uncontracted, light-heavy) &
\begin{fmffile}{bb-vZd-ulh}
\begin{fmfgraph*}(40,10)
\fmfleft{i1,i2}
\fmfright{o1,o2}
\fmf{dashes,label=$i'$,l.s=left,l.d=2pt}{i1,vZ}
\fmf{plain,label=$j$,l.s=right,l.d=1pt}{o1,vZ}
\fmfv{decor.shape=square,decor.filled=60,decor.size=3thick}{vZ}
\end{fmfgraph*}
\end{fmffile}
& $Z_{HL\,i'j}^{\dagger\mu} P_\mu$ \\
$Z^\dagger$ insertion (uncontracted, light-light) &
\begin{fmffile}{bb-vZd-ull}
\begin{fmfgraph*}(40,10)
\fmfleft{i1,i2}
\fmfright{o1,o2}
\fmf{dashes,label=$i'$,l.s=left,l.d=2pt}{i1,vZ}
\fmf{dashes,label=$j'$,l.s=right,l.d=1pt}{o1,vZ}
\fmfv{decor.shape=square,decor.filled=60,decor.size=3thick}{vZ}
\end{fmfgraph*}
\end{fmffile}
& $Z_{L\,i'j'}^{\dagger\mu} P_\mu$ \\
$Z^\dagger$ insertion (contracted, heavy-heavy) &
\begin{fmffile}{bb-vZd-chh}
\begin{fmfgraph*}(32,20)
\fmfleft{i1,i2,i3}
\fmfright{o1,o2,o3}
\fmf{plain,label=$i$,l.s=left,l.d=2pt}{i2,vZ}
\fmf{plain,label=$j$,l.s=right,l.d=1pt}{o2,vZ}
\fmf{phantom}{i1,b,o1}
\fmf{dots,width=thick,tension=0}{b,vZ}
\fmfv{decor.shape=square,decor.filled=60,decor.size=3thick}{vZ}
\end{fmfgraph*}
\end{fmffile}
& $-Z_{H\,ij}^{\dagger\mu}$ \\
$Z^\dagger$ insertion (contracted, heavy-light) &
\begin{fmffile}{bb-vZd-chl}
\begin{fmfgraph*}(32,20)
\fmfleft{i1,i2,i3}
\fmfright{o1,o2,o3}
\fmf{plain,label=$i$,l.s=left,l.d=2pt}{i2,vZ}
\fmf{dashes,label=$j'$,l.s=right,l.d=1pt}{o2,vZ}
\fmf{phantom}{i1,b,o1}
\fmf{dots,width=thick,tension=0}{b,vZ}
\fmfv{decor.shape=square,decor.filled=60,decor.size=3thick}{vZ}
\end{fmfgraph*}
\end{fmffile}
& $-Z_{LH\,ij'}^{\dagger\mu}$ \\
$Z^\dagger$ insertion (contracted, light-heavy) &
\begin{fmffile}{bb-vZd-clh}
\begin{fmfgraph*}(32,20)
\fmfleft{i1,i2,i3}
\fmfright{o1,o2,o3}
\fmf{dashes,label=$i'$,l.s=left,l.d=2pt}{i2,vZ}
\fmf{plain,label=$j$,l.s=right,l.d=1pt}{o2,vZ}
\fmf{phantom}{i1,b,o1}
\fmf{dots,width=thick,tension=0}{b,vZ}
\fmfv{decor.shape=square,decor.filled=60,decor.size=3thick}{vZ}
\end{fmfgraph*}
\end{fmffile}
& $-Z_{HL\,i'j}^{\dagger\mu}$ \\
$Z^\dagger$ insertion (contracted, light-light) &
\begin{fmffile}{bb-vZd-cll}
\begin{fmfgraph*}(32,20)
\fmfleft{i1,i2,i3}
\fmfright{o1,o2,o3}
\fmf{dashes,label=$i'$,l.s=left,l.d=2pt}{i2,vZ}
\fmf{dashes,label=$j'$,l.s=right,l.d=1pt}{o2,vZ}
\fmf{phantom}{i1,b,o1}
\fmf{dots,width=thick,tension=0}{b,vZ}
\fmfv{decor.shape=square,decor.filled=60,decor.size=3thick}{vZ}
\end{fmfgraph*}
\end{fmffile}
& $-Z_{L\,i'j'}^{\dagger\mu}$ \\
\hline
\end{tabular}
\caption{Additional building blocks of covariant diagrams in the presence of open covariant derivatives in the $\bf X$ matrix, as derived in Section~\ref{sec:cd-ocd}, up to one-open-covariant-derivative terms $P_\mu{\bf Z}^\mu +{\bf Z}^{\dagger\mu}P_\mu$. Example applications can be found in Section~\ref{sec:triplet-gauge}.
\label{tab:ocd}}
\end{table}


\begin{table}[tbp]
\centering
\begin{tabular}{|l|c|c|}
\hline
Element of diagram & Symbol & Expression \\
\hline
heavy propagator (fermionic, uncontracted) &
\begin{fmffile}{bb-pH-fu}
\begin{fmfgraph*}(40,10)
\fmfleft{i1,i2}
\fmfright{o1,o2}
\fmf{plain,label=$i$,l.s=left,l.d=2pt}{i1,o1}
\end{fmfgraph*}
\end{fmffile}
& $M_i$ \\
heavy propagator (fermionic, contracted) &
\begin{fmffile}{bb-pH-fc}
\begin{fmfgraph*}(32,20)
\fmfleft{i1,i2,i3}
\fmfright{o1,o2,o3}
\fmf{plain}{i2,v}
\fmf{plain}{v,o2}
\fmf{phantom}{i1,b,o1}
\fmf{dots,width=thick,tension=0}{b,v}
\fmfv{label=$i$,l.a=90,l.d=2pt}{v}
\end{fmfgraph*}
\end{fmffile}
& $-\gamma^\mu$ \\
light propagator (fermionic) &
\begin{fmffile}{bb-pL-fc}
\begin{fmfgraph*}(32,20)
\fmfleft{i1,i2,i3}
\fmfright{o1,o2,o3}
\fmf{dashes}{i2,v}
\fmf{dashes}{o2,v}
\fmf{phantom}{i1,b,o1}
\fmf{dots,width=thick,tension=0}{b,v}
\fmfv{label=$i'$,l.a=90,l.d=2pt}{v}
\end{fmfgraph*}
\end{fmffile}
& $-\gamma^\mu$ \\
light mass insertion (fermionic) &
\begin{fmffile}{bb-vm-f}
\begin{fmfgraph*}(40,10)
\fmfleft{i1,i2}
\fmfright{o1,o2}
\fmf{dashes,label=$i'$,l.s=left,l.d=2pt}{i1,vm}
\fmf{dashes,label=$j'$,l.s=right,l.d=1pt}{o1,vm}
\fmfv{decor.shape=cross,decor.size=3thick}{vm}
\end{fmfgraph*}
\end{fmffile}
& $ m_{i'}\delta_{i'j'}$ \\
$P$ insertion (fermionic, heavy) &
\begin{fmffile}{bb-vP-fh}
\begin{fmfgraph*}(40,10)
\fmfleft{i1,i2}
\fmfright{o1,o2}
\fmf{plain,label=$i$,l.s=left,l.d=2pt}{i1,vP}
\fmf{plain,label=$j$,l.s=left,l.d=1pt}{vP,o1}
\fmfv{decor.shape=circle,decor.filled=full,decor.size=3thick}{vP}
\end{fmfgraph*}
\end{fmffile}
& $-\Psl \delta_{ij}$ \\
$P$ insertion (fermionic, light) &
\begin{fmffile}{bb-vP-fl}
\begin{fmfgraph*}(40,10)
\fmfleft{i1,i2}
\fmfright{o1,o2}
\fmf{dashes,label=$i'$,l.s=left,l.d=2pt}{i1,vP}
\fmf{dashes,label=$j'$,l.s=right,l.d=1pt}{o1,vP}
\fmfv{decor.shape=circle,decor.filled=full,decor.size=3thick}{vP}
\end{fmfgraph*}
\end{fmffile}
& $-\Psl \delta_{i'j'}$ \\
\hline
\end{tabular}
\caption{Additional building blocks of covariant diagrams when Dirac fermions are involved in matching, as derived in Section~\ref{sec:cd-ferm}. These are used when the quadratic operator for fermionic fields is not squared like in Eq.~\eqref{SEFTloopferm}. Example applications can be found in Sections~\ref{sec:vlf} and~\ref{sec:singlet}.
\label{tab:ferm}}
\end{table}

All derivations from Section~\ref{sec:functionalmatching} to Section~\ref{sec:cd-ferm} are done {\it once and for all}. Now we summarize the results obtained into a recipe that can be easily followed without repeating the derivations.

Starting from an UV Lagrangian $\L_\text{UV}[\Phi,\phi]$ involving heavy fields $\Phi$ of masses $\{M_i\}$ and light fields $\phi$ of masses $\{m_{i'}\}\ll\{M_i\}$, the low-energy EFT can be obtained up to one loop level with the following procedure: 
\begin{enumerate}
\item Solve the classical equation of motion $\frac{\delta\L_\text{UV}}{\delta\Phi}\bigl[\Phi_\c[\phi], \phi\bigr]=0$ for $\Phi_\c[\phi]$ as an expansion of local operators\,\footnote{From here on we omit the hat in $\hat\Phi_\c[\phi]$ and simply write $\Phi_\c[\phi]$. The distinction between the two was important in our derivation in Section~\ref{sec:functionalmatching}, but will not be relevant in the rest of the paper.}. The tree-level effective Lagrangian is given by $\L_\text{EFT}^\text{tree}[\phi] = \L_\text{UV}\bigl[\Phi_\c[\phi], \phi\bigr]$.
\item Expand all fields about classical backgrounds, $\Phi=\Phi_\b+\Phi'$, $\phi=\phi_\b+\phi'$, and {\it extract the $\bf X$ matrix} from terms in $\L_\text{UV}$ that are quadratic in the quantum fluctuations,
\beq
\L_\text{UV,\,quad.} = 
-\frac{1}{2} \bigl( \Phi^{\prime\dagger}\,,\; \phi^{\prime\dagger} \bigr) \,
\bigl({\bf K}+
{\bf X}[\Phi_\b, \phi_\b]
\bigr)
\left(
\begin{matrix}
\Phi' \\ \phi'
\end{matrix}
\right)
\quad
\text{with}
\quad
{\bf X} =
\left(
\begin{matrix}
X_H & X_{HL} \\
X_{LH} & X_L
\end{matrix}
\right),
\eeq{LUVquad}
where $\bf K$ is the diagonal kinetic operator with elements $-P^2+M_i^2$ ($-P^2+m_{i'}^2$) for heavy (light) bosons and $-\Psl+M_i$ ($-\Psl+m_{i'}$) for heavy (light) fermions. Note that the notation $P_\mu\equiv iD_\mu$ is introduced, which is a hermitian operator. A field whose kinetic term has prefactor $-1$ rather than $-\frac{1}{2}$, such as a complex scalar or a Dirac fermion, is usually represented by two fields in the field multiplet (e.g.\ itself and its appropriately-defined conjugate), so that Eq.~\eqref{LUVquad} still holds. For gauge boson fields, add gauge-fixing terms and use the Feynman gauge ($\xi=1$). If the (hermitian) $\bf X$ matrix contains open covariant derivatives ($P_\mu$'s acting openly to the right instead of appearing in commutators), cast it in the following form,
\beq
{\bf X} = {\bf U} +P_\mu{\bf Z}^\mu +{\bf Z}^{\dagger\mu}P_\mu +\dots
\eeqn
with $\bf U$ and $\bf Z$ matrices containing no open covariant derivatives.
\item Draw one-loop diagrams consisting of {\it propagators} and {\it vertex insertions}. In the simplest case of pure heavy bosonic loops with no open covariant derivatives in $\bf X$ (Section~\ref{sec:cd-heavy}), only those listed in Table~\ref{tab:heavy} are needed. Additional elements needed for mixed heavy-light loops (Section~\ref{sec:cd-mixed}), open covariant derivatives (up to $P_\mu{\bf Z}^\mu +{\bf Z}^{\dagger\mu}P_\mu$ terms, Section~\ref{sec:cd-ocd}), and loops with Dirac fermions (Section~\ref{sec:cd-ferm}) are listed in Tables~\ref{tab:mixed}, \ref{tab:ocd} and~\ref{tab:ferm}, respectively. These will be sufficient for the example calculations that we show in the next section. In each diagram, at least one heavy propagator must be present, and dotted lines emanating from all ``contracted'' propagators and vertex insertions must be connected in pairs.
\item The value of a diagram is given by
\beq
-i c_s \,\frac{1}{S}\,\I[q^{2n_c}]_{ij\dots 0}^{n_i n_j \dots n_L} \,\tr\,\O \,.
\eeqn
\begin{itemize}
\item $\frac{1}{S}$ is a {\it symmetry factor} that is present if the diagram has a $\mathbb{Z}_S$ symmetry under rotation. 
\item $n_i, n_j,$ etc., $n_L$ and $n_c$ are the numbers of heavy propagators of type $i,j,$ etc., light propagators and (dotted) contraction lines, respectively. The {\it master integrals} are defined by
\beq
\int\frac{d^dq}{(2\pi)^d} \frac{q^{\mu_1}\cdots q^{\mu_{2n_c}}}{(q^2-M_i^2)^{n_i}(q^2-M_j^2)^{n_j}\cdots (q^2)^{n_L}}
\,\equiv\, g^{\mu_1\dots\mu_{2n_c}} \,\I[q^{2n_c}]_{ij\dots 0}^{n_i n_j\dots n_L} .
\eeq{MIdef}
where $g^{\mu_1\dots\mu_{2n_c}}$ is the completely symmetric tensor, e.g.\ $g^{\mu\nu\rho\sigma}=g^{\mu\nu}g^{\rho\sigma} +g^{\mu\rho}g^{\nu\sigma} +g^{\mu\sigma}g^{\nu\rho}$. These master integrals can be worked out and tabulated as in Appendix~\ref{app:MI}. For simplicity, we will omit the argument ``$[q^{2n_c}]$'' when $n_c=0$.
\item The {\it operator structure} $\O$ is obtained by starting from any propagator on the loop and reading off expressions of propagators and vertex insertions (see Tables~\ref{tab:heavy}-\ref{tab:ferm}) clockwise, with Lorentz indices contracted between elements connected by a dotted line.
\item The {\it spin factor} $c_s$, discussed in the first bullet point below Eq.~\eqref{LEFTloopferm}, is determined by the propagator one starts from when reading the diagram. There are no extra tricky minus signs as in conventional Feynman diagrams.
\end{itemize}
Note that in our formalism, no functional manipulations nor loop integrations are needed --- one simply reads off the elements of a diagram and look up the tabulated master integrals.
\item Add up all diagrams contributing to the effective operators of interest.
\begin{itemize}
\item One may wish to obtain all operators up to some dimension (e.g.\ six) --- this will be the case in Section~\ref{sec:UOLEA} below. Enumeration of diagrams is straightforward, since all operator dimensions are carried by {\it vertex insertions}. In particular, each $P$ insertion has operator dimension 1, while the operator dimensions of a $U$ insertion, a contracted $Z$ insertion, and an uncontracted $Z$ insertion are UV theory-dependent, with lower bounds 1, 1, 2, respectively.
\item Alternatively, for specific applications one may wish to study just a few effective operators rather than the entire effective Lagrangian --- this will be the case in Sections~\ref{sec:triplet-scalar}-\ref{sec:singlet} below. An easy way to determine what diagrams to compute is to write out the field content of various vertex insertions (as in e.g.\ Eqs.~\eqref{Utriplet}, \eqref{Ztriplet}), and enumerate combinations of them that can make up the specific operators of interest (as in e.g.\ Eqs.~\eqref{tripletenumerate}, \eqref{tripletgaugeterms}). 
\end{itemize}
At this step, diagrams giving rise to $\tr\,\O=\tr(\dots P^2\dots)$ can be omitted, as we discussed in Sections~\ref{sec:cd-heavy} and~\ref{sec:cd-ferm}. These include, e.g.\ those with contractions between adjacent bosonic $P$ insertions, or (when the loop integral is convergent) between fermionic propagators separated by two fermionic $P$ insertions and one uncontracted fermionic heavy propagator. Also note that diagrams which are mirror images of each other are related by hermitian conjugation, so only one in such a pair needs to be explicitly computed.
\item The $\tr(\dots P^2\dots)$ terms omitted in the previous step can be recovered by requiring the operator structures obtained organize into gauge-invariant operator traces where $P_\mu$'s only appear in commutators. However, instead of working out these extra terms explicitly, it is often easier in practice to first write down all independent operator traces expected in the final result, and then expand the commutators and match the result of the previous step to solve for their coefficients.
\item Finally, to obtain $\L_\text{EFT}^\text{1-loop}[\phi]$ for a specific $\L_\text{UV}[\Phi,\phi]$, evaluate the operator traces by plugging in specific forms of the $\bf U$ and $\bf Z$ matrices, with $\Phi$ set to $\Phi_\c[\phi]$. The traces are over internal indices of the fields, including Lorentz indices carried by vector bosons which should be contracted using $-g_{\alpha\beta}$ as discussed in Section~\ref{sec:cd-heavy}.
\end{enumerate}
It should be emphasized that while the procedure above has been stated in the context of matching a specific UV theory to an EFT, Steps 3-6 are actually universal and independent of UV model details. The only assumption made about the UV Lagrangian is the (quite general) form of its quadratic terms (see Step 2). Therefore, {\it Steps 3-6 above also constitute a recipe for deriving universal results of one-loop matching.}

\section{Examples}
\label{sec:example}

\subsection{Universal One-Loop Effective Action (UOLEA) simplified}
\label{sec:UOLEA}

As a first application of the covariant diagrams techniques introduced in the previous section, we reproduce the Universal One-Loop Effective Action (UOLEA) reported in~\cite{DEQY} (and~\cite{HLM14} for the degenerate limit) with a simpler derivation. Recall that the UOLEA is a universal master formula for one-loop matching up to dimension six level in the absence of mixed heavy-light contributions and open covariant derivatives in the ${\bf X}$ matrix. We will show that this master formula can be obtained as a {\it sum of covariant diagrams} easily built from the ingredients in Table~\ref{tab:heavy}.

We begin by writing down all independent operator traces involving $P_\mu$ and $U_H$ which may contain terms with operator dimensions up to six. To do so, recall $\text{dim}(P_\mu)=1$, $\text{dim}(U_H)\ge1$. Writing $U\equiv U_H$ for simplicity, we have
\beqa
\L_\text{UOLEA} &=& -i c_s \,\tr\, \Bigl\{ 
f_2^i\, U_{ii} 
+f_3^i\, G^{\prime\mu\nu}_i G^\prime_{\mu\nu,i} 
+f_4^{ij}\, U_{ij} U_{ji} \CR 
&&\qquad\quad
+f_5^i\, [P^\mu, G^\prime_{\mu\nu,i}] [P_\rho, G^{\prime\rho\nu}_i]
+f_6^i\, G^{\prime\mu}_{\;\;\,\nu,i} G^{\prime\nu}_{\;\;\,\rho,i} G^{\prime\rho}_{\;\;\,\mu,i} \CR 
&&\qquad\quad
+f_7^{ij}\, [P^\mu, U_{ij}] [P_\mu, U_{ji}]
+f_8^{ijk}\, U_{ij} U_{jk} U_{ki}
+f_9^i\, U_{ii} G^{\prime\mu\nu}_i G^\prime_{\mu\nu,i} \CR 
&&\qquad\quad
+f_{10}^{ijkl}\, U_{ij} U_{jk} U_{kl} U_{li}
+f_{11}^{ijk}\, U_{ij} [P^\mu, U_{jk}] [P_\mu, U_{ki}] \CR 
&&\qquad\quad
+f_{12}^{ij}\, \bigl[P^\mu, [P_\mu, U_{ij}]\bigr] \bigl[P^\nu, [P_\nu, U_{ji}]\bigr]
+f_{13}^{ij}\, U_{ij} U_{ji} G^{\prime\mu\nu}_i G^\prime_{\mu\nu,i} \CR 
&&\qquad\quad
+f_{14}^{ij}\, [P^\mu, U_{ij}] [P^\nu, U_{ji}] G^\prime_{\nu\mu, i}
+f_{15}^{ij}\, \bigl( U_{ij} [P^\mu, U_{ji}] -[P^\mu, U_{ij}] U_{ji} \bigr) [P^\nu, G^\prime_{\nu\mu,i}] \CR 
&&\qquad\quad
+f_{16}^{ijklm}\, U_{ij} U_{jk} U_{kl} U_{lm} U_{mi} \CR 
&&\qquad\quad
+f_{17}^{ijkl}\, U_{ij} U_{jk} [P^\mu, U_{kl}] [P_\mu, U_{li}]
+f_{18}^{ijkl}\, U_{ij} [P^\mu, U_{jk}] U_{kl} [P_\mu, U_{li}] \CR 
&&\qquad\quad
+f_{19}^{ijklmn}\, U_{ij} U_{jk} U_{kl} U_{lm} U_{mn} U_{ni}
\Bigr\} ,
\eeqa{LUOLEA}
where 
$G'_{\mu\nu}\equiv -[P_\mu, P_\nu] = -igG_{\mu\nu}$. Note that $G'_{\mu\nu}$, like $P_\mu$, is a diagonal matrix in the field multiplet space, and we use $G'_{\mu\nu,i}$ to denote its diagonal elements. We have adopted the notation in~\cite{DEQY} for the {\it universal coefficients} $f_N\,(N=2,\dots,19)$\,\footnote{Some redundancies in the parameterization in~\cite{DEQY} have been removed here. In particular, the terms $f_{12,a}^{ij}\bigl[P^\mu, [P^\nu, U_{ij}]\bigr] \bigl[P_\mu, [P_\nu, U_{ji}]\bigr] +f_{12,b}^{ij}\bigl[P^\mu, [P^\nu, U_{ij}]\bigr] \bigl[P_\nu, [P_\mu, U_{ji}]\bigr]$ written out in~\cite{DEQY} can be set to zero because $f_{12,a/b}^{ij}=-f_{12,a/b}^{ji}$ while the operator traces are symmetric in $i,j$. Also, $f_{15,a}^{ijk}$ and $f_{15,b}^{ijk}$ introduced in~\cite{DEQY}, which are associated with $U_{ij} [P^\mu, U_{jk}] [P^\nu, G^\prime_{\nu\mu,ki}]$ and $-[P^\mu, U_{ij}] U_{jk} [P^\nu, G^\prime_{\nu\mu,ki}]$, respectively, are equal when $k=i$ (as dictated by $G'_{\mu\nu}$ being diagonal).}. In the following, we compute in turn terms in Eq.~\eqref{LUOLEA} with 0, 2, 4, 6 covariant derivatives, from which the universal coefficients can be extracted.

\paragraph{$\O(P^0)$ terms ($f_{2,4,8,10,16,19}$).}

Diagrams with no $P$ insertions all share a similar structure, from which six universal coefficients can be derived, each in terms of a single master integral:
\bseq
\beqa
\label{diag:U}
\begin{gathered}
\begin{fmffile}{UOLEA-U}
\begin{fmfgraph}(40,40)
\fmfsurround{v0,vU}
\fmf{plain,left=1}{vU,v0,vU}
\fmfv{decor.shape=circle,decor.filled=empty,decor.size=3thick}{vU}
\end{fmfgraph}
\end{fmffile}
\end{gathered}
\;\; &=& -i c_s \, \I_i^1 \, \tr\, U_{ii} \quad \Rightarrow \quad f_2^i = \,\I_i^1 \,, \\[4pt]
\label{diag:U2}
\begin{gathered}
\begin{fmffile}{UOLEA-U2}
\begin{fmfgraph}(40,40)
\fmfsurround{vU2,vU1}
\fmf{plain,left=1}{vU1,vU2,vU1}
\fmfv{decor.shape=circle,decor.filled=empty,decor.size=3thick}{vU1,vU2}
\end{fmfgraph}
\end{fmffile}
\end{gathered}
\;\; &=& -i c_s \frac{1}{2} \, \I_{ij}^{11} \, \tr (U_{ij}U_{ji}) \quad \Rightarrow \quad f_4^{ij} = \frac{1}{2}\, \I_{ij}^{11} \,, \\[4pt]
\label{diag:U3}
\begin{gathered}
\begin{fmffile}{UOLEA-U3}
\begin{fmfgraph}(40,40)
\fmfsurround{vU2,vU1,vU3}
\fmf{plain,left=0.6}{vU1,vU2,vU3,vU1}
\fmfv{decor.shape=circle,decor.filled=empty,decor.size=3thick}{vU1,vU2,vU3}
\end{fmfgraph}
\end{fmffile}
\end{gathered}
\;\; &=& -i c_s \frac{1}{3} \, \I_{ijk}^{111} \, \tr (U_{ij}U_{jk}U_{ki}) \quad\Rightarrow \quad f_8^{ijk} = \frac{1}{3}\, \I_{ijk}^{111} \,,\\[8pt]
\label{diag:U4}
\begin{gathered}
\begin{fmffile}{UOLEA-U4}
\begin{fmfgraph}(40,40)
\fmfsurround{vU3,vU2,vU1,vU4}
\fmf{plain,left=0.4}{vU1,vU2,vU3,vU4,vU1}
\fmfv{decor.shape=circle,decor.filled=empty,decor.size=3thick}{vU1,vU2,vU3,vU4}
\end{fmfgraph}
\end{fmffile}
\end{gathered}
\;\; &=& -i c_s \frac{1}{4} \, \I_{ijkl}^{1111} \, \tr (U_{ij}U_{jk}U_{kl}U_{li}) \quad \Rightarrow \quad f_{10}^{ijkl} = \frac{1}{4}\, \I_{ijkl}^{1111} \,, \\[8pt]
\label{diag:U5}
\begin{gathered}
\begin{fmffile}{UOLEA-U5}
\begin{fmfgraph}(40,40)
\fmfsurround{vU3,vU2,vU1,vU5,vU4}
\fmf{plain,left=0.3}{vU1,vU2,vU3,vU4,vU5,vU1}
\fmfv{decor.shape=circle,decor.filled=empty,decor.size=3thick}{vU1,vU2,vU3,vU4,vU5}
\end{fmfgraph}
\end{fmffile}
\end{gathered}
\;\; &=& -i c_s \frac{1}{5} \, \I_{ijklm}^{11111} \, \tr (U_{ij}U_{jk}U_{kl}U_{lm}U_{mi}) \quad \Rightarrow \quad f_{16}^{ijklm} = \frac{1}{5}\, \I_{ijklm}^{11111} \,, \\[8pt]
\label{diag:U6}
\begin{gathered}
\begin{fmffile}{UOLEA-U6}
\begin{fmfgraph}(40,40)
\fmfsurround{vU4,vU3,vU2,vU1,vU6,vU5}
\fmf{plain,left=0.25}{vU1,vU2,vU3,vU4,vU5,vU6,vU1}
\fmfv{decor.shape=circle,decor.filled=empty,decor.size=3thick}{vU1,vU2,vU3,vU4,vU5,vU6}
\end{fmfgraph}
\end{fmffile}
\end{gathered}
\;\; &=& -i c_s \frac{1}{6} \, \I_{ijklmn}^{111111} \, \tr (U_{ij}U_{jk}U_{kl}U_{lm}U_{mn}U_{ni}) \quad \Rightarrow \quad f_{19}^{ijklmn} = \frac{1}{6}\, \I_{ijklmn}^{111111} \,.\quad
\eeqan
\eseq{trP0}
We have omitted propagator labels $i,j,\dots$ in the diagrams above for simplicity, which can be trivially restored. 
Note the symmetry factor $\frac{1}{S}$ with $S$ being the number of $U$ insertions.

\paragraph{$\O(P^2)$ terms ($f_{7,11,17,18}$).}  
The two $P$ insertions must be contracted with each other. To avoid adjacent contraction, at least two $U$ insertions are needed:
\vspace{4pt}
\beqa
\begin{gathered}
\begin{fmffile}{UOLEA-P2U2}
\begin{fmfgraph*}(40,40)
\fmfsurround{vP2,vU1,vP1,vU2}
\fmf{plain,left=0.4,label=$i$,l.d=3pt}{vU2,vP1,vU1}
\fmf{plain,left=0.4,label=$j$,l.d=3pt}{vU1,vP2,vU2}
\fmfv{decor.shape=circle,decor.filled=full,decor.size=3thick}{vP1,vP2}
\fmfv{decor.shape=circle,decor.filled=empty,decor.size=3thick}{vU1,vU2}
\fmf{dots,width=thick}{vP1,vP2}
\end{fmfgraph*}
\end{fmffile}
\end{gathered}
\;\; &=& -i c_s \frac{2^2}{2} \, \I[q^2]_{ij}^{22} \,\tr (P^\mu U_{ij} P_\mu U_{ji}) \subset -i c_s \, \I[q^2]_{ij}^{22} \,\tr\bigl( [P^\mu, U_{ij}] [P_\mu, U_{ji}]\bigr) \CR
&\Rightarrow& \quad f_7^{ij} =\, \I[q^2]_{ij}^{22} \,.
\eeqa{diag:P2U2}
This diagram was in fact already worked out in Eq.~\eqref{HloopEx}. The meaning of ``$\subset$'' is that with the addition of terms involving $\tr(\dots P^2\dots)$, the RHS can be obtained from the LHS; in other words, the RHS is the only independent gauge-invariant operator (or operator combination) with all $P_\mu$'s appearing in commutators which can contain the structure on the LHS.

With three $U$ insertions, still only a single diagram contributes:
\vspace{8pt}
\beq
\begin{gathered}
\begin{fmffile}{UOLEA-P2U3}
\begin{fmfgraph*}(40,40)
\fmfsurround{vU2,vP1,vU1,vU3,vP2}
\fmf{plain,left=0.3,label=$i$,l.d=3pt}{vU3,vU1}
\fmf{plain,left=0.3,label=$j$,l.d=3pt}{vU1,vP1,vU2}
\fmf{plain,left=0.3,label=$k$,l.d=3pt}{vU2,vP2,vU3}
\fmfv{decor.shape=circle,decor.filled=full,decor.size=3thick}{vP1,vP2}
\fmfv{decor.shape=circle,decor.filled=empty,decor.size=3thick}{vU1,vU2,vU3}
\fmf{dots,width=thick,right=0.2}{vP1,vP2}
\end{fmfgraph*}
\end{fmffile}
\end{gathered}
\;\; =\, -i c_s 2^2 \, \I[q^2]_{ijk}^{122} \,\tr (U_{ij} P^\mu U_{jk} P_\mu U_{ki}) \,.
\vspace{4pt}
\eeq{diag:P2U3}
To derive the corresponding universal coefficient $f_{11}$ in the UOLEA, note that
\beqa
f_{11}^{ijk} \,\tr\bigl( U_{ij} [P^\mu, U_{jk}] [P_\mu, U_{ki}]\bigr) &\supset& f_{11}^{ijk} \,\tr ( U_{ij} P^\mu U_{jk} P_\mu U_{ki} + U_{jk} P^\mu U_{ki} P_\mu U_{ij} - U_{ki} P^\mu U_{ij} P_\mu U_{jk} ) \CR
&=& \bigl( f_{11}^{ijk} +f_{11}^{kij} -f_{11}^{jki} \bigr) \,\tr (U_{ij} P^\mu U_{jk} P_\mu U_{ki}) \CR
\Rightarrow f_{11}^{ijk} +f_{11}^{kij} -f_{11}^{jki} &=& 4\,\I[q^2]_{ijk}^{122} \,,
\eeqan
which can be solved simply by permuting the indices $i\to j\to k$ and adding to the original equation. We thus obtain $f_{11}$ in terms of two master integrals,
\beq
f_{11}^{ijk} = 2\bigl( \,\I[q^2]_{ijk}^{122}\, + \,\I[q^2]_{ijk}^{212}\, \bigr).
\eeqn

Finally, with four $U$ insertions, there are two possible diagrams:
\vspace{12pt}
\bseq
\beqa
\begin{gathered}
\begin{fmffile}{UOLEA-P2U4-a}
\begin{fmfgraph*}(40,40)
\fmfsurround{vP2,vU2,vU1,vP1,vU4,vU3}
\fmf{plain,left=0.25,label=$i$,l.d=3pt}{vU4,vP1,vU1}
\fmf{plain,left=0.25,label=$j$,l.d=3pt}{vU1,vU2}
\fmf{plain,left=0.25,label=$k$,l.d=3pt}{vU2,vP2,vU3}
\fmf{plain,left=0.25,label=$l$,l.d=3pt}{vU3,vU4}
\fmfv{decor.shape=circle,decor.filled=full,decor.size=3thick}{vP1,vP2}
\fmfv{decor.shape=circle,decor.filled=empty,decor.size=3thick}{vU1,vU2,vU3,vU4}
\fmf{dots,width=thick}{vP1,vP2}
\end{fmfgraph*}
\end{fmffile}
\end{gathered}
\;\; &=& -i c_s \frac{2^2}{2} \, \I[q^2]_{ijkl}^{2121} \,\tr (P^\mu U_{ij} U_{jk} P_\mu U_{kl} U_{li}) \,, \\[20pt]
\begin{gathered}
\begin{fmffile}{UOLEA-P2U4-b}
\begin{fmfgraph*}(40,40)
\fmfsurround{vU1,vP1,vU4,vU3,vU2,vP2}
\fmf{plain,left=0.25,label=$i$,l.d=3pt}{vU4,vP1,vU1}
\fmf{plain,left=0.25,label=$j$,l.d=3pt}{vU1,vP2,vU2}
\fmf{plain,left=0.25,label=$k$,l.d=3pt}{vU2,vU3}
\fmf{plain,left=0.25,label=$l$,l.d=3pt}{vU3,vU4}
\fmfv{decor.shape=circle,decor.filled=full,decor.size=3thick}{vP1,vP2}
\fmfv{decor.shape=circle,decor.filled=empty,decor.size=3thick}{vU1,vU2,vU3,vU4}
\fmf{dots,width=thick,right=0.4}{vP1,vP2}
\end{fmfgraph*}
\end{fmffile}
\end{gathered}
\;\; &=& -i c_s 2^2 \, \I[q^2]_{ijkl}^{2211} \,\tr (P^\mu U_{ij} P_\mu U_{jk} U_{kl} U_{li}) \,.
\eeqan
\eseq{diag:P2U4}
They organize into two independent operator traces, which we have chosen to be
\beqa
&& f_{17}^{ijkl} \,\tr\bigl( U_{ij} U_{jk} [P^\mu, U_{kl}] [P_\mu, U_{li}]\bigr) + f_{18}^{ijkl} \,\tr\bigl( U_{ij} [P^\mu, U_{jk}] U_{kl} [P_\mu, U_{li}]\bigr) \CR
&\supset& \bigl( -f_{17}^{ijkl} +f_{18}^{ijkl} +f_{18}^{jkli} \bigr) \,\tr (P^\mu U_{ij} U_{jk} P_\mu U_{kl} U_{li}) \CR
&& + \bigl( f_{17}^{klij} +f_{17}^{jkli} -f_{18}^{ijkl} -f_{18}^{klij} \bigr) \,\tr (P^\mu U_{ij} P_\mu U_{jk} U_{kl} U_{li})\,.
\eeqan
We therefore obtain the following two equations,
\beq
-f_{17}^{ijkl} +f_{18}^{ijkl} +f_{18}^{jkli} = 2\, \I[q^2]_{ijkl}^{2121} \,, \quad
f_{17}^{klij} +f_{17}^{jkli} -f_{18}^{ijkl} -f_{18}^{klij} = 4 \, \I[q^2]_{ijkl}^{2211} \,.
\eeqn
which are solved by
\bseq
\beqa
f_{17}^{ijkl} &=& 2 \bigl(\, \I[q^2]_{ijkl}^{2112} +\, \I[q^2]_{ijkl}^{1212} +\, \I[q^2]_{ijkl}^{1122} \,\bigr)\,, \\
f_{18}^{ijkl} &=& \, \I[q^2]_{ijkl}^{2121} +\, \I[q^2]_{ijkl}^{2112} +\, \I[q^2]_{ijkl}^{1221} +\, \I[q^2]_{ijkl}^{1212} +\,\I[q^2]_{ijkl}^{1122} -\,\I[q^2]_{ijkl}^{2211} \CR
&\to& \, \I[q^2]_{ijkl}^{2121} +\, \I[q^2]_{ijkl}^{2112} +\, \I[q^2]_{ijkl}^{1221} +\, \I[q^2]_{ijkl}^{1212}\,.
\eeqan
\eseqn
We have dropped terms in $f_{18}^{ijkl}$ that are antisymmetric under $ij\leftrightarrow kl$, since the associated operator trace is symmetric. We see that $f_{17}$ and $f_{18}$ together depend on only five master integrals.

\paragraph{$\O(P^4)$ terms ($f_{3,9,12,13,14,15}$).} 

The four $P$ insertions can be contracted among themselves without $U$ insertions:
\vspace{4pt}
\beqa
\begin{gathered}
\begin{fmffile}{UOLEA-P4}
\begin{fmfgraph}(40,40)
\fmfsurround{vP3,vP2,vP1,vP4}
\fmf{plain,left=0.4}{vP1,vP2,vP3,vP4,vP1}
\fmfv{decor.shape=circle,decor.filled=full,decor.size=3thick}{vP1,vP2,vP3,vP4}
\fmf{dots,width=thick}{vP1,vP3}
\fmf{dots,width=thick}{vP2,vP4}
\end{fmfgraph}
\end{fmffile}
\end{gathered}
\;\; &=& -i c_s \frac{2^4}{4} \, \I[q^4]_i^4 \,\tr (P^\mu P^\nu P_\mu P_\nu) \subset -i c_s 2\, \I[q^4]_i^4 \,\tr\bigl( [P^\mu, P^\nu] [P_\mu, P_\nu]\bigr) \CR
&\Rightarrow& \quad f_3^i = 2\, \I[q^4]_i^4 \,.
\eeqa{diag:P4}

Similarly, with one $U$ insertion,
\vspace{4pt}
\beqa
\begin{gathered}
\begin{fmffile}{UOLEA-P4U}
\begin{fmfgraph}(40,40)
\fmfsurround{vU,vP4,vP3,vP2,vP1}
\fmf{plain,left=0.3}{vU,vP1,vP2,vP3,vP4,vU}
\fmfv{decor.shape=circle,decor.filled=full,decor.size=3thick}{vP1,vP2,vP3,vP4}
\fmfv{decor.shape=circle,decor.filled=empty,decor.size=3thick}{vU}
\fmf{dots,width=thick,right=0.2}{vP1,vP3}
\fmf{dots,width=thick,right=0.2}{vP2,vP4}
\end{fmfgraph}
\end{fmffile}
\end{gathered}
\;\; &=& -i c_s 2^4 \, \I[q^4]_i^5 \,\tr (U_{ii} P^\mu P^\nu P_\mu P_\nu) \subset -i c_s 8\, \I[q^4]_i^5 \,\tr\bigl( U_{ii}[P^\mu, P^\nu] [P_\mu, P_\nu]\bigr) \CR
&\Rightarrow& \quad f_9^i = 8\, \I[q^4]_i^5 \,.
\eeqa{diag:P4U}

With two $U$ insertions, four diagrams can be drawn:
\vspace{6pt}
\bseq
\beqa
\begin{gathered}
\begin{fmffile}{UOLEA-P4U2-a}
\begin{fmfgraph*}(40,40)
\fmfsurround{vP1,vU2,vU1,vP4,vP3,vP2}
\fmf{plain,left=0.25,label=$i$,l.d=3pt}{vU2,vP1,vP2,vP3,vP4,vU1}
\fmf{plain,left=0.25,label=$j$,l.d=3pt}{vU1,vU2}
\fmfv{decor.shape=circle,decor.filled=full,decor.size=3thick}{vP1,vP2,vP3,vP4}
\fmfv{decor.shape=circle,decor.filled=empty,decor.size=3thick}{vU1,vU2}
\fmf{dots,width=thick,right=0.3}{vP1,vP3}
\fmf{dots,width=thick,right=0.3}{vP2,vP4}
\end{fmfgraph*}
\end{fmffile}
\end{gathered}
\;\; &=& -i c_s 2^4 \, \I[q^4]_{ij}^{51} \,\tr (P^\mu P^\nu P_\mu P_\nu U_{ij} U_{ji}) \,, \\[20pt]
\begin{gathered}
\begin{fmffile}{UOLEA-P4U2-b}
\begin{fmfgraph*}(40,40)
\fmfsurround{vP2,vP1,vU2,vP4,vU1,vP3}
\fmf{plain,left=0.25,label=$i$,l.d=3pt}{vU2,vP1,vP2,vP3,vU1}
\fmf{plain,left=0.25,label=$j$,l.d=3pt}{vU1,vP4,vU2}
\fmfv{decor.shape=circle,decor.filled=full,decor.size=3thick}{vP1,vP2,vP3,vP4}
\fmfv{decor.shape=circle,decor.filled=empty,decor.size=3thick}{vU1,vU2}
\fmf{dots,width=thick,right=0.3}{vP1,vP3}
\fmf{dots,width=thick}{vP2,vP4}
\end{fmfgraph*}
\end{fmffile}
\end{gathered}
\;\; &=& -i c_s 2^4 \, \I[q^4]_{ij}^{42} \,\tr (P^\mu P^\nu P_\mu U_{ij} P_\nu U_{ji}) \,, \\[20pt]
\begin{gathered}
\begin{fmffile}{UOLEA-P4U2-c}
\begin{fmfgraph*}(40,40)
\fmfsurround{vU1,vP2,vP1,vU2,vP4,vP3}
\fmf{plain,left=0.25,label=$i$,l.d=3pt}{vU2,vP1,vP2,vU1}
\fmf{plain,left=0.25,label=$j$,l.d=3pt}{vU1,vP3,vP4,vU2}
\fmfv{decor.shape=circle,decor.filled=full,decor.size=3thick}{vP1,vP2,vP3,vP4}
\fmfv{decor.shape=circle,decor.filled=empty,decor.size=3thick}{vU1,vU2}
\fmf{dots,width=thick}{vP1,vP3}
\fmf{dots,width=thick}{vP2,vP4}
\end{fmfgraph*}
\end{fmffile}
\end{gathered}
\;\; &=& -i c_s \frac{2^4}{2} \, \I[q^4]_{ij}^{33} \,\tr (P^\mu P^\nu U_{ij} P_\mu P_\nu U_{ji}) \,, \\[20pt]
\begin{gathered}
\begin{fmffile}{UOLEA-P4U2-d}
\begin{fmfgraph*}(40,40)
\fmfsurround{vU1,vP2,vP1,vU2,vP4,vP3}
\fmf{plain,left=0.25,label=$i$,l.d=3pt}{vU2,vP1,vP2,vU1}
\fmf{plain,left=0.25,label=$j$,l.d=3pt}{vU1,vP3,vP4,vU2}
\fmfv{decor.shape=circle,decor.filled=full,decor.size=3thick}{vP1,vP2,vP3,vP4}
\fmfv{decor.shape=circle,decor.filled=empty,decor.size=3thick}{vU1,vU2}
\fmf{dots,width=thick,left=0.3}{vP1,vP4}
\fmf{dots,width=thick,right=0.3}{vP2,vP3}
\end{fmfgraph*}
\end{fmffile}
\end{gathered}
\;\; &=& -i c_s \frac{2^4}{2} \, \I[q^4]_{ij}^{33} \,\tr (P^\mu P^\nu U_{ij} P_\nu P_\mu U_{ji}) \,.
\vspace{4pt}
\eeqan
\eseq{diag:P4U2}
These terms are contained in four independent operator traces, which we have chosen to be
\beqa
&& f_{12}^{ij} \,\tr \bigl( \bigl[P^\mu, [P_\mu, U_{ij}] \bigr] \bigl[P^\nu, [P_\nu, U_{ji}] \bigr]\bigr) +f_{13}^{ij} \,\tr\bigl( U_{ij} U_{ji} [P^\mu, P^\nu] [P_\mu, P_\nu]\bigr) \CR
&& +f_{14}^{ij} \,\tr \bigl([P^\mu, U_{ij}] [P^\nu, U_{ji}] [P_\mu, P_\nu]\bigr) +f_{15}^{ij} \,\tr \bigl(( U_{ij} [P^\mu, U_{ji}] -[P^\mu, U_{ij}] U_{ji} ) \bigl[ P^\nu, [P_\mu, P_\nu] \bigr]\bigr) \CR
&\supset& \bigl( 2f_{13}^{ij} -f_{14}^{ij} -4f_{15}^{ij} \bigr) \,\tr (P^\mu P^\nu P_\mu P_\nu U_{ij} U_{ji}) 
+\bigl( 2f_{14}^{ij} +4f_{15}^{ij} \bigr) \,\tr (P^\mu P^\nu P_\mu U_{ij} P_\nu U_{ji}) \CR
&& -f_{14}^{ij} \,\tr (P^\mu P^\nu U_{ij} P_\mu P_\nu U_{ji}) 
+\bigl( 4f_{12}^{ij} +f_{14}^{ij} \bigr) \,\tr (P^\mu P^\nu U_{ij} P_\nu P_\mu U_{ji}) \,.
\eeqan
Solving the set of four equations,
\beqa
&& 2f_{13}^{ij} -f_{14}^{ij} -4f_{15}^{ij} = 16 \, \I[q^4]_{ij}^{51} \,, \quad
2f_{14}^{ij} +4f_{15}^{ij} = 16 \, \I[q^4]_{ij}^{42} \,, \CR
&& -f_{14}^{ij} = 8 \, \I[q^4]_{ij}^{33} \,, \quad
4f_{12}^{ij} +f_{14}^{ij} = 8 \, \I[q^4]_{ij}^{33} \,,
\eeqan
we obtain the four universal coefficients $f_{12,13,14,15}$ in terms of just three master integrals:
\bseq
\beqa
f_{12}^{ij} &=& 4 \, \I[q^4]_{ij}^{33} \,, \\
f_{13}^{ij} &=& 4 \bigl( \, \I[q^4]_{ij}^{33} +2 \, \I[q^4]_{ij}^{42} +2 \, \I[q^4]_{ij}^{51}\bigr) \,,\\
f_{14}^{ij} &=& -8 \, \I[q^4]_{ij}^{33} \,,\\
f_{15}^{ij} &=& 4 \bigl( \, \I[q^4]_{ij}^{33} +\, \I[q^4]_{ij}^{42} \bigr) .
\eeqan
\eseqn

\paragraph{$\O(P^6)$ terms ($f_{5,6}$).} Only pure gauge pieces are of interest here, since $P^6$ already has operator dimension six. There are two diagrams contributing, which differ by Lorentz contraction:
\bseq
\beqa
\begin{gathered}
\begin{fmffile}{UOLEA-P6-a}
\begin{fmfgraph}(40,40)
\fmfsurround{vP4,vP3,vP2,vP1,vP6,vP5}
\fmf{plain,left=0.25}{vP1,vP2,vP3,vP4,vP5,vP6,vP1}
\fmfv{decor.shape=circle,decor.filled=full,decor.size=3thick}{vP1,vP2,vP3,vP4,vP5,vP6}
\fmf{dots,width=thick}{vP1,vP4}
\fmf{dots,width=thick}{vP2,vP5}
\fmf{dots,width=thick}{vP3,vP6}
\end{fmfgraph}
\end{fmffile}
\end{gathered}
\;\; &=& -i c_s \frac{2^6}{6} \, \I[q^6]_i^6 \,\tr (P^\mu P^\nu P^\rho P_\mu P_\nu P_\rho) \,, \\[8pt]
\begin{gathered}
\begin{fmffile}{UOLEA-P6-b}
\begin{fmfgraph}(40,40)
\fmfsurround{vP4,vP3,vP2,vP1,vP6,vP5}
\fmf{plain,left=0.25}{vP1,vP2,vP3,vP4,vP5,vP6,vP1}
\fmfv{decor.shape=circle,decor.filled=full,decor.size=3thick}{vP1,vP2,vP3,vP4,vP5,vP6}
\fmf{dots,width=thick}{vP1,vP4}
\fmf{dots,width=thick,left=0.3}{vP2,vP6}
\fmf{dots,width=thick,right=0.3}{vP3,vP5}
\end{fmfgraph}
\end{fmffile}
\end{gathered}
\;\; &=& -i c_s \frac{2^6}{2} \, \I[q^6]_i^6 \,\tr (P^\mu P^\nu P^\rho P_\nu P_\mu P_\rho) \,.
\eeqan
\eseq{diag:P6}
They follow from two independent operators, which are chosen as
\beqa
&& f_5^i \,\tr \bigl( \bigl[ P^\mu, [P_\mu, P_\nu] \bigr] \bigl[ P_\rho, [P^\rho, P^\nu] \bigr] \bigr) 
- f_6^i \,\tr \bigl( [P_\mu, P^\nu] [P_\nu, P^\rho] [P_\rho, P^\mu] \bigr) \CR
&\supset& f_6^i \,\tr (P^\mu P^\nu P^\rho P_\mu P_\nu P_\rho) 
+\bigl( 4f_5^i -3f_6^i \bigr) \,\tr (P^\mu P^\nu P^\rho P_\nu P_\mu P_\rho) \,.
\eeqan
As a result, we have
\beq
f_6^i = \frac{32}{3} \, \I[q^6]_i^6 \,, \quad
4f_5^i -3f_6^i = 32 \, \I[q^6]_i^6 \,,
\eeqn
which yield
\beq
f_5^i = 16\, \I[q^6]_i^6 \,, \quad
f_6^i = \frac{32}{3} \, \I[q^6]_i^6 \,.
\eeqn

\bigskip
We summarize the results of the four paragraphs above in Table~\ref{tab:UOLEA}. Complete agreement is found between our explicit expressions of the universal coefficients in terms of heavy particle masses (listed in Appendix~\ref{app:coefficients}) and those reported in~\cite{DEQY}, upon proper symmetrizations allowed by symmetries of operator traces under exchanging particle labels (e.g.\ our $f_8^{ijk}$ is equal to $\frac{1}{3}(f_8^{ijk}+f_8^{jki}+f_8^{kij})$ in~\cite{DEQY}). Note, however, that we have obtained the universal coefficients in terms of much fewer master integrals, and many of their explicit expressions are also simpler than those in~\cite{DEQY}.

\begin{table}[tbp]
\centering
\begin{tabular}{|l|l|l|}
\hline
Universal coefficient & Operator & Diagram(s) \\
\hline
$f_2^i = \,\I_i^1$ & $U_{ii}$ & Eq.~\eqref{diag:U} \\
\hline
$f_3^i = 2\, \I[q^4]_i^4$ & $G^{\prime\mu\nu}_i G^\prime_{\mu\nu,i}$ & Eq.~\eqref{diag:P4} \\
\hline
$f_4^{ij} = \frac{1}{2}\, \I_{ij}^{11}$ & $U_{ij} U_{ji}$ & Eq.~\eqref{diag:U2} \\
\hline
$f_5^i = 16\, \I[q^6]_i^6$ & $[P^\mu, G^\prime_{\mu\nu,i}] [P_\rho, G^{\prime\rho\nu}_i]$ & \multirow{2}{*}{Eq.~\eqref{diag:P6}} \\
\cline{1-2}
$f_6^i = \frac{32}{3} \, \I[q^6]_i^6$ & $G^{\prime\mu}_{\;\;\,\nu,i} G^{\prime\nu}_{\;\;\,\rho,i} G^{\prime\rho}_{\;\;\,\mu,i}$ & \\
\hline
$f_7^{ij} = \,\I[q^2]_{ij}^{22}$ & $[P^\mu, U_{ij}] [P_\mu, U_{ji}]$ & Eq.~\eqref{diag:P2U2} \\
\hline
$f_8^{ijk} = \frac{1}{3}\, \I_{ijk}^{111}$ & $U_{ij} U_{jk} U_{ki}$ & Eq.~\eqref{diag:U3} \\
\hline
$f_9^i = 8\, \I[q^4]_i^5$ & $U_{ii} G^{\prime\mu\nu}_i G^\prime_{\mu\nu,i}$ & Eq.~\eqref{diag:P4U} \\
\hline
$f_{10}^{ijkl} = \frac{1}{4}\, \I_{ijkl}^{1111}$ & $U_{ij} U_{jk} U_{kl} U_{li}$ & Eq.~\eqref{diag:U4} \\
\hline
$f_{11}^{ijk} = 2\bigl( \,\I[q^2]_{ijk}^{122}\, + \,\I[q^2]_{ijk}^{212}\, \bigr)$ & $U_{ij} [P^\mu, U_{jk}] [P_\mu, U_{ki}]$ & Eq.~\eqref{diag:P2U3} \\
\hline
$f_{12}^{ij} = 4 \, \I[q^4]_{ij}^{33}$ & $\bigl[P^\mu, [P_\mu, U_{ij}]\bigr] \bigl[P^\nu, [P_\nu, U_{ji}]\bigr]$ & \multirow{5}{*}{Eq.~\eqref{diag:P4U2}} \\
\cline{1-2}
$f_{13}^{ij} = 4 \bigl( \, \I[q^4]_{ij}^{33} $ & \multirow{2}{*}{$U_{ij} U_{ji} G^{\prime\mu\nu}_i G^\prime_{\mu\nu,i}$} & \\
\hspace{6ex}$+2 \, \I[q^4]_{ij}^{42} +2 \, \I[q^4]_{ij}^{51}\bigr)$ & & \\
\cline{1-2}
$f_{14}^{ij} = -8 \, \I[q^4]_{ij}^{33}$ & $[P^\mu, U_{ij}] [P^\nu, U_{ji}] G^\prime_{\nu\mu, i}$ & \\
\cline{1-2}
$f_{15}^{ij} = 4\bigl( \, \I[q^4]_{ij}^{33} +\, \I[q^4]_{ij}^{42} \bigr)$ & $\bigl( U_{ij} [P^\mu, U_{ji}] -[P^\mu, U_{ij}] U_{ji} \bigr) [P^\nu, G^\prime_{\nu\mu,i}]$ & \\
\hline
$f_{16}^{ijklm} = \frac{1}{5}\, \I_{ijklm}^{11111}$ & $U_{ij} U_{jk} U_{kl} U_{lm} U_{mi}$ & Eq.~\eqref{diag:U5} \\
\hline
$f_{17}^{ijkl} = 2 \bigl(\, \I[q^2]_{ijkl}^{2112} $ & \multirow{2}{*}{$U_{ij} U_{jk} [P^\mu, U_{kl}] [P_\mu, U_{li}]$} & \multirow{4}{*}{Eq.~\eqref{diag:P2U4}} \\
\hspace{8ex}$+\, \I[q^2]_{ijkl}^{1212} +\, \I[q^2]_{ijkl}^{1122} \,\bigr)$ & & \\
\cline{1-2}
$f_{18}^{ijkl} = \,\I[q^2]_{ijkl}^{2121} +\, \I[q^2]_{ijkl}^{2112} $ & \multirow{2}{*}{$U_{ij} [P^\mu, U_{jk}] U_{kl} [P_\mu, U_{li}]$} & \\
\hspace{8ex}$+\, \I[q^2]_{ijkl}^{1221} +\, \I[q^2]_{ijkl}^{1212}$ &  & \\
\hline
$f_{19}^{ijklmn} = \frac{1}{6}\, \I_{ijklmn}^{111111}$ & $U_{ij} U_{jk} U_{kl} U_{lm} U_{mn} U_{ni}$ & Eq.~\eqref{diag:U6} \\
\hline
\end{tabular}
\caption{
List of universal coefficients in terms of the master integrals defined in Eq.~\eqref{MIdef} (Column 1). The UOLEA master formula for one-loop matching reported in~\cite{DEQY} is reproduced by adding up traces of the operators in Column 2 with the corresponding universal coefficients, and multiplying the overall factor $-i c_s$; see Eq.~\eqref{LUOLEA}. The covariant diagrams used to compute each universal coefficient are listed in Column 3. See Appendix~\ref{app:coefficients} for expressions of the universal coefficients in terms of heavy particle masses.
\label{tab:UOLEA}}
\end{table}

\subsection{Integrating out a scalar triplet: the scalar sector}
\label{sec:triplet-scalar}

We next consider more specific examples where additional ingredients in Tables~\ref{tab:mixed}, \ref{tab:ocd} and~\ref{tab:ferm} are involved in covariant diagrams. Our goal is to demonstrate the techniques, instead of deriving complete universal master formulas. The latter task is left to future publications.

As a standard test case, a simple extension of the SM by a heavy electroweak scalar triplet was used in several recent papers~\cite{HLM16,EQYZ,FPR} to illustrate various functional approaches to mixed heavy-light matching at work. The scalar sector of the model is given by
\beqa
\mathcal{L}_\text{UV} \,&\supset&\, |D_\mu \phi|^2 - m^2|\phi|^2 - \lambda|\phi|^4 + \frac{1}{2} (D_\mu\Phi^a)^2 - \frac{1}{2}M^2\Phi^a\Phi^a -\frac{1}{4}\lambda_\Phi(\Phi^a\Phi^a)^2 \CR
&&+ \kappa \phi^\dagger\sigma^a \phi \Phi^a - \eta |\phi|^2 \Phi^a\Phi^a \, ,
\eeqa{LUVtriplet}
where $\Phi$ is a heavy $SU(2)_L$ triplet with zero hypercharge, and $\phi$ is the light SM Higgs doublet with mass squared $m^2<0$. We shall focus on the following subset of dimension-six effective operators\,\footnote{We will not make any field or parameter redefinitions unless otherwise specified, so that the operator coefficients are unambiguous.} generated by integrating out $\Phi$,
\beq
\O_T = \frac{1}{2} \bigl( \phi^\dagger \overleftrightarrow{D}_\mu \phi \bigr)^2 \, , \quad
\O_H = \frac{1}{2} \bigl( \partial_\mu|\phi|^2 \bigr)^2 \, , \quad
\O_R = |\phi|^2|D_\mu \phi|^2 \, ,
\eeq{OTOHOR}
where $\phi^\dagger \overleftrightarrow{D}_\mu \phi = \phi^\dagger (D_\mu \phi) -(D_\mu \phi^\dagger)\phi$. Pure heavy contributions to the operator coefficients can be easily obtained by applying the degenerate limit of the UOLEA, which is illustrated in~\cite{HLM14}. We will thus be interested in computing mixed heavy-light contributions. We first reproduce, in the present subsection, the results in~\cite{HLM16,EQYZ} for terms independent of the SM gauge couplings. Terms that depend on the SM gauge couplings, which involve treatment of open covariant derivatives and were not obtained in~\cite{HLM16,EQYZ}, will be computed in the next subsection.

To begin with, we solve for $\Phi_\c[\phi]$ up to the order needed [counting $\kappa$ as $\O(M)$],
\beq
\Phi_\c^a [\phi] = \frac{\kappa}{M^2}\phi^\dagger \sigma^a \phi - \frac{\kappa}{M^4}\Bigl[2\eta|\phi|^2 \bigl(\phi^\dagger \sigma^a \phi \bigr) +D^2\bigl(\phi^\dagger \sigma^a \phi \bigr)\Bigr] +\O (M^{-5}) \, ,
\eeqn
and extract the $\bf U$ matrix from the quadratic terms of Eq.~\eqref{LUVtriplet},
\beq
\L_\text{UV, quad.} \supset 
-\frac{1}{2}
\bigl( \Phi^{\prime a} \; \phi^{\prime\dagger} \; \tilde\phi^{\prime\dagger} \bigr)
\bigl( -P^2 +{\bf M}^2 + {\bf U}[\Phi_\b, \phi_\b, \tilde\phi_\b] \bigr)
\left(
\begin{matrix}
\Phi^{\prime b} \\
\phi^\prime \\
\tilde\phi^\prime
\end{matrix}
\right),
\eeqn
where
\beqa
{\bf M}^2 &=& \text{diag}(M^2 \delta^{ab}, m^2, m^2) \,,\\[4pt]
{\bf U} &=&
\left(
\begin{matrix}
U_H & (U_{HL})_{1\times2} \\
(U_{LH})_{2\times1} & (U_L)_{2\times2}
\end{matrix}
\right)
=
\left(
\begin{matrix}
U_\Phi^{ab} & (U_{\phi\Phi}^{\dagger a})_{1\times2} \\
(U_{\phi\Phi}^b)_{2\times1} & (U_\phi)_{2\times2}
\end{matrix}
\right).
\eeqan
The internal index ``$b$'' (italicized) should not be confused with the subscript label ``b'' (for background). 
The components of the $\bf U$ matrix, with $\Phi$ set to $\Phi_\c[\phi]$, read
\bseq
\beqa
U_\Phi^{ab} &=& 2\eta\, |\phi|^2 \delta^{ab} + \lambda_\Phi \bigl( \Phi_\c^d\Phi_\c^d \, \delta^{ab} + 2\, \Phi_\c^a \Phi_\c^b \bigr) 
\sim \O(\phi^2,\, \phi^4,\, P^2 \phi^4, \dots), \\
U_{\phi\Phi}^b &=& 
\left(
\begin{matrix}
-\kappa\, \sigma^b \phi + 2\eta\, \phi \, \Phi_\c^b \\
\kappa\, \sigma^b \tilde\phi + 2\eta \, \tilde\phi\, \Phi_\c^b
\end{matrix}
\right)
\sim \O(\phi,\, \phi^3,\, P^2 \phi^3,\dots), \\
U_\phi &=&
\begin{small}
\left(
\begin{matrix}
2\lambda\, (|\phi|^2\, \identity_2 + \phi\,\phi^\dagger) - \kappa\, \Phi_\c^d\sigma^d + \eta\, \Phi_\c^d\Phi_\c^d \,\identity_2 &
2\lambda\, \phi\,\tilde\phi^\dagger \\
2\lambda\, \tilde\phi\,\phi^\dagger &
2\lambda\, (|\phi|^2\, \identity_2 + \tilde\phi\, \tilde\phi^\dagger) + \kappa\, \Phi_\c^d\sigma^d + \eta\, \Phi_\c^d\Phi_\c^d \,\identity_2
\end{matrix}
\right)
\end{small} \CR
&&\sim \O(\phi^2,\, \phi^4,\, P^2 \phi^2,\, P^2 \phi^4,\dots).
\eeqan
\eseq{Utriplet}
Note that the two real components of the complex scalar $\phi$ should be written out separately in the field multiplet. In practice, it is convenient to use $\phi$ and $\tilde\phi\equiv i\sigma^2 \phi^*$, since $\tilde\phi$ transforms in the same way as $\phi$ under $SU(2)_L$.

From Eq.~\eqref{Utriplet} it is clear that to obtain mixed heavy-light contributions to the operators $\O_T, \O_H, \O_R$ in Eq.~\eqref{OTOHOR}, all of which contain four $\phi$'s and two covariant derivatives, we need to compute one-loop covariant diagrams that are proportional to
\beq
U_{HL} U_{LH},\; 
U_{HL} U_L U_{LH},\; 
P^2 U_{HL} U_{LH},\; 
P^2 U_{HL} U_{LH} U_H,\; 
P^2 U_{HL} U_L U_{LH},\; 
P^2 (U_{HL} U_{LH})^2.
\eeq{tripletenumerate}
Using the rules in Tables~\ref{tab:heavy} and~\ref{tab:mixed}, we have (with $M_i=M$ in the master integrals from here on)
\bseq
\beqa
&&
\begin{gathered}
\begin{fmffile}{triplet-UhlUlh}
\begin{fmfgraph}(40,40)
\fmfsurround{vULH,vUHL}
\fmf{plain,right=1}{vULH,vUHL}
\fmf{dashes,right=1}{vUHL,vULH}
\fmfv{decor.shape=circle,decor.filled=empty,decor.size=3thick}{vUHL,vULH}
\end{fmfgraph}
\end{fmffile}
\end{gathered}
\;\;\, =\, -i c_s \,\I_{i0}^{11}\,\tr (U_{HL} U_{LH}) , \\[4pt]
&&
\begin{gathered}
\begin{fmffile}{triplet-UhlUlUlh}
\begin{fmfgraph}(40,40)
\fmfsurround{vUL,vUHL,vULH}
\fmf{plain,left=0.6}{vULH,vUHL}
\fmf{dashes,left=0.6}{vUHL,vUL,vULH}
\fmfv{decor.shape=circle,decor.filled=empty,decor.size=3thick}{vUHL,vUL,vULH}
\end{fmfgraph}
\end{fmffile}
\end{gathered}
\;\;\, =\, -i c_s \,\I_{i0}^{12}\,\tr (U_{HL} U_L U_{LH}) , \\
&&
\begin{gathered}
\begin{fmffile}{triplet-P2UhlUlh}
\begin{fmfgraph}(40,40)
\fmfsurround{vP2,vUHL,vP1,vULH}
\fmf{plain,left=0.4}{vULH,vP1,vUHL}
\fmf{dashes,left=0.4}{vUHL,vP2,vULH}
\fmfv{decor.shape=circle,decor.filled=full,decor.size=3thick}{vP1,vP2}
\fmfv{decor.shape=circle,decor.filled=empty,decor.size=3thick}{vUHL,vULH}
\fmf{dots,width=thick}{vP1,vP2}
\end{fmfgraph}
\end{fmffile}
\end{gathered}
\;\;\, =\, -i c_s \,2^2 \,\I[q^2]_{i0}^{22}\,\tr (P^\mu U_{HL} P_\mu U_{LH}) 
\subset -i c_s \,2 \,\I[q^2]_{i0}^{22}\,\tr\bigl( [P^\mu, U_{HL}] [P_\mu, U_{LH}] \bigr) \,, \CR\\
&&
\begin{fmffile}{triplet-P2UhlUlhUh}
\begin{gathered}
\begin{fmfgraph}(40,40)
\fmfsurround{vUH,vP2,vULH,vUHL,vP1}
\fmf{plain,left=0.3}{vULH,vP2,vUH,vP1,vUHL}
\fmf{dashes,left=0.3}{vUHL,vULH}
\fmfv{decor.shape=circle,decor.filled=full,decor.size=3thick}{vP1,vP2}
\fmfv{decor.shape=circle,decor.filled=empty,decor.size=3thick}{vUHL,vULH,vUH}
\fmf{dots,width=thick,left=0.2}{vP1,vP2}
\end{fmfgraph}
\end{gathered}
\;\;+\;\,
\begin{gathered}
\begin{fmfgraph}(40,40)
\fmfsurround{vULH,vP2,vUHL,vUH,vP1}
\fmf{plain,left=0.3}{vULH,vP1,vUH,vUHL}
\fmf{dashes,left=0.3}{vUHL,vP2,vULH}
\fmfv{decor.shape=circle,decor.filled=full,decor.size=3thick}{vP1,vP2}
\fmfv{decor.shape=circle,decor.filled=empty,decor.size=3thick}{vUHL,vULH,vUH}
\fmf{dots,width=thick,left=0.2}{vP1,vP2}
\end{fmfgraph}
\end{gathered}
\;\;+\;\,
\begin{gathered}
\begin{fmfgraph}(40,40)
\fmfsurround{vUHL,vP1,vUH,vULH,vP2}
\fmf{plain,left=0.3}{vULH,vUH,vP1,vUHL}
\fmf{dashes,left=0.3}{vUHL,vP2,vULH}
\fmfv{decor.shape=circle,decor.filled=full,decor.size=3thick}{vP1,vP2}
\fmfv{decor.shape=circle,decor.filled=empty,decor.size=3thick}{vUHL,vULH,vUH}
\fmf{dots,width=thick,right=0.2}{vP1,vP2}
\end{fmfgraph}
\end{gathered}
\end{fmffile}
\CR[6pt]
&& = -i c_s 2^2 
\bigl\{
\I[q^2]_{i0}^{41} \tr (P_\mu U_{HL} U_{LH} P^\mu U_H) \CR
&&\qquad\qquad
+\I[q^2]_{i0}^{32} \,\tr(P^\mu U_H U_{HL} P_\mu U_{LH} +P^\mu U_{HL} P_\mu U_{LH} U_H)
\bigr\} \CR[2pt]
&& \subset -i c_s \bigl\{ 4\,\I[q^2]_{i0}^{32}\,\tr\bigl( [P^\mu, U_{HL}] [P_\mu, U_{LH}] U_H \bigr) \CR
&&\qquad\qquad
+2\,(\,\I[q^2]_{i0}^{41}+\I[q^2]_{i0}^{32})\,\tr\bigl( [P^\mu, U_{HL} U_{LH}] [P_\mu, U_{H}] \bigr) \bigr\},\\[10pt]
&&
\begin{fmffile}{triplet-P2UhlUlUlh}
\begin{gathered}
\begin{fmfgraph}(40,40)
\fmfsurround{vUL,vP2,vUHL,vULH,vP1}
\fmf{plain,left=0.3}{vULH,vUHL}
\fmf{dashes,left=0.3}{vUHL,vP2,vUL,vP1,vULH}
\fmfv{decor.shape=circle,decor.filled=full,decor.size=3thick}{vP1,vP2}
\fmfv{decor.shape=circle,decor.filled=empty,decor.size=3thick}{vUHL,vUL,vULH}
\fmf{dots,width=thick,left=0.2}{vP1,vP2}
\end{fmfgraph}
\end{gathered}
\;\;+\;\,
\begin{gathered}
\begin{fmfgraph}(40,40)
\fmfsurround{vUHL,vP1,vULH,vUL,vP2}
\fmf{plain,left=0.3}{vULH,vP1,vUHL}
\fmf{dashes,left=0.3}{vUHL,vP2,vUL,vULH}
\fmfv{decor.shape=circle,decor.filled=full,decor.size=3thick}{vP1,vP2}
\fmfv{decor.shape=circle,decor.filled=empty,decor.size=3thick}{vUHL,vUL,vULH}
\fmf{dots,width=thick,right=0.2}{vP1,vP2}
\end{fmfgraph}
\end{gathered}
\;\;+\;\,
\begin{gathered}
\begin{fmfgraph}(40,40)
\fmfsurround{vULH,vP2,vUL,vUHL,vP1}
\fmf{plain,left=0.3}{vULH,vP1,vUHL}
\fmf{dashes,left=0.3}{vUHL,vUL,vP2,vULH}
\fmfv{decor.shape=circle,decor.filled=full,decor.size=3thick}{vP1,vP2}
\fmfv{decor.shape=circle,decor.filled=empty,decor.size=3thick}{vUHL,vUL,vULH}
\fmf{dots,width=thick,left=0.2}{vP1,vP2}
\end{fmfgraph}
\end{gathered}
\end{fmffile}
\CR[6pt]
&& = -i c_s \,2^2 
\bigl\{
\I[q^2]_{i0}^{14} \,\tr (P_\mu U_{LH} U_{HL} P^\mu U_L) \CR
&&\qquad\qquad
+\I[q^2]_{i0}^{23} \,\tr(P^\mu U_L U_{LH} P_\mu U_{HL} +P^\mu U_{LH} P_\mu U_{HL} U_L)
\bigr\} \CR[2pt]
&& \subset -i c_s \bigl\{ 4\,\I[q^2]_{i0}^{23}\,\tr\bigl( [P^\mu, U_{LH}] [P_\mu, U_{HL}] U_L\bigr) \CR
&&\qquad\qquad
+2\,(\I[q^2]_{i0}^{14}+\I[q^2]_{i0}^{23})\,\tr\bigl( [P^\mu, U_{LH} U_{HL}] [P_\mu, U_L] \bigr) \bigr\},\CR\\
&&
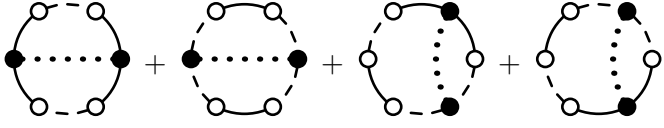
\begin{fmffile}{triplet-P2UhlUlh2}
\begin{gathered}
\begin{fmfgraph}(40,40)
\fmfsurround{vP2,vULH1,vUHL1,vP1,vULH2,vUHL2}
\fmf{plain,left=0.25}{vULH2,vP1,vUHL1}
\fmf{dashes,left=0.25}{vUHL1,vULH1}
\fmf{plain,left=0.25}{vULH1,vP2,vUHL2}
\fmf{dashes,left=0.25}{vUHL2,vULH2}
\fmfv{decor.shape=circle,decor.filled=full,decor.size=3thick}{vP1,vP2}
\fmfv{decor.shape=circle,decor.filled=empty,decor.size=3thick}{vUHL1,vULH1,vUHL2,vULH2}
\fmf{dots,width=thick}{vP1,vP2}
\end{fmfgraph}
\end{gathered}
\;\;+\;\,
\begin{gathered}
\begin{fmfgraph}(40,40)
\fmfsurround{vP2,vUHL1,vULH1,vP1,vUHL2,vULH2}
\fmf{dashes,left=0.25}{vUHL2,vP1,vULH1}
\fmf{plain,left=0.25}{vULH1,vUHL1}
\fmf{dashes,left=0.25}{vUHL1,vP2,vULH2}
\fmf{plain,left=0.25}{vULH2,vUHL2}
\fmfv{decor.shape=circle,decor.filled=full,decor.size=3thick}{vP1,vP2}
\fmfv{decor.shape=circle,decor.filled=empty,decor.size=3thick}{vUHL1,vULH1,vUHL2,vULH2}
\fmf{dots,width=thick}{vP1,vP2}
\end{fmfgraph}
\end{gathered}
\;\;+\;\,
\begin{gathered}
\begin{fmfgraph}(40,40)
\fmfsurround{vUHL1,vP1,vULH2,vUHL2,vULH1,vP2}
\fmf{plain,left=0.25}{vULH2,vP1,vUHL1}
\fmf{dashes,left=0.25}{vUHL1,vP2,vULH1}
\fmf{plain,left=0.25}{vULH1,vUHL2}
\fmf{dashes,left=0.25}{vUHL2,vULH2}
\fmfv{decor.shape=circle,decor.filled=full,decor.size=3thick}{vP1,vP2}
\fmfv{decor.shape=circle,decor.filled=empty,decor.size=3thick}{vUHL1,vULH1,vUHL2,vULH2}
\fmf{dots,width=thick,right=0.3}{vP1,vP2}
\end{fmfgraph}
\end{gathered}
\;\;+\;\,
\begin{gathered}
\begin{fmfgraph}(40,40)
\fmfsurround{vULH1,vP1,vUHL2,vULH2,vUHL1,vP2}
\fmf{dashes,left=0.25}{vUHL2,vP1,vULH1}
\fmf{plain,left=0.25}{vULH1,vP2,vUHL1}
\fmf{dashes,left=0.25}{vUHL1,vULH2}
\fmf{plain,left=0.25}{vULH2,vUHL2}
\fmfv{decor.shape=circle,decor.filled=full,decor.size=3thick}{vP1,vP2}
\fmfv{decor.shape=circle,decor.filled=empty,decor.size=3thick}{vUHL1,vULH1,vUHL2,vULH2}
\fmf{dots,width=thick,right=0.3}{vP1,vP2}
\end{fmfgraph}
\end{gathered}
\end{fmffile}
\CR[6pt]
&& = -i c_s \,2^2 \Bigl\{
\frac{1}{2}\,\I[q^2]_{i0}^{42}\,\tr (P^\mu U_{HL} U_{LH} P_\mu U_{HL} U_{LH}) +
\frac{1}{2}\,\I[q^2]_{i0}^{24}\,\tr (P^\mu U_{LH} U_{HL} P_\mu U_{LH} U_{HL}) \CR
&&\qquad\qquad + \,\I[q^2]_{i0}^{33}\,\tr (P^\mu U_{HL} P_\mu U_{LH} U_{HL} U_{LH} + P^\mu U_{LH} P_\mu U_{HL} U_{LH} U_{HL} )
\Bigr\} \CR[2pt]
&& \subset -i c_s \bigl\{
(2\,\I[q^2]_{i0}^{24}+4\,\I[q^2]_{i0}^{33})\,\tr\bigl( [P^\mu, U_{HL}] [P_\mu, U_{LH}] U_{HL} U_{LH}\bigr) \CR
&&\qquad\qquad +(2\,\I[q^2]_{i0}^{42}+4\,\I[q^2]_{i0}^{33}) \,\tr\bigl( [P^\mu, U_{LH}] [P_\mu, U_{HL}] U_{LH} U_{HL}\bigr) \CR
&&\qquad\qquad +(\I[q^2]_{i0}^{42}+\I[q^2]_{i0}^{24} +2\,\I[q^2]_{i0}^{33})\CR
&&\qquad\qquad\quad \tr\,\bigl( [P^\mu, U_{HL}] U_{LH} [P_\mu, U_{HL}] U_{LH} + U_{HL} [P^\mu, U_{LH}] U_{HL} [P_\mu, U_{LH}] \bigr)
\bigr\}.
\eeqan
\eseq{tripletdiagrams}
Note that diagrams with $m^2$ insertions are of higher order and therefore not considered. The results in the equations above are summarized in Table~\ref{tab:triplet}, where explicit expressions for the coefficients and operators are also worked out. Summing up all terms in the table, we obtain (with $c_s=\frac{1}{2}$ and $\mu$ set to $M$)
\beqa
\L_\text{EFT}^\text{1-loop}[\phi] \supset \frac{1}{16\pi^2}\frac{3\kappa^2}{2M^2}|D_\mu \phi|^2 + \frac{1}{16\pi^2}\frac{\kappa^2}{M^4}\Bigl[\Bigl(\frac{\kappa^2}{2M^2} -8\eta + 3\lambda \Bigr)\O_T  \CR
+ \Bigl(- \frac{9\kappa^2}{2M^2} -6\eta + 10\lambda \Bigr)\O_H+ \Bigl(- \frac{21\kappa^2}{2M^2} -21\eta + 25\lambda \Bigr)\O_R \Bigr] \, ,
\eeqan
in agreement with~\cite{DKS,HLM16,EQYZ}\,\footnote{There is an additional contribution to $\L_\text{EFT}^\text{1-loop}[\phi]$ from $\L_\text{EFT}^\text{tree}[\phi] \supset \frac{\kappa^2}{M^4}(\O_T+2\O_R) \to (1-\frac{1}{16\pi^2}\frac{3\kappa^2}{M^2})\frac{\kappa^2}{M^4}(\O_T+2\O_R)$ if one rescales the SM Higgs field $\phi \to (1-\frac{1}{16\pi^2}\frac{3\kappa^2}{4M^2}) \phi$ to render its kinetic term canonically normalized.}.

\begin{table}[tbp]
\centering
\begin{small}
\begin{tabular}{|l|l|}
\hline
Coefficient & Operator \\
\hline
\multirow{2}{*}{$-ic_s\,\I_{i0}^{11} = \pref \bigl( 1 -\logm{M^2} \bigr)$} & $\tr (U_{HL} U_{LH})$ \\
& $\to U_{\phi\Phi}^{\dagger a} U_{\phi\Phi}^a \supset -\frac{16\kappa^2\eta}{M^4}(\O_T +2 \O_R )$ \\
\hline
\multirow{2}{*}{$-i c_s \,\I_{i0}^{12} = \pref \frac{1}{M^2} \bigl( 1 -\logm{M^2} \bigr)$} & $\tr (U_{HL} U_L U_{LH})$ \\
& $\to U_{\phi\Phi}^{\dagger a} U_\phi U_{\phi\Phi}^a \supset \frac{4\kappa^4}{M^4} (\O_T +2\O_R)$ \\
\hline
\multirow{3}{*}{$-ic_s\,2\,\I[q^2]_{i0}^{22} = \pref \bigl(-\frac{1}{2M^2}\bigr)$} & $\tr\bigl([P^\mu, U_{HL}] [P_\mu, U_{LH}]\bigr)$ \\
& $\to [P^\mu, U_{\phi\Phi}^{\dagger a}] [P_\mu, U_{\phi\Phi}^a]$ \\
& $\quad\supset -6\kappa^2 |D_\mu\phi|^2 + \frac{8\kappa^2\eta}{M^2}\left(\O_H + \O_R\right) $ \\
\hline
\multirow{2}{*}{$-ic_s\,4\,\I[q^2]_{i0}^{32} = \pref\frac{1}{2M^4}$} & $\tr\bigl([P^\mu, U_{HL}] [P_\mu, U_{LH}] U_H\bigr)$ \\
& $\to [P^\mu, U_{\phi\Phi}^{\dagger a}] [P_\mu, U_{\phi\Phi}^b] U_\Phi^{ba} \supset -12\kappa^2 \eta \O_R$ \\
\hline
\multirow{2}{*}{$-ic_s\,2\,(\I[q^2]_{i0}^{41}+\I[q^2]_{i0}^{32}) = \pref\frac{1}{3M^4}$} & $\tr\bigl([P^\mu, U_{HL} U_{LH}] [P_\mu, U_{H}]\bigr)$ \\
& $\to [P^\mu, U_{\phi\Phi}^{\dagger a} U_{\phi\Phi}^b]  [P_\mu, U_\Phi^{ba}]\supset -24\kappa^2\eta \O_H$ \\
\hline
\multirow{4}{*}{$-ic_s\,4\,\I[q^2]_{i0}^{23} = \pref\frac{1}{M^4}\bigl(-\frac{5}{2} +\logm{M^2}\bigr)$} & $\tr\bigl([P^\mu, U_{LH}] [P_\mu, U_{HL}] U_L\bigr)$ \\
& $\to [P^\mu, U_{\phi\Phi}^a] [P_\mu, U_{\phi\Phi}^{\dagger a}] U_\phi$ \\
& $\quad\supset 2\kappa^2 \bigl[ \bigl(\frac{\kappa^2}{M^2}-2\lambda\bigr)\O_T - \frac{\kappa^2}{M^2}\O_H$ \\
& $\qquad\quad + \bigl(\frac{\kappa^2}{M^2} - 10\lambda\bigr) \O_R \bigr]$ \\
\hline
\multirow{4}{*}{$-ic_s\,2\,(\I[q^2]_{i0}^{14}+\I[q^2]_{i0}^{23}) = \pref \bigl(-\frac{1}{2M^4}\bigr)$} & $\tr\bigl([P^\mu, U_{LH} U_{HL}] [P_\mu, U_{L}]\bigr)$ \\
& $\to [P^\mu, U_{\phi\Phi}^a U_{\phi\Phi}^{\dagger a}]  [P_\mu, U_\phi]$ \\
& $\quad\supset 4\kappa^2 \bigl[ \bigl( -\frac{\kappa^2}{M^2} + 2\lambda\bigr)\O_T$ \\
& $\qquad\quad - 10\lambda\O_H - \frac{2\kappa^2}{M^2}\O_R \bigr]$ \\
\hline
\multirow{2}{*}{$-ic_s (2\,\I[q^2]_{i0}^{24}+4\,\I[q^2]_{i0}^{33}) = \pref\frac{1}{M^6}$} & $\tr\bigl([P^\mu, U_{HL}] [P_\mu, U_{LH}] U_{HL} U_{LH}\bigr)$ \\
& $\to [P^\mu, U_{\phi\Phi}^{\dagger a}] [P_\mu, U_{\phi\Phi}^b] U_{\phi\Phi}^{\dagger b} U_{\phi\Phi}^a \supset -12\kappa^4 \O_R$ \\
\hline
\multirow{3}{*}{$-ic_s (2\,\I[q^2]_{i0}^{42}+4\,\I[q^2]_{i0}^{33}) = \pref\frac{1}{M^6}\bigl(\frac{17}{6}-\logm{M^2}\bigr)$} & $\tr\bigl([P^\mu, U_{LH}] [P_\mu, U_{HL}] U_{LH} U_{HL}\bigr)$ \\
& $\to [P^\mu, U_{\phi\Phi}^a] [P_\mu, U_{\phi\Phi}^{\dagger a}] U_{\phi\Phi}^b U_{\phi\Phi}^{\dagger b}$ \\
& $\quad \supset -2\kappa^4(\O_H +4\O_R)$ \\
\hline
\multirow{5}{*}{$-ic_s(\I[q^2]_{i0}^{42}+\I[q^2]_{i0}^{24}+2\,\I[q^2]_{i0}^{33}) = \pref\frac{5}{12M^6}$} & $\tr\bigl( [P^\mu, U_{HL}] U_{LH} [P_\mu, U_{HL}] U_{LH}$ \\
& $\quad + U_{HL} [P^\mu, U_{LH}] U_{HL} [P_\mu, U_{LH}]\bigr)$ \\
& $\to [P^\mu, U_{\phi\Phi}^{\dagger a}] U_{\phi\Phi}^b [P_\mu, U_{\phi\Phi}^{\dagger b}] U_{\phi\Phi}^a$ \\
& $\quad +U_{\phi\Phi}^{\dagger a} [P^\mu, U_{\phi\Phi}^b] U_{\phi\Phi}^{\dagger b} [P_\mu, U_{\phi\Phi}^a]$ \\
& $\quad \supset 4\kappa^4 (-5\O_H +4\O_R)$ \\
\hline
\end{tabular}
\end{small}
\caption{
Summary of the results in Eq.~\eqref{tripletdiagrams} for mixed heavy-light contributions to one-loop matching for the scalar triplet model. The SM gauge coupling-independent terms for the three operators $\O_T, \O_H, \O_R$ in Eq.~\eqref{OTOHOR} are computed (in the $\overline{\text{MS}}$ scheme).
\label{tab:triplet}}
\end{table}

Two comments are in order:
\begin{itemize}
\item The calculation above parallels that in~\cite{EQYZ}. In particular, it is the same calculation in the ``Operator'' column of Table~\ref{tab:triplet} that is done in~\cite{EQYZ}; the coefficients part, however, follows from a more straightforward computation here than in~\cite{EQYZ}.
\item While the calculation in this subsection was done in the context of the scalar triplet model, most of the results obtained are universal. In fact, the only model-dependent part is the expression after each ``$\to$'' in the ``Operator'' column of Table~\ref{tab:triplet}. In this respect, Eq.~\eqref{tripletdiagrams} constitutes part of the derivation of a master formula for mixed heavy-light matching (with degenerate heavy particle masses), which we plan to complete in future work.
\end{itemize}

\subsection{Integrating out a scalar triplet: the gauge sector}
\label{sec:triplet-gauge}

Now we move on to the gauge sector of the scalar triplet model. To account for mixed heavy-light contributions to one-loop matching that involve SM gauge interactions, we need to extend the field multiplet to include the electroweak gauge bosons. The relevant quadratic pieces of the UV theory Lagrangian then read
\beq
\L_\text{UV, quad.} \supset 
-\frac{1}{2}
\bigl( \Phi^{\prime a} \; \phi^{\prime\dagger} \; \tilde\phi^{\prime\dagger} \; W^{\prime a}_\alpha \; B^\prime_\alpha \bigr)
\bigl( -P^2 +{\bf M}^2 + {\bf U} +P_\mu {\bf Z}^\mu + {\bf Z}^{\dagger\mu} P_\mu \bigr)
\left(
\begin{matrix}
\Phi^{\prime b} \\
\phi^\prime \\
\tilde\phi^\prime \\
W^{\prime b}_\beta \\
B^\prime_\beta
\end{matrix}
\right),
\eeqn
where the arguments $[\Phi_\b, \phi_\b, \tilde\phi_\b, W_\b, B_\b]$ of the $\bf U$ and $\bf Z$ matrices have been dropped for simplicity, and
\beqa
{\bf M}^2 &=& \text{diag}(M^2, m^2, m^2, 0, 0) \,,\\
{\bf U} &=&
\left(
\begin{matrix}
U_H & (U_{HL})_{1\times4} \\
(U_{LH})_{4\times1} & (U_L)_{4\times4}
\end{matrix}
\right)
=
\left(
\begin{matrix}
U_\Phi^{ab} & (U_{\phi\Phi}^{\dagger a})_{1\times2} & U_{\Phi W}^{ab\beta} & 0 \\
(U_{\phi\Phi}^b)_{2\times1} & (U_\phi)_{2\times2} & (U_{\phi W}^{b\beta})_{2\times1} & (U_{\phi B}^\beta)_{2\times1} \\
U_{\Phi W}^{\dagger ab\alpha} & (U_{\phi W}^{\dagger a\alpha})_{1\times2} & U_W^{ab\alpha\beta} & U_{BW}^{a\alpha\beta} \\
0 & (U_{\phi B}^{\dagger \alpha})_{1\times2} & U_{BW}^{b\alpha\beta} &  U_B^{\alpha\beta}
\end{matrix}
\right),\quad \\
{\bf Z}^\mu &=&
\left(
\begin{matrix}
Z_H^\mu & (Z_{HL}^\mu)_{1\times4} \\
(Z_{LH}^\mu)_{4\times1} & (Z_L^\mu)_{4\times4}
\end{matrix}
\right)
=
\left(
\begin{matrix}
0 & {\bf 0}_{1\times2} & Z_{\Phi W}^{\mu\,ab\beta} & 0 \\
{\bf 0}_{2\times1} & {\bf 0}_{2\times2} & (Z_{\phi W}^{\mu\, b\beta})_{2\times1} & (Z_{\phi B}^{\mu\,\beta})_{2\times1} \\
0 & {\bf 0}_{1\times2} & 0 & 0 \\
0 & {\bf 0}_{1\times2} & 0 &  0
\end{matrix}
\right).
\eeqan
Note that $W$ and $B$ vector bosons are massless in the $SU(2)_L\times U(1)_Y$ symmetric phase and, as discussed in Section~\ref{sec:cd-mixed}, there is no need to retain their masses in the calculation as IR regulators. Also, Lorentz indices $\alpha, \beta$ of the vector bosons are treated on the same footing as internal indices. With $\Phi$ set to $\Phi_\c[\phi]$, the relevant components of the $\bf U$ and $\bf Z$ matrices are, in addition to those in Eq.~\eqref{Utriplet},
\bseq
\beqa
&& Z_{\Phi W}^{\mu\,ab\beta} = g^{\mu\beta} ig\epsilon^{adb}\Phi_\c^d
\sim \O(g\phi^2,\, gP^2\phi^2,\, g\phi^4, \dots)\,, \quad
U_{\Phi W} = [P_\mu, Z_{\Phi W}^\mu]\,, \\
&& Z_{\phi W}^{\mu\, b\beta} = -g^{\mu\beta}\frac{g}{2}
\left(\begin{matrix}
\sigma^b \phi \\
\sigma^b \tilde\phi
\end{matrix}\right)
\sim \O(g\phi) \,, \quad
U_{\phi W} = [P_\mu, Z_{\phi W}^\mu] \,, \\
&& Z_{\phi B}^{\mu\,\beta} = -g^{\mu\beta}\frac{g'}{2}
\left(\begin{matrix}
\phi \\
-\tilde\phi
\end{matrix}\right)
\sim \O(g'\phi) \,,\quad
U_{\phi B} = [P_\mu, Z_{\phi B}^\mu] \,. 
\eeqan
\eseq{Ztriplet}
We are interested in terms in $\L_\text{EFT}^\text{1-loop}$ from mixed heavy-light matching that are $\O(g^2 P^2\phi^4)$ or $\O(g'^2 P^2\phi^4)$\,\footnote{Higher powers of $g$ or $g'$ are not possible at one loop, which can be easily seen by $\hbar$ dimension counting.}, which can come from, schematically,
\bseq
\beqa
&&
Z_{\Phi W} Z_{\Phi W}^\dagger \subset Z_{HL} Z_{HL}^\dagger,\quad
P^2 Z_{\Phi W} Z_{\Phi W}^\dagger \subset P^2 Z_{HL} Z_{HL}^\dagger,\CR
&&
P\, Z_{\Phi W} U_{\Phi W}^\dagger +\text{h.c.} \subset P\, Z_{HL} U_{LH} +\text{h.c.},\quad
U_{\Phi W} U_{\Phi W}^\dagger \subset U_{HL} U_{LH} ; \\
&&
Z_{\Phi W} Z_{\phi W}^\dagger U_{\phi\Phi}+\text{h.c.} \subset Z_{HL} Z_L^\dagger U_{LH} +\text{h.c.},\CR
&&
P^2 U_{\phi\Phi}^\dagger Z_{\phi W} Z_{\Phi W}^\dagger +\text{h.c.} \subset P^2 U_{HL} Z_L Z_{HL}^\dagger +\text{h.c.},\CR
&&
P\, Z_{\Phi W} U_{\phi W}^\dagger U_{\phi\Phi} +\text{h.c.} \subset P\, Z_{HL} U_L U_{LH} +\text{h.c.},\CR
&&
P\, U_{\phi\Phi}^\dagger Z_{\phi W} U_{\Phi W}^\dagger +\text{h.c.} \subset P\, U_{HL} Z_L U_{LH} +\text{h.c.},\CR
&&
U_{\Phi W} U_{\phi W}^\dagger U_{\phi\Phi} +\text{h.c.} \subset U_{HL} U_L U_{LH}; \\
&&
P^2 U_{\phi\Phi}^\dagger Z_{\phi V} Z_{\phi V}^\dagger U_{\phi\Phi} \subset P^2 U_{HL} Z_L Z_L^\dagger U_{LH} ,\CR
&&
P\, U_{\phi\Phi}^\dagger Z_{\phi V} U_{\phi V}^\dagger U_{\phi\Phi} +\text{h.c.} \subset P\, U_{HL} Z_L U_L U_{LH} +\text{h.c.},\CR
&&
U_{\phi\Phi}^\dagger U_{\phi V} U_{\phi V}^\dagger U_{\phi\Phi} \subset U_{HL} U_L^2 U_{LH} ,
\eeqan
\eseq{tripletgaugeterms}
where $V=W,B$. Note that the vector boson block of the $\bf U$ matrix (not explicitly written out above) does not contribute, since each of $U_{W, WB, BW, B}$ already contains two powers of SM gauge couplings, and additional insertions of $U$ or $Z$, which are necessary in order to have at least one heavy propagator in the loop, will bring in more powers of $g$ or $g'$.

In Eq.~\eqref{tripletgaugeterms}, we have organized the operator structures by the total number of $Z$ and $U$ insertions, which makes the enumeration straightforward. To proceed, however, it is more convenient to group the terms in Eq.~\eqref{tripletgaugeterms} by the powers of $P$ and $Z^{(\dagger)}$. We will do so in the following paragraphs, and compute each group in turn using the rules in Tables~\ref{tab:heavy}, \ref{tab:mixed}, and~\ref{tab:ocd}. We will derive universal results before working out explicit forms of effective operators for the scalar triplet model.

\paragraph{$\O(P^0 Z^0)$ terms.}
Two of the three terms are readily available from the first two rows of Table~\ref{tab:triplet},
\beq
\L_\text{EFT}^\text{1-loop} \supset \pref \Bigl( 1 -\logm{M^2} \Bigr) \Bigl\{ \tr (U_{HL} U_{LH}) +\frac{1}{M^2}\,\tr (U_{HL} U_L U_{LH}) \Bigr\}.
\eeq{P0Z0ab}
The remaining term in this group easily follows from a single diagram,
\beq
\begin{gathered}
\begin{fmffile}{triplet-P0Z0-c}
\begin{fmfgraph}(40,40)
\fmfsurround{vUL1,vUHL,vULH,vUL2}
\fmf{plain,left=0.4}{vULH,vUHL}
\fmf{dashes,left=0.4}{vUHL,vUL1,vUL2,vULH}
\fmfv{decor.shape=circle,decor.filled=empty,decor.size=3thick}{vUHL,vUL1,vUL2,vULH}
\end{fmfgraph}
\end{fmffile}
\end{gathered}
\;\;\, =\, -i c_s \,\I_{i0}^{13}\,\tr (U_{HL} U_L^2 U_{LH})
= \pref \Bigl( 1 -\logm{M^2} \Bigr)\frac{1}{M^4} \,\tr (U_{HL} U_L^2 U_{LH}).
\eeq{P0Z0c}

\paragraph{$\O(P^0 Z^2)$ terms.}
Both terms in this group are also straightforward to compute, with the $Z^\mu$ and $Z^{\dagger\mu}$ contracted so that no $P_\mu$'s are picked up from vertex insertions:
\bseq
\beqa
&&
\begin{gathered}
\begin{fmffile}{triplet-P0Z2-a}
\begin{fmfgraph}(40,40)
\fmfsurround{vZHLd,vZHL}
\fmf{plain,left=1}{vZHLd,vZHL}
\fmf{dashes,left=1}{vZHL,vZHLd}
\fmfv{decor.shape=square,decor.filled=20,decor.size=3thick}{vZHL}
\fmfv{decor.shape=square,decor.filled=60,decor.size=3thick}{vZHLd}
\fmf{dots,width=thick}{vZHL,vZHLd}
\end{fmfgraph}
\end{fmffile}
\end{gathered}
\;\;\, =\, -i c_s \,\I[q^2]_{i0}^{11} \,\tr (Z_{HL}^\mu Z_{HL\,\mu}^\dagger) = \pref \Bigl( \frac{3}{8} -\frac{1}{4}\logm{M^2} \Bigr) M^2\,\tr (Z_{HL}^\mu Z_{HL\,\mu}^\dagger)\,, \CR[-8pt]\\[-4pt]
&&
\begin{gathered}
\begin{fmffile}{triplet-P0Z2-b}
\begin{fmfgraph}(40,40)
\fmfsurround{vULH,vZLd,vZHL}
\fmf{plain,left=0.6}{vULH,vZHL}
\fmf{dashes,left=0.6}{vZHL,vZLd,vULH}
\fmfv{decor.shape=square,decor.filled=20,decor.size=3thick}{vZHL}
\fmfv{decor.shape=square,decor.filled=60,decor.size=3thick}{vZLd}
\fmfv{decor.shape=circle,decor.filled=empty,decor.size=3thick}{vULH}
\fmf{dots,width=thick,right=0.4}{vZHL,vZLd}
\end{fmfgraph}
\end{fmffile}
\end{gathered}
\;\;+\text{h.c.} \,=\, -i c_s \,\I[q^2]_{i0}^{12} \,\tr (Z_{HL}^\mu Z_{L\,\mu}^\dagger U_{LH}) +\text{h.c.} \CR
&&\qquad\qquad\qquad\quad\;\;\; 
= \pref \Bigl( \frac{3}{8} -\frac{1}{4}\logm{M^2} \Bigr) \,\tr( Z_{HL}^\mu Z_{L\,\mu}^\dagger U_{LH} +\text{h.c.})\,.
\eeqan
\eseq{P0Z2}
\vspace{-12pt}
\paragraph{$\O(P^1 Z^1)$ terms.}
More diagrams contribute in this case, since the covariant derivative can either come from an uncontracted $Z^{(\dagger)}$ insertion, or be directly inserted. In the latter case, the $P$ and $Z^{(\dagger)}$ insertions should be contracted. The four terms in this group are calculated as follows:
\bseq
\beqa
&&
\begin{fmffile}{triplet-P1Z1-a}
\begin{gathered}
\begin{fmfgraph}(40,40)
\fmfsurround{vZHL,vULH}
\fmf{plain,left=1}{vULH,vZHL}
\fmf{dashes,left=1}{vZHL,vULH}
\fmfv{decor.shape=circle,decor.filled=empty,decor.size=3thick}{vULH}
\fmfv{decor.shape=square,decor.filled=20,decor.size=3thick}{vZHL}
\end{fmfgraph}
\end{gathered}
\;\;+\;\,
\begin{gathered}
\begin{fmfgraph}(40,40)
\fmfsurround{vZHL,vP,vULH,v0}
\fmf{plain,left=0.4}{vULH,vP,vZHL}
\fmf{dashes,left=0.4}{vZHL,v0,vULH}
\fmfv{decor.shape=circle,decor.filled=empty,decor.size=3thick}{vULH}
\fmfv{decor.shape=square,decor.filled=20,decor.size=3thick}{vZHL}
\fmfv{decor.shape=circle,decor.filled=full,decor.size=3thick}{vP}
\fmf{dots,width=thick,right=0.6}{vP,vZHL}
\end{fmfgraph}
\end{gathered}
\;\;+\;\,
\begin{gathered}
\begin{fmfgraph}(40,40)
\fmfsurround{vZHL,v0,vULH,vP}
\fmf{plain,left=0.4}{vULH,v0,vZHL}
\fmf{dashes,left=0.4}{vZHL,vP,vULH}
\fmfv{decor.shape=circle,decor.filled=empty,decor.size=3thick}{vULH}
\fmfv{decor.shape=square,decor.filled=20,decor.size=3thick}{vZHL}
\fmfv{decor.shape=circle,decor.filled=full,decor.size=3thick}{vP}
\fmf{dots,width=thick,right=0.6}{vZHL,vP}
\end{fmfgraph}
\end{gathered}
\;\;+\text{h.c.}
\end{fmffile}
\CR[4pt]
&& = -i c_s \bigl\{ (\,\I_{i0}^{11} -2\,\I[q^2]_{i0}^{21}) \,\tr (P_\mu Z_{HL}^\mu U_{LH}) -2\,\I[q^2]_{i0}^{12}\,\tr (Z_{HL}^\mu P_\mu U_{LH}) \bigr\} +\text{h.c.} \CR
&& = \pref \Bigl( \frac{3}{4} -\frac{1}{2}\logm{M^2} \Bigr) \,\tr (P_\mu Z_{HL}^\mu U_{LH} -Z_{HL}^\mu P_\mu U_{LH}) +\text{h.c.} \CR
&& = \pref \Bigl( \frac{3}{4} -\frac{1}{2}\logm{M^2} \Bigr) \,\tr\bigl( \,[P_\mu, Z_{HL}^\mu] \,U_{LH} +\text{h.c.} \bigr), \\[10pt]
&&
\begin{fmffile}{triplet-P1Z1-b}
\begin{gathered}
\begin{fmfgraph}(40,40)
\fmfsurround{vZHL,vULH,vUL}
\fmf{plain,left=0.6}{vULH,vZHL}
\fmf{dashes,left=0.6}{vZHL,vUL,vULH}
\fmfv{decor.shape=circle,decor.filled=empty,decor.size=3thick}{vULH,vUL}
\fmfv{decor.shape=square,decor.filled=20,decor.size=3thick}{vZHL}
\end{fmfgraph}
\end{gathered}
\;\;+\;\,
\begin{gathered}
\begin{fmfgraph}(40,40)
\fmfsurround{vZHL,vP,vULH,v0,vUL,v1}
\fmf{plain,left=0.25}{vULH,vP,vZHL}
\fmf{dashes,left=0.25}{vZHL,v1,vUL,v0,vULH}
\fmfv{decor.shape=circle,decor.filled=empty,decor.size=3thick}{vULH,vUL}
\fmfv{decor.shape=square,decor.filled=20,decor.size=3thick}{vZHL}
\fmfv{decor.shape=circle,decor.filled=full,decor.size=3thick}{vP}
\fmf{dots,width=thick,right=0.75}{vP,vZHL}
\end{fmfgraph}
\end{gathered}
\;\;+\;\,
\begin{gathered}
\begin{fmfgraph}(40,40)
\fmfsurround{vZHL,v0,vULH,v1,vUL,vP}
\fmf{plain,left=0.25}{vULH,v0,vZHL}
\fmf{dashes,left=0.25}{vZHL,vP,vUL,v1,vULH}
\fmfv{decor.shape=circle,decor.filled=empty,decor.size=3thick}{vULH,vUL}
\fmfv{decor.shape=square,decor.filled=20,decor.size=3thick}{vZHL}
\fmfv{decor.shape=circle,decor.filled=full,decor.size=3thick}{vP}
\fmf{dots,width=thick,right=0.75}{vZHL,vP}
\end{fmfgraph}
\end{gathered}
\;\;+\;\,
\begin{gathered}
\begin{fmfgraph}(40,40)
\fmfsurround{vZHL,v0,vULH,vP,vUL,v1}
\fmf{plain,left=0.25}{vULH,v0,vZHL}
\fmf{dashes,left=0.25}{vZHL,v1,vUL,vP,vULH}
\fmfv{decor.shape=circle,decor.filled=empty,decor.size=3thick}{vULH,vUL}
\fmfv{decor.shape=square,decor.filled=20,decor.size=3thick}{vZHL}
\fmfv{decor.shape=circle,decor.filled=full,decor.size=3thick}{vP}
\fmf{dots,width=thick}{vP,vZHL}
\end{fmfgraph}
\end{gathered}
\;\;+\text{h.c.}
\end{fmffile}
\CR[4pt]
&& = -i c_s \bigl\{ (\,\I_{i0}^{12} -2\,\I[q^2]_{i0}^{22}) \,\tr (P_\mu Z_{HL}^\mu U_L U_{LH}) \CR
&&\qquad\qquad -2\,\I[q^2]_{i0}^{13}\,\tr (Z_{HL}^\mu P_\mu U_L U_{LH} +Z_{HL}^\mu U_L P_\mu U_{LH}) \bigr\} +\text{h.c.} \CR
&& = \pref\frac{1}{M^2} \Bigl(\frac{3}{4} -\frac{1}{2}\logm{M^2}\Bigr) \CR
&&\qquad\qquad \tr( 2 P_\mu Z_{HL}^\mu U_L U_{LH} -Z_{HL}^\mu P_\mu U_L U_{LH} -Z_{HL}^\mu U_L P_\mu U_{LH} ) +\text{h.c.} \CR
&& = \pref\frac{1}{M^2} \Bigl(\frac{3}{4} -\frac{1}{2}\logm{M^2}\Bigr) \,\tr\bigl( [P_\mu, Z_{HL}^\mu] U_L U_{LH} -Z_{HL}^\mu U_L [P_\mu, U_{LH}] +\text{h.c.}\bigr) ,\\[10pt]
&&
\begin{fmffile}{triplet-P1Z1-c}
\begin{gathered}
\begin{fmfgraph}(40,40)
\fmfsurround{vZL,vUHL,vULH}
\fmf{plain,left=0.6}{vULH,vUHL}
\fmf{dashes,left=0.6}{vUHL,vZL,vULH}
\fmfv{decor.shape=circle,decor.filled=empty,decor.size=3thick}{vUHL,vULH}
\fmfv{decor.shape=square,decor.filled=20,decor.size=3thick}{vZL}
\end{fmfgraph}
\end{gathered}
\;\;+\;\,
\begin{gathered}
\begin{fmfgraph}(40,40)
\fmfsurround{vZL,vP,vUHL,v1,vULH,v0}
\fmf{plain,left=0.25}{vULH,v1,vUHL}
\fmf{dashes,left=0.25}{vUHL,vP,vZL,v0,vULH}
\fmfv{decor.shape=circle,decor.filled=empty,decor.size=3thick}{vUHL,vULH}
\fmfv{decor.shape=square,decor.filled=20,decor.size=3thick}{vZL}
\fmfv{decor.shape=circle,decor.filled=full,decor.size=3thick}{vP}
\fmf{dots,width=thick,right=0.75}{vP,vZL}
\end{fmfgraph}
\end{gathered}
\;\;+\;\,
\begin{gathered}
\begin{fmfgraph}(40,40)
\fmfsurround{vZL,v0,vUHL,v1,vULH,vP}
\fmf{plain,left=0.25}{vULH,v1,vUHL}
\fmf{dashes,left=0.25}{vUHL,v0,vZL,vP,vULH}
\fmfv{decor.shape=circle,decor.filled=empty,decor.size=3thick}{vUHL,vULH}
\fmfv{decor.shape=square,decor.filled=20,decor.size=3thick}{vZL}
\fmfv{decor.shape=circle,decor.filled=full,decor.size=3thick}{vP}
\fmf{dots,width=thick,right=0.75}{vZL,vP}
\end{fmfgraph}
\end{gathered}
\;\;+\;\,
\begin{gathered}
\begin{fmfgraph}(40,40)
\fmfsurround{vZL,v0,vUHL,vP,vULH,v1}
\fmf{plain,left=0.25}{vULH,vP,vUHL}
\fmf{dashes,left=0.25}{vUHL,v0,vZL,v1,vULH}
\fmfv{decor.shape=circle,decor.filled=empty,decor.size=3thick}{vUHL,vULH}
\fmfv{decor.shape=square,decor.filled=20,decor.size=3thick}{vZL}
\fmfv{decor.shape=circle,decor.filled=full,decor.size=3thick}{vP}
\fmf{dots,width=thick}{vP,vZL}
\end{fmfgraph}
\end{gathered}
\;\;+\text{h.c.}
\end{fmffile}
\CR[4pt]
&& = -i c_s \bigl\{ (\,\I_{i0}^{12} -2\,\I[q^2]_{i0}^{13}) \,\tr (U_{HL} P_\mu Z_L^\mu U_{LH})
-2\,\I[q^2]_{i0}^{13} \,\tr (U_{HL} Z_L^\mu P_\mu U_{LH}) \CR
&&\qquad\qquad -2\,\I[q^2]_{i0}^{22} \,\tr (U_{HL} Z_L^\mu U_{LH} P_\mu) \bigr\} +\text{h.c.} \CR
&& = \pref\frac{1}{M^2} \Bigl\{ \Bigl(\frac{1}{4} -\frac{1}{2}\logm{M^2}\Bigr)\,\tr (U_{HL} P_\mu Z_L^\mu U_{LH})
+ \Bigl(-\frac{3}{4} +\frac{1}{2}\logm{M^2} \Bigr) \,\tr (U_{HL} Z_L^\mu P_\mu U_{LH}) \CR
&&\qquad\qquad +\frac{1}{2}\,\tr (U_{HL} Z_L^\mu U_{LH} P_\mu) +\text{h.c.} \Bigr\} \CR
&& = \pref\frac{1}{M^2} \Bigl\{ \Bigl(\frac{3}{4} -\frac{1}{2}\logm{M^2}\Bigr)\,\tr\bigl( U_{HL} [P_\mu, Z_L^\mu] U_{LH}\bigr)  +\frac{1}{2}\,\tr\bigl( [P_\mu, U_{HL}] Z_L^\mu U_{LH}\bigr) +\text{h.c.} \Bigr\},\CR \\
&&
\begin{fmffile}{triplet-P1Z1-d}
\begin{gathered}
\begin{fmfgraph}(40,40)
\fmfsurround{vZL,vUHL,vULH,vUL}
\fmf{plain,left=0.4}{vULH,vUHL}
\fmf{dashes,left=0.4}{vUHL,vZL,vUL,vULH}
\fmfv{decor.shape=circle,decor.filled=empty,decor.size=3thick}{vUHL,vUL,vULH}
\fmfv{decor.shape=square,decor.filled=20,decor.size=3thick}{vZL}
\end{fmfgraph}
\end{gathered}
\;\;+\;\,
\begin{gathered}
\begin{fmfgraph}(40,40)
\fmfsurround{vZL,vP,vUHL,v2,vULH,v1,vUL,v0}
\fmf{plain,left=0.25}{vULH,v2,vUHL}
\fmf{dashes,left=0.25}{vUHL,vP,vZL,v0,vUL,v1,vULH}
\fmfv{decor.shape=circle,decor.filled=empty,decor.size=3thick}{vUHL,vUL,vULH}
\fmfv{decor.shape=square,decor.filled=20,decor.size=3thick}{vZL}
\fmfv{decor.shape=circle,decor.filled=full,decor.size=3thick}{vP}
\fmf{dots,width=thick,right=0.75}{vP,vZL}
\end{fmfgraph}
\end{gathered}
\;\;+\;\,
\begin{gathered}
\begin{fmfgraph}(40,40)
\fmfsurround{vZL,v0,vUHL,v2,vULH,v1,vUL,vP}
\fmf{plain,left=0.25}{vULH,v2,vUHL}
\fmf{dashes,left=0.25}{vUHL,v0,vZL,vP,vUL,v1,vULH}
\fmfv{decor.shape=circle,decor.filled=empty,decor.size=3thick}{vUHL,vUL,vULH}
\fmfv{decor.shape=square,decor.filled=20,decor.size=3thick}{vZL}
\fmfv{decor.shape=circle,decor.filled=full,decor.size=3thick}{vP}
\fmf{dots,width=thick,right=0.75}{vZL,vP}
\end{fmfgraph}
\end{gathered}
\;\;+\;\,
\begin{gathered}
\begin{fmfgraph}(40,40)
\fmfsurround{vZL,v0,vUHL,v2,vULH,vP,vUL,v1}
\fmf{plain,left=0.25}{vULH,v2,vUHL}
\fmf{dashes,left=0.25}{vUHL,v0,vZL,v1,vUL,vP,vULH}
\fmfv{decor.shape=circle,decor.filled=empty,decor.size=3thick}{vUHL,vUL,vULH}
\fmfv{decor.shape=square,decor.filled=20,decor.size=3thick}{vZL}
\fmfv{decor.shape=circle,decor.filled=full,decor.size=3thick}{vP}
\fmf{dots,width=thick,right=0.3}{vZL,vP}
\end{fmfgraph}
\end{gathered}
\;\;+\;\,
\begin{gathered}
\begin{fmfgraph}(40,40)
\fmfsurround{vZL,v0,vUHL,vP,vULH,v2,vUL,v1}
\fmf{plain,left=0.25}{vULH,vP,vUHL}
\fmf{dashes,left=0.25}{vUHL,v0,vZL,v1,vUL,v2,vULH}
\fmfv{decor.shape=circle,decor.filled=empty,decor.size=3thick}{vUHL,vUL,vULH}
\fmfv{decor.shape=square,decor.filled=20,decor.size=3thick}{vZL}
\fmfv{decor.shape=circle,decor.filled=full,decor.size=3thick}{vP}
\fmf{dots,width=thick,right=0.3}{vP,vZL}
\end{fmfgraph}
\end{gathered}
\;\;+\text{h.c.}
\end{fmffile}
\CR[4pt]
&& = -i c_s \bigl\{ (\,\I_{i0}^{13} -2\,\I[q^2]_{i0}^{14})\,\tr (U_{HL} P_\mu Z_L^\mu U_L U_{LH}) \CR
&&\qquad\qquad -2\,\I[q^2]_{i0}^{14} \,\tr(U_{HL} Z_L^\mu P_\mu U_L U_{LH} +U_{HL} Z_L^\mu U_L P_\mu U_{LH}) \CR
&&\qquad\qquad -2\,\I[q^2]_{i0}^{23} \,\tr (U_{HL} Z_L^\mu U_L U_{LH} P_\mu)
\bigr\} +\text{h.c.} \CR
&& = \pref \frac{1}{M^4} \Bigl\{ \Bigl(\frac{1}{4} -\frac{1}{2}\logm{M^2}\Bigr) \,\tr (U_{HL} P_\mu Z_L^\mu U_L U_{LH}) \CR
&&\qquad\qquad +\Bigl(-\frac{3}{4} +\frac{1}{2}\logm{M^2}\Bigr) \,\tr (U_{HL} Z_L^\mu P_\mu U_L U_{LH} +U_{HL} Z_L^\mu U_L P_\mu U_{LH}) \CR
&&\qquad\qquad +\Bigl(\frac{5}{4} -\frac{1}{2}\logm{M^2}\Bigr) \,\tr (U_{HL} Z_L^\mu U_L U_{LH} P_\mu)
 +\text{h.c.}\Bigr\} \CR
&& = \pref \frac{1}{M^4} \Bigl\{ \Bigl(\frac{3}{4} -\frac{1}{2}\logm{M^2}\Bigr) \,\tr \bigl( U_{HL} [P_\mu, Z_L^\mu] U_L U_{LH} -U_{HL} Z_L^\mu U_L [P_\mu, U_{LH}] \bigr) \CR
&&\qquad\qquad +\frac{1}{2} \,\tr\bigl([P_\mu, U_{HL}] Z_L^\mu U_L U_{LH}\bigr)  +\text{h.c.}\Bigr\} .
\eeqan
\eseq{P1Z1}

\paragraph{$\O(P^2 Z^2)$ terms.} The number of diagrams increases further, but the calculation is still quite manageable even if done by hand. Since the procedure should be clear by now, we refrain from enumerating all the diagrams for the three terms in this group, but simply report the final results:
\bseq
\beqa
&&
\begin{gathered}
\begin{fmffile}{triplet-P2Z2-a}
\begin{fmfgraph}(40,40)
\fmfsurround{vZHLd,vZHL}
\fmf{plain,left=1}{vZHLd,vZHL}
\fmf{dashes,left=1}{vZHL,vZHLd}
\fmfv{decor.shape=square,decor.filled=20,decor.size=3thick}{vZHL}
\fmfv{decor.shape=square,decor.filled=60,decor.size=3thick}{vZHLd}
\end{fmfgraph}
\end{fmffile}
\end{gathered}
\;\;+\text{$(2+6)$ more}\CR
&&\subset \pref \Bigl\{ \Bigl(\frac{5}{72} -\frac{1}{12}\logm{M^2}\Bigr) \,\tr\bigl([P^\mu, Z_{HL}^\nu] [P_\mu, Z_{HL\,\nu}^\dagger]\bigr) \CR
&&\quad +\Bigl(-\frac{11}{18} +\frac{1}{3}\logm{M^2}\Bigr) \,\tr\bigl([P_\mu, Z_{HL}^\mu] [P_\nu, Z_{HL}^{\dagger\nu}]\bigr) \CR
&&\quad +\Bigl(\frac{1}{18} -\frac{1}{6}\logm{M^2}\Bigr) \,\tr\bigl( Z_{HL}^\mu Z_{HL}^{\dagger\nu} [P_\mu, P_\nu]\bigr)
+\Bigl(-\frac{11}{36} +\frac{1}{6}\logm{M^2}\Bigr) \,\tr\bigl(Z_{HL}^\mu [P_\mu, P_\nu] Z_{HL}^{\dagger\nu}\bigr) \Bigr\} , \CR\\
&&
\begin{gathered}
\begin{fmffile}{triplet-P2Z2-b}
\begin{fmfgraph}(40,40)
\fmfsurround{vUHL,vZHLd,vZL}
\fmf{plain,left=0.6}{vZHLd,vUHL}
\fmf{dashes,left=0.6}{vUHL,vZL,vZHLd}
\fmfv{decor.shape=square,decor.filled=20,decor.size=3thick}{vZL}
\fmfv{decor.shape=square,decor.filled=60,decor.size=3thick}{vZHLd}
\fmfv{decor.shape=circle,decor.filled=empty,decor.size=3thick}{vUHL}
\end{fmfgraph}
\end{fmffile}
\end{gathered}
\;\;+\text{h.c.} \;+ \text{$(6+15)$ more} \CR[2pt]
&& \subset \pref \frac{1}{M^2} 
\CR &&\quad 
\Bigl\{ \Bigl(-\frac{5}{72} +\frac{1}{12}\logm{M^2}\Bigr) \,\tr\bigl(U_{HL} \bigl[P^\mu, [P_\mu, Z_L^\nu]\bigr] Z_{HL\,\nu}^\dagger\bigr) 
-\frac{1}{6}\,\tr\bigl( [P^\mu, U_{HL}] Z_L^\nu [P_\mu, Z_{HL\,\nu}^\dagger] \bigr)
\CR &&\quad 
+\Bigl(-\frac{7}{9} +\frac{1}{3}\logm{M^2}\Bigr) \,\tr\bigl( U_{HL} [P_\mu, Z_L^\mu] [P_\nu, Z_{HL}^{\dagger\nu}]\bigr)
+\frac{1}{6} \,\tr\bigl( U_{HL} [P_\nu, Z_L^\mu] [P_\mu, Z_{HL}^{\dagger\nu}]\bigr)
\CR &&\quad 
+\Bigl(\frac{5}{36} -\frac{1}{6}\logm{M^2}\Bigr) \,\tr\bigl( [P_\nu, U_{HL}] [P_\mu, Z_L^\mu] Z_{HL}^{\dagger\nu} \bigr)
-\frac{1}{3} \,\tr\bigl( [P_\mu, U_{HL}] Z_L^\mu [P_\nu, Z_{HL}^{\dagger\nu}] \bigr)
\CR &&\quad 
-\frac{1}{12} \,\tr\bigl( U_{HL} Z_L^\mu Z_{HL}^{\dagger\nu} [P_\mu, P_\nu]\bigr)
+\Bigl(-\frac{17}{36} +\frac{1}{6}\logm{M^2}\Bigr) \,\tr\bigl( U_{HL} Z_L^\mu [P_\mu, P_\nu] Z_{HL}^{\dagger\nu} \bigr)
\CR &&\quad 
+\Bigl(\frac{11}{36} -\frac{1}{6}\logm{M^2}\Bigr) \,\tr\bigl( U_{HL} [P_\mu, P_\nu] Z_L^\mu Z_{HL}^{\dagger\nu} \bigr)
+\text{h.c.}\Bigr\}, \\[10pt]
&&
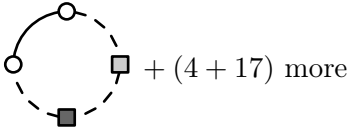
\begin{gathered}
\begin{fmffile}{triplet-P2Z2-c}
\begin{fmfgraph}(40,40)
\fmfsurround{vZL,vUHL,vULH,vZLd}
\fmf{plain,left=0.4}{vULH,vUHL}
\fmf{dashes,left=0.4}{vUHL,vZL,vZLd,vULH}
\fmfv{decor.shape=circle,decor.filled=empty,decor.size=3thick}{vUHL,vULH}
\fmfv{decor.shape=square,decor.filled=20,decor.size=3thick}{vZL}
\fmfv{decor.shape=square,decor.filled=60,decor.size=3thick}{vZLd}
\end{fmfgraph}
\end{fmffile}
\end{gathered}
\;\; + \text{$(4+17)$ more} \CR[2pt]
&&\subset \pref \frac{1}{M^4} \Bigl\{ \Bigl(\frac{5}{72} -\frac{1}{12}\logm{M^2}\Bigr) \cdot \CR
&&\qquad\quad \tr\Bigl( U_{HL} [P^\mu, Z_L^\nu] [P_\mu, Z_{L\,\nu}^\dagger] U_{LH}
+ \bigl( [P^\mu, U_{HL}] [P_\mu, Z_L^\nu] Z_{L\,\nu}^\dagger U_{LH} +\text{h.c.}\bigr) \Bigr) \CR
&&\qquad -\frac{1}{6} \,\tr\bigl( U_{HL} [P^\mu, Z_L^\nu] Z_{L\,\nu}^\dagger [P_\mu, U_{LH}] +\text{h.c.}\bigr) \CR
&&\qquad +\Bigl(-\frac{17}{24} +\frac{1}{4}\logm{M^2}\Bigr)\,\tr\bigl( [P^\mu, U_{HL}] Z_L^\nu Z_{L\,\nu}^\dagger [P_\mu, U_{LH}]\bigr) \CR
&&\qquad +\Bigl(-\frac{11}{36} +\frac{1}{6}\logm{M^2}\Bigr) \,\tr\bigl( U_{HL} [P_\mu, Z_L^\mu] [P_\nu, Z_L^{\dagger\nu}] U_{LH} + U_{HL} [P_\nu, Z_L^\mu] [P_\mu, Z_L^{\dagger\nu}] U_{LH} \bigr) \CR
&&\qquad +\Bigl(\frac{11}{18} -\frac{1}{3}\logm{M^2}\Bigr) \,\tr\bigl( [P_\nu, U_{HL}] [P_\mu, Z_L^\mu] Z_L^{\dagger\nu} U_{LH} +\text{h.c.}\bigr) \CR
&&\qquad +\Bigl(-\frac{17}{36} +\frac{1}{6}\logm{M^2}\Bigr) \,\tr\bigl( [P_\mu, U_{HL}] [P_\nu, Z_L^\mu] Z_L^{\dagger\nu} U_{LH} +\text{h.c.}\bigr) \CR
&&\qquad -\frac{1}{3} \,\tr\bigl( U_{HL} [P_\mu, Z_L^\mu] Z_L^{\dagger\nu} [P_\nu, U_{LH}] +\text{h.c.}\bigr) 
-\frac{1}{6} \,\tr\bigl([P_\mu, U_{HL}] Z_L^\mu Z_L^{\dagger\nu} [P_\nu, U_{LH}]\bigr) \CR
&&\qquad +\frac{1}{12} \,\tr\bigl(U_{HL} Z_L^\mu Z_L^{\dagger\nu} U_{LH} [P_\mu, P_\nu] \bigr)
-\frac{1}{6} \,\tr\bigl( U_{HL} Z_L^\mu Z_L^{\dagger\nu} [P_\mu, P_\nu] U_{LH}  +\text{h.c.}\bigr)
\Bigr\}.
\eeqan
\eseq{P2Z2}
In the equations above, we have shown, for each term, the one diagram with zero $P$ insertions, and the number of diagrams with one and two $P$ insertions (e.g.~2 and~6, respectively, for the $P^2 Z_{HL}Z_{HL}^\dagger$ term). The counting excludes hermitian conjugation and adjacent $P_\mu$ contractions. Following the rules in Section~\ref{sec:cd-sum}, the reader should be able to easily draw all the diagrams, and fill in the intermediate steps (which are straightforward though perhaps a bit lengthy) that lead to the final results in Eq.~\eqref{P2Z2}.

\bigskip
{\it All results presented in the four paragraphs above, namely Eqs.~\eqref{P0Z0ab}, \eqref{P0Z0c}, \eqref{P0Z2}, \eqref{P1Z1} and~\eqref{P2Z2}, are universal and model-independent.} Now we focus on the scalar triplet model, and work out the traces involved in these equations that yield the three effective operators in Eq.~\eqref{OTOHOR}:
\bseq
\beqa
&& \tr (U_{HL} U_{LH}),\; 
\tr\bigl( [P_\mu, Z_{HL}^\mu] U_{LH}\bigr) \text{ and h.c.},\; 
-\tr\bigl( [P_\mu, Z_{HL}^\mu] [P_\nu, Z_{HL}^{\dagger\nu}]\bigr) \CR
&&\quad \to -g_{\alpha\beta}U_{\Phi W}^{ab\alpha} U_{\Phi W}^{\dagger ba\beta} 
\supset -\frac{4\kappa^2}{M^4} g^2 (\O_T+2\O_R) \,;\\[4pt]
&& \tr (U_{HL} U_L U_{LH}),\; 
\tr\bigl( [P_\mu, Z_{HL}^\mu] U_L U_{LH} +\text{h.c.}\bigr),\CR && 
\tr\bigl( U_{HL} [P_\mu, Z_L^\mu] U_{LH} +\text{h.c.}\bigr),\; 
- \tr\bigl(U_{HL} [P_\mu, Z_L^\mu] [P_\nu, Z_{HL}^{\dagger\nu}]  +\text{h.c.}\bigr) \CR
&&\quad \to -g_{\alpha\beta}U_{\Phi W}^{ab\alpha} U_{\phi W}^{\dagger b\beta} U_{\phi\Phi}^a +\text{h.c.}
\supset \frac{4\kappa^2}{M^2} g^2 (\O_T+2\O_R) \,; \\[4pt]
&& \tr (U_{HL} U_L^2 U_{LH}) ,\;
\tr\bigl( U_{HL} [P_\mu, Z_L^\mu] U_L U_{LH}\bigr) \text{ and h.c.} ,\;
-\tr\bigl( U_{HL} [P_\mu, Z_L^\mu] [P_\nu, Z_L^{\dagger\nu}] U_{LH}\bigr)
\CR
&&\quad \to -g_{\alpha\beta} \bigl( U_{\phi\Phi}^{\dagger a} U_{\phi W}^{b\alpha} U_{\phi W}^{\dagger b\beta} U_{\phi\Phi}^a +U_{\phi\Phi}^{\dagger a} U_{\phi B}^\alpha U_{\phi B}^{\dagger\beta} U_{\phi\Phi}^a \bigr) \CR
&&\quad \supset \frac{\kappa^2}{2} \bigl[ g^2 (\O_T -4\O_R) +g'^2 (\O_H -2\O_R) \bigr] \,;\\[4pt]
&& \tr (Z_{HL}^\mu Z_{HL\,\mu}^\dagger) \CR
&&\quad \to -g_{\alpha\beta} Z_{\Phi W}^{\mu\,ab\alpha} Z_{\Phi W\,\mu}^{\dagger ba\beta} 
\supset -\Bigl(1-\frac{\epsilon}{4}\Bigr) \frac{32\kappa^2}{M^6} g^2 (\O_T+2\O_R) \,; \\[4pt]
&& \tr (Z_{HL}^\mu Z_{L\,\mu}^\dagger U_{LH} +\text{h.c.}) \CR
&&\quad \to -g_{\alpha\beta} Z_{\Phi W}^{\mu\,ab\alpha} Z_{\phi W\,\mu}^{\dagger b\beta} U_{\phi\Phi}^a +\text{h.c.}
\supset \Bigl(1-\frac{\epsilon}{4}\Bigr) \frac{32\kappa^2}{M^4} g^2 (\O_T+2\O_R) \,; \\[4pt]
&& \tr\bigl( Z_{HL}^\mu U_L [P_\mu, U_{LH}] +\text{h.c.}\bigr),\;
-\tr\bigl( [P_\nu, U_{HL}] [P_\mu, Z_L^\mu] Z_{HL}^{\dagger\nu} +\text{h.c.}\bigr) \CR
&&\quad \to -g_{\alpha\beta} Z_{\Phi W}^{\mu\,ab\alpha} U_{\phi W}^{\dagger b\beta} [P_\mu, U_{\phi\Phi}^a] +\text{h.c.}
\supset -\frac{4\kappa^2}{M^2} g^2 (\O_T-\O_H+\O_R) \,; \\[4pt]
&& \tr\bigl( [P_\mu, U_{HL}] Z_L^\mu U_{LH} +\text{h.c.}\bigr),\;
-\tr \bigl( [P_\mu, U_{HL}] Z_L^\mu [P_\nu, Z_{HL}^{\dagger\nu}] +\text{h.c.} \bigr)\CR
&&\quad \to -g_{\alpha\beta} [P_\mu, U_{\phi\Phi}^{\dagger a}] Z_{\phi W}^{\mu\,b\alpha} U_{\Phi W}^{\dagger ba\beta} +\text{h.c.} 
\supset \frac{4\kappa^2}{M^2} g^2 (\O_T+2\O_R) \,;\\[4pt]
&& \tr\bigl( U_{HL} Z_L^\mu U_L [P_\mu, U_{LH}] +\text{h.c.}\bigr) ,\;
-\tr\bigl( [P_\nu, U_{HL}] [P_\mu, Z_L^\mu] Z_L^{\dagger\nu} U_{LH} +\text{h.c.} \bigr) \CR
&&\quad \to -g_{\alpha\beta} \bigl( U_{\phi\Phi}^{\dagger a} Z_{\phi W}^{\mu\,b\alpha} U_{\phi W}^{\dagger b\beta} [P_\mu, U_{\phi\Phi}^a] +U_{\phi\Phi}^{\dagger a} Z_{\phi B}^{\mu\,\alpha} U_{\phi B}^{\dagger \beta} [P_\mu, U_{\phi\Phi}^a] \bigr) +\text{h.c.} \CR
&&\quad \supset 4\kappa^2 g^2 (\O_T-\O_H+\O_R) \,; \\[4pt]
&& \tr\bigl( [P_\mu, U_{HL}] Z_L^\mu U_L U_{LH} +\text{h.c.}\bigr) ,\;
-\tr \bigl( U_{HL} [P_\mu, Z_L^\mu] Z_L^{\dagger\nu} [P_\nu, U_{LH}] +\text{h.c.}\bigr) \CR
&&\quad \to -g_{\alpha\beta} \bigl( [P_\mu, U_{\phi\Phi}^{\dagger a}] Z_{\phi W}^{\mu\,b\alpha} U_{\phi W}^{\dagger b\beta} U_{\phi\Phi}^a +[P_\mu, U_{\phi\Phi}^{\dagger a}] Z_{\phi B}^{\mu\,\alpha} U_{\phi B}^{\dagger \beta} U_{\phi\Phi}^a \bigr) +\text{h.c.} \CR
&&\quad \supset -\kappa^2 \bigl[g^2(5\O_T+4\O_R) +g'^2(\O_H-2\O_R)\bigr] \,; \\[4pt]
&& \tr\bigl( [P^\mu, Z_{HL}^\nu] [P_\mu, Z_{HL\,\nu}^\dagger]\bigr) \CR
&&\quad \to -g_{\alpha\beta} [P^\mu, Z_{\Phi W}^{\nu\,ab\alpha}] [P_\mu, Z_{\Phi W\,\nu}^{\dagger ba\beta}] 
\supset \Bigl(1-\frac{\epsilon}{4}\Bigr) \frac{16\kappa^2}{M^4} g^2 (\O_T+2\O_R) \,; \\[4pt]
&& \tr\bigl( U_{HL} \bigl[P^\mu, [P_\mu, Z_L^\nu]\bigr] Z_{HL\,\nu}^\dagger +\text{h.c.} \bigr) \CR
&&\quad \to -g_{\alpha\beta} U_{\phi\Phi}^{\dagger a} \bigl[P^\mu, [P_\mu, Z_{\phi W}^{\nu\,b\alpha}]\bigr] Z_{\Phi W\,\nu}^{\dagger ba\beta} +\text{h.c.}
\supset \Bigl(1-\frac{\epsilon}{4}\Bigr) \frac{16\kappa^2}{M^2} g^2 (\O_H+\O_R) \,; \\[4pt]
&& \tr\bigl( [P^\mu, U_{HL}] Z_L^\nu [P_\mu, Z_{HL\,\nu}^\dagger] +\text{h.c.}\bigr) \CR
&&\quad \to -g_{\alpha\beta} [P^\mu, U_{\phi\Phi}^{\dagger a}] Z_{\phi W}^{\nu\,b\alpha} [P_\mu, Z_{\Phi W\,\nu}^{\dagger ba\beta}] +\text{h.c.} 
\supset -\Bigl(1-\frac{\epsilon}{4}\Bigr) \frac{16\kappa^2}{M^2} g^2 (\O_T+2\O_R) \,; \\[4pt]
&& \tr\bigl( U_{HL} [P_\nu, Z_L^\mu] [P_\mu, Z_{HL}^{\dagger\nu}] +\text{h.c.}\bigr) \CR
&&\quad \to -g_{\alpha\beta} U_{\phi\Phi}^{\dagger a} [P^\mu, Z_{\phi W}^{\nu\,b\alpha}] [P_\mu, Z_{\Phi W\,\nu}^{\dagger ba\beta}] +\text{h.c.}
\supset -\frac{4\kappa^2}{M^2}g^2 (\O_T+2\O_R) \,; \\[4pt]
&& \tr\bigl( U_{HL} [P^\mu, Z_L^\nu] [P_\mu, Z_{L\,\nu}^\dagger] U_{LH}\bigr) \CR
&&\quad \to -g_{\alpha\beta} \bigl( U_{\phi\Phi}^{\dagger a} [P^\mu, Z_{\phi W}^{\nu\,b\alpha}] [P_\mu, Z_{\phi W\,\nu}^{\dagger b\beta}] U_{\phi\Phi}^a + U_{\phi\Phi}^{\dagger a} [P^\mu, Z_{\phi B}^{\nu\,\alpha}] [P_\mu, Z_{\phi B\,\nu}^{\dagger \beta}] U_{\phi\Phi}^a \bigr) \CR
&&\quad \supset -\Bigl(1-\frac{\epsilon}{4}\Bigr) 2\kappa^2 \bigl[ g^2(\O_T-4\O_R) +g'^2(\O_H-2\O_R) \bigr] \,; \\[4pt]
&& \tr\bigl( [P^\mu, U_{HL}] [P_\mu, Z_L^\nu] Z_{L\,\nu}^\dagger U_{LH} +\text{h.c.}\bigr) \CR
&&\quad \to -g_{\alpha\beta} \bigl([P^\mu, U_{\phi\Phi}^{\dagger a}] [P_\mu, Z_{\phi W}^{\nu\,b\alpha}] Z_{\phi W\,\nu}^{\dagger b\beta} U_{\phi\Phi}^a +[P^\mu, U_{\phi\Phi}^{\dagger a}] [P_\mu, Z_{\phi B}^{\nu\,\alpha}] Z_{\phi B\,\nu}^{\dagger \beta} U_{\phi\Phi}^a \bigr) +\text{h.c.} \CR
&&\quad \supset -\Bigl(1-\frac{\epsilon}{4}\Bigr) 16\kappa^2 g^2(\O_T-\O_H+\O_R) \,; \\[4pt]
&& \tr\bigl( U_{HL} [P^\mu, Z_L^\nu] Z_{L\,\nu}^\dagger [P_\mu, U_{LH}] +\text{h.c.}\bigr) \CR
&&\quad \to -g_{\alpha\beta} \bigl( U_{\phi\Phi}^{\dagger a} [P^\mu, Z_{\phi W}^{\nu\,b\alpha}] Z_{\phi W\,\nu}^{\dagger b\beta} [P_\mu, U_{\phi\Phi}^a] +U_{\phi\Phi}^{\dagger a} [P^\mu, Z_{\phi B}^{\nu\,\alpha}] Z_{\phi B\,\nu}^{\dagger \beta} [P_\mu, U_{\phi\Phi}^a]\bigr) +\text{h.c.} \CR
&&\quad \supset \Bigl(1-\frac{\epsilon}{4}\Bigr) 4\kappa^2 \bigl[g^2(5\O_T+4\O_R) +g'^2(\O_H-2\O_R)\bigr] \,; \\[4pt]
&& \tr\bigl( [P^\mu, U_{HL}] Z_L^\nu Z_{L\,\nu}^\dagger [P_\mu, U_{LH}]\bigr) \CR
&&\quad \to -g_{\alpha\beta} \bigl([P^\mu, U_{\phi\Phi}^{\dagger a}] Z_{\phi W}^{\nu\,b\alpha} Z_{\phi W\,\nu}^{\dagger b\beta} [P_\mu, U_{\phi\Phi}^a] +[P^\mu, U_{\phi\Phi}^{\dagger a}] Z_{\phi B}^{\nu\,\alpha} Z_{\phi B\,\nu}^{\dagger \beta} [P_\mu, U_{\phi\Phi}^a]\bigr) \CR
&&\quad \supset -\Bigl(1-\frac{\epsilon}{4}\Bigr) 2\kappa^2 \bigl[ g^2(\O_T-4\O_R) +g'^2(\O_H-2\O_R) \bigr] \,; \\[4pt]
&& \tr\bigl( U_{HL} [P_\nu, Z_L^\mu] [P_\mu, Z_L^{\dagger\nu}] U_{LH}\bigr) \CR
&&\quad \to -g_{\alpha\beta} \bigl( U_{\phi\Phi}^{\dagger a} [P_\nu, Z_{\phi W}^{\mu\,b\alpha}] [P_\mu, Z_{\phi W}^{\dagger\nu\, b\beta}] U_{\phi\Phi}^a +U_{\phi\Phi}^{\dagger a} [P_\nu, Z_{\phi B}^{\mu\,\alpha}] [P_\mu, Z_{\phi B}^{\dagger\nu\, \beta}] U_{\phi\Phi}^a\bigr) \CR
&&\quad \supset -\frac{\kappa^2}{2} \bigl[ g^2(\O_T-4\O_R) +g'^2(\O_H-2\O_R) \bigr] \,; \\[4pt]
&& \tr\bigl( [P_\mu, U_{HL}] [P_\nu, Z_L^\mu] Z_L^{\dagger\nu} U_{LH} +\text{h.c.}\bigr) \CR
&&\quad \to -g_{\alpha\beta} \bigl( [P_\mu, U_{\phi\Phi}^{\dagger a}] [P_\nu, Z_{\phi W}^{\mu\,b\alpha}] Z_{\phi W}^{\dagger\nu\, b\beta} U_{\phi\Phi}^a +[P_\mu, U_{\phi\Phi}^{\dagger a}] [P_\nu, Z_{\phi B}^{\mu\,\alpha}] Z_{\phi B}^{\dagger\nu\, \beta} U_{\phi\Phi}^a\bigr) +\text{h.c.} \CR
&&\quad \supset -4\kappa^2 g^2(\O_T-\O_H+\O_R) \,;\\[4pt]
&& \tr\bigl( [P_\mu, U_{HL}] Z_L^\mu Z_L^{\dagger\nu} [P_\nu, U_{LH}]\bigr) \CR
&&\quad \to -g_{\alpha\beta} \bigl( [P_\mu, U_{\phi\Phi}^{\dagger a}] Z_{\phi W}^{\mu\,b\alpha} Z_{\phi W}^{\dagger\nu\, b\beta} [P_\nu, U_{\phi\Phi}^a] +[P_\mu, U_{\phi\Phi}^{\dagger a}] Z_{\phi B}^{\mu\,\alpha} Z_{\phi B}^{\dagger\nu\, \beta} [P_\nu, U_{\phi\Phi}^a] \bigr) \CR
&&\quad \supset -\frac{\kappa^2}{2} \bigl[ g^2(\O_T-4\O_R) +g'^2(\O_H-2\O_R) \bigr] \,.
\eeqan
\eseq{tripletgaugetr}
Note that Lorentz indices of the gauge boson fields $\alpha, \beta$ should be contracted with $-g_{\alpha\beta}$, as discussed in Section~\ref{sec:cd-heavy}. Also, $\O(\epsilon)$ terms from $g_{\alpha\beta}g^{\alpha\beta}=d=4-\epsilon$ must be kept in cases where the master integrals have $\frac{1}{\epsilon}$ poles. The latter were not written out explicitly above, but can be easily recovered by
\beq
-\logm{M^2} \to \frac{2}{\epsilon} -\logm{M^2} \,.
\eeq{recoverpole}
Adding up all terms in Eqs.~\eqref{P0Z0ab}, \eqref{P0Z0c}, \eqref{P0Z2}, \eqref{P1Z1} and~\eqref{P2Z2} with the replacement Eq.~\eqref{recoverpole}, plugging in Eq.~\eqref{tripletgaugetr}, and finally dropping $\frac{1}{\epsilon}$ poles, we obtain the final result (with $c_s=\frac{1}{2}$ and $\mu$ set to $M$ in the $\overline{\text{MS}}$ scheme),
\beq
\L_\text{EFT}^\text{1-loop} [\phi] \supset \frac{1}{16\pi^2} \frac{5\kappa^2}{8M^4}\bigl[\,g^2 \,\O_T + g'^2 \,\O_H -( 4g^2+2g'^2) \,\O_R \,\bigr] \,.
\eeqn
This agrees with the result in~\cite{DKS} obtained by Feynman diagram calculations.

\subsection{Integrating out a vectorlike fermion: pure gauge operators}
\label{sec:vlf}

Our final two examples demonstrate treatment of fermions in our covariant diagram approach. In the present subsection, we consider a simple but quite general setup of a vectorlike fermion of mass $M$ charged under some gauge symmetry. We will compute pure gauge effective operators up to dimension six which are generated by integrating out the heavy vectorlike fermion, independent of possible presence of light matter fields. The results are familiar in various contexts, including integrating out a heavy quark flavor in QCD, and integrating out a heavy vectorlike fermion that may arise in many beyond-SM scenarios. We also note that the same results are obtained in~\cite{HLM14} following the alternative approach to integrating out heavy fermions discussed at the beginning of Section~\ref{sec:cd-ferm}.

\paragraph{$\O(P^4)$ terms.}
We first consider diagrams with four (fermionic) $P$ insertions. Five diagrams can be drawn which differ by whether and how the heavy fermionic propagators are contracted with each other. One of them can be dropped where fermionic propagators separated by two $P$ insertions are contracted (while the loop integral $\,\I[q^2]_i^4$ is finite), because it only gives rise to $\tr(\dots P^2\dots)$. The remaining four diagrams are, by the rules in Table~\ref{tab:ferm},
\vspace{4pt}
\beqa
&&
\begin{fmffile}{vlf-P4}
\begin{gathered}
\begin{fmfgraph}(40,40)
\fmfsurround{v3,v2,v1,v4}
\fmf{plain,left=0.4}{v1,v2,v3,v4,v1}
\fmfv{decor.shape=circle,decor.filled=full,decor.size=3thick}{v1,v2,v3,v4}
\end{fmfgraph}
\end{gathered}
\;\;+\;\,
\begin{gathered}
\begin{fmfgraph}(40,40)
\fmfsurround{v3,v23,v2,v12,v1,v41,v4,v34}
\fmf{plain,left=0.25}{v1,v12,v2,v23,v3,v34,v4,v41,v1}
\fmfv{decor.shape=circle,decor.filled=full,decor.size=3thick}{v1,v2,v3,v4}
\fmf{dots,width=thick,right=0.6}{v12,v23}
\end{fmfgraph}
\end{gathered}
\;\;+\;\,
\begin{gathered}
\begin{fmfgraph}(40,40)
\fmfsurround{v3,v23,v2,v12,v1,v41,v4,v34}
\fmf{plain,left=0.25}{v1,v12,v2,v23,v3,v34,v4,v41,v1}
\fmfv{decor.shape=circle,decor.filled=full,decor.size=3thick}{v1,v2,v3,v4}
\fmf{dots,width=thick,right=0.6}{v12,v23}
\fmf{dots,width=thick,right=0.6}{v34,v41}
\end{fmfgraph}
\end{gathered}
\;\;+\;\,
\begin{gathered}
\begin{fmfgraph}(40,40)
\fmfsurround{v3,v23,v2,v12,v1,v41,v4,v34}
\fmf{plain,left=0.25}{v1,v12,v2,v23,v3,v34,v4,v41,v1}
\fmfv{decor.shape=circle,decor.filled=full,decor.size=3thick}{v1,v2,v3,v4}
\fmf{dots,width=thick}{v12,v34}
\fmf{dots,width=thick}{v23,v41}
\end{fmfgraph}
\end{gathered}
\end{fmffile} \CR[4pt]
&& =\, -i c_s \Bigl\{\, \frac{1}{4} M^4 \,\I_i^4\, \tr (\Psl^4)
+M^2 \,\I[q^2]_i^4 \,\tr (\gamma^\alpha \Psl \gamma_\alpha \Psl^3) \CR
&&\qquad\qquad +\,\I[q^4]_i^4 \Bigl( \frac{1}{2}\tr (\gamma^\alpha\Psl\gamma_\alpha\Psl\gamma^\beta\Psl\gamma_\beta\Psl) +\frac{1}{4}\tr (\gamma^\alpha\Psl\gamma^\beta\Psl\gamma_\alpha\Psl\gamma_\beta\Psl) \Bigr) \Bigr\}.
\eeqa{trP4fdiag}
Evaluation of the gamma matrix traces is standard and straightforward,
\bseq
\beqa
&& 
\tr(\Psl^4) 
= \tr(\gamma^\mu\gamma^\nu\gamma^\rho\gamma^\sigma) \tr( P_\mu P_\nu P_\rho P_\sigma) 
\supset -4\,\tr(P^\mu P^\nu P_\mu P_\nu) \,,\\[2pt]
&&
\tr (\gamma^\alpha \Psl \gamma_\alpha \Psl^3) 
= -2\, \tr(\Psl^4) +\O(\epsilon)
\supset 8\,\tr(P^\mu P^\nu P_\mu P_\nu) +\O(\epsilon) \,,\qquad\\[2pt]
&&
\tr (\gamma^\alpha\Psl\gamma_\alpha\Psl\gamma^\beta\Psl\gamma_\beta\Psl) 
= 4(1-\epsilon) \,\tr(\Psl^4) 
\supset -16(1-\epsilon)\,\tr(P^\mu P^\nu P_\mu P_\nu) \,, \\[2pt]
&&
\tr (\gamma^\alpha\Psl\gamma^\beta\Psl\gamma_\alpha\Psl\gamma_\beta\Psl) 
= \tr (\gamma^\alpha\gamma^\mu\gamma^\beta\gamma^\nu \gamma_\alpha\gamma^\rho \gamma_\beta\gamma^\sigma) \tr(P_\mu P_\nu P_\rho P_\sigma) \CR
&&\quad
= \bigl\{ -2\,\tr (\gamma^\nu\gamma^\beta\gamma^\mu \gamma^\rho \gamma_\beta\gamma^\sigma) +\epsilon \,\tr (\gamma^\mu\gamma^\beta\gamma^\nu \gamma^\rho \gamma_\beta\gamma^\sigma) \bigr\} \tr(P_\mu P_\nu P_\rho P_\sigma) \CR
&&\quad
= \bigl\{ -8 g^{\mu\rho}\,\tr (\gamma^\nu\gamma^\sigma) +2\epsilon\,\tr (\gamma^\nu\gamma^\mu \gamma^\rho\gamma^\sigma) +4\epsilon\, g^{\nu\rho}\,\tr (\gamma^\mu\gamma^\sigma) \bigr\} \tr(P_\mu P_\nu P_\rho P_\sigma) \CR
&&\quad
= \bigl\{ 8\epsilon\, g^{\mu\nu}g^{\rho\sigma} -(32-8\epsilon)g^{\mu\rho}g^{\nu\sigma} +8\epsilon\, g^{\mu\sigma}g^{\nu\rho} \bigr\} \tr(P_\mu P_\nu P_\rho P_\sigma) \CR
&&\quad
\supset -32 \Bigl(1-\frac{\epsilon}{4}\Bigr)\tr(P^\mu P^\nu P_\mu P_\nu) \,,
\eeqan
\eseq{trP4fcalc}
where terms involving $\tr(\dots P^2 \dots)$ have been dropped. Note that $\O(\epsilon)$ terms must be kept for the last two traces, since they are multiplied by $\,\I[q^4]_i^4$ which contains a $\frac{1}{\epsilon}$ pole. Plugging Eq.~\eqref{trP4fcalc} into \eqref{trP4fdiag}, we have
\beqa
\L_\text{EFT}^\text{1-loop} \supset && -i c_s \bigl\{ -M^4\,\I_i^4 +8M^2\,\I[q^2]_i^4 +(-16+10\epsilon)\,\I[q^4]_i^4 \bigr\} \,\tr(P^\mu P^\nu P_\mu P_\nu) \CR
= &&\pref \frac{2}{3} \logm{M^2} \,\tr(P^\mu P^\nu P_\mu P_\nu) 
\subset -\frac{1}{16\pi^2} \frac{1}{3} \logm{M^2} \,\tr\bigl([P^\mu, P^\nu] [P_\mu, P_\nu]\bigr) \CR
=&& -\frac{1}{16\pi^2} \frac{1}{3} \logm{M^2} \,\tr(G'^{\mu\nu}G'_{\mu\nu}) 
= \frac{g^2}{16\pi^2} T(R) \Bigl(-\frac{4}{3} \logm{M^2}\Bigr)\Bigl(-\frac{1}{4} G^{a\mu\nu} G^a_{\mu\nu}\Bigr) \,,\qquad
 \eeqa{LEFTloopVL}
 where $T(R)$ is the Dynkin index for the representation $R$ of the heavy vectorlike fermion, defined by $\tr(t_R^a t_R^b) = T(R)\delta^{ab}$ with $t_R^a$ being the generators in representation $R$; for example, $T(R)=\frac{1}{2}$ and $N$ for the fundamental and adjoint representations of $SU(N)$, respectively. Also, recall $c_s=-1$ for Dirac fermions\,\footnote{Unlike in Eq.~\eqref{LUVquad}, here $\L_\text{UV,\,quad.}$ can be written with prefactor $-1$, with only the vectorlike fermion field in the field multiplet of interest, and it is not necessary to represent this single Dirac fermion field by two fields as mentioned below Eq.~\eqref{LUVquad}. Of course the latter is OK to do, in which case the two fields would effectively have $c_s=-\frac{1}{2}$ each and contribute equally to $\L_\text{EFT}^\text{1-loop}$, leading to the same final result as Eq.~\eqref{LEFTloopVL}.}, and $G'_{\mu\nu}= -[P_\mu, P_\nu] = -igG_{\mu\nu} = -igG^a_{\mu\nu}t_R^a$. One can rescale the gauge fields to canonically normalize the kinetic terms while keeping $gG_{\mu\nu}$ unchanged. The result is the familiar one-loop matching formula for the gauge coupling across a heavy vectorlike fermion mass threshold (see e.g.~\cite{PichEFT}),
\beq
\frac{g^2_\text{eff}(\mu)}{g^2(\mu)} = 1+ \frac{g^2}{16\pi^2} T(R) \Bigl(\frac{4}{3} \logm{M^2}\Bigr).
\eeqn

\paragraph{$\O(P^6)$ terms.}
Diagrams with six $P$ insertions can be computed similarly. We enumerate them in the following, using $\gamma^\alpha\gamma^\mu\gamma_\alpha=-2\gamma^\mu$ to simplify the operator trace. Again, diagrams only giving rise to $\tr(\dots P^2\dots)$ are dropped.
\vspace{4pt}
\bseq
\beqa
&&
\begin{fmffile}{vlf-P6-a}
\begin{gathered}
\begin{fmfgraph}(40,40)
\fmfsurround{v4,v3,v2,v1,v6,v5}
\fmf{plain,left=0.25}{v1,v2,v3,v4,v5,v6,v1}
\fmfv{decor.shape=circle,decor.filled=full,decor.size=3thick}{v1,v2,v3,v4,v5,v6}
\end{fmfgraph}
\end{gathered}
\;\;+\;\,
\begin{gathered}
\begin{fmfgraph}(40,40)
\fmfsurround{v4,v34,v3,v23,v2,v12,v1,v61,v6,v56,v5,v45}
\fmf{plain,left=0.12}{v1,v12,v2,v23,v3,v34,v4,v45,v5,v56,v6,v61,v1}
\fmfv{decor.shape=circle,decor.filled=full,decor.size=3thick}{v1,v2,v3,v4,v5,v6}
\fmf{dots,width=thick,right=0.75}{v61,v12}
\end{fmfgraph}
\end{gathered}
\;\;+\;\,
\begin{gathered}
\begin{fmfgraph}(40,40)
\fmfsurround{v4,v34,v3,v23,v2,v12,v1,v61,v6,v56,v5,v45}
\fmf{plain,left=0.12}{v1,v12,v2,v23,v3,v34,v4,v45,v5,v56,v6,v61,v1}
\fmfv{decor.shape=circle,decor.filled=full,decor.size=3thick}{v1,v2,v3,v4,v5,v6}
\fmf{dots,width=thick,right=0.75}{v12,v23}
\fmf{dots,width=thick,right=0.75}{v56,v61}
\end{fmfgraph}
\end{gathered}
\;\;+\;\,
\begin{gathered}
\begin{fmfgraph}(40,40)
\fmfsurround{v4,v34,v3,v23,v2,v12,v1,v61,v6,v56,v5,v45}
\fmf{plain,left=0.12}{v1,v12,v2,v23,v3,v34,v4,v45,v5,v56,v6,v61,v1}
\fmfv{decor.shape=circle,decor.filled=full,decor.size=3thick}{v1,v2,v3,v4,v5,v6}
\fmf{dots,width=thick,right=0.75}{v61,v12}
\fmf{dots,width=thick,right=0.75}{v34,v45}
\end{fmfgraph}
\end{gathered}
\;\;+\;\,
\begin{gathered}
\begin{fmfgraph}(40,40)
\fmfsurround{v4,v34,v3,v23,v2,v12,v1,v61,v6,v56,v5,v45}
\fmf{plain,left=0.12}{v1,v12,v2,v23,v3,v34,v4,v45,v5,v56,v6,v61,v1}
\fmfv{decor.shape=circle,decor.filled=full,decor.size=3thick}{v1,v2,v3,v4,v5,v6}
\fmf{dots,width=thick,right=0.75}{v61,v12}
\fmf{dots,width=thick,right=0.75}{v23,v34}
\fmf{dots,width=thick,right=0.75}{v45,v56}
\end{fmfgraph}
\end{gathered}
\end{fmffile}
\CR[2pt]
&&\quad = -i c_s \Bigl\{ \frac{1}{6}M^6\,\I_i^6 +(-2)M^4\,\I[q^2]_i^6 +(-2)^2 \Bigl(1+\frac{1}{2}\Bigr)M^2\,\I[q^4]_i^6 +(-2)^3 \frac{1}{3} \,\I[q^6]_i^6 \Bigr\} \,\tr(\Psl^6) \,,\CR\\
&&
\begin{fmffile}{vlf-P6-b}
\begin{gathered}
\begin{fmfgraph}(40,40)
\fmfsurround{v4,v34,v3,v23,v2,v12,v1,v61,v6,v56,v5,v45}
\fmf{plain,left=0.12}{v1,v12,v2,v23,v3,v34,v4,v45,v5,v56,v6,v61,v1}
\fmfv{decor.shape=circle,decor.filled=full,decor.size=3thick}{v1,v2,v3,v4,v5,v6}
\fmf{dots,width=thick}{v23,v56}
\end{fmfgraph}
\end{gathered}
\;\;+\;\,
\begin{gathered}
\begin{fmfgraph}(40,40)
\fmfsurround{v4,v34,v3,v23,v2,v12,v1,v61,v6,v56,v5,v45}
\fmf{plain,left=0.12}{v1,v12,v2,v23,v3,v34,v4,v45,v5,v56,v6,v61,v1}
\fmfv{decor.shape=circle,decor.filled=full,decor.size=3thick}{v1,v2,v3,v4,v5,v6}
\fmf{dots,width=thick}{v23,v56}
\fmf{dots,width=thick,right=0.75}{v61,v12}
\end{fmfgraph}
\end{gathered}
\;\;+\;\,
\begin{gathered}
\begin{fmfgraph}(40,40)
\fmfsurround{v4,v34,v3,v23,v2,v12,v1,v61,v6,v56,v5,v45}
\fmf{plain,left=0.12}{v1,v12,v2,v23,v3,v34,v4,v45,v5,v56,v6,v61,v1}
\fmfv{decor.shape=circle,decor.filled=full,decor.size=3thick}{v1,v2,v3,v4,v5,v6}
\fmf{dots,width=thick}{v23,v56}
\fmf{dots,width=thick,right=0.75}{v61,v12}
\fmf{dots,width=thick,right=0.75}{v34,v45}
\end{fmfgraph}
\end{gathered}
\end{fmffile}
\CR[2pt]
&&\quad = -i c_s \Bigl\{ \frac{1}{2} M^4\,\I[q^2]_i^6 +(-2)M^2\,\I[q^4]_i^6 +(-2)^2\frac{1}{2}\,\I[q^6]_i^6 \Bigr\} \,\tr (\gamma^\alpha \Psl^3 \gamma_\alpha \Psl^3) \,,\\[8pt]
&&
\begin{fmffile}{vlf-P6-c}
\begin{gathered}
\begin{fmfgraph}(40,40)
\fmfsurround{v4,v34,v3,v23,v2,v12,v1,v61,v6,v56,v5,v45}
\fmf{plain,left=0.12}{v1,v12,v2,v23,v3,v34,v4,v45,v5,v56,v6,v61,v1}
\fmfv{decor.shape=circle,decor.filled=full,decor.size=3thick}{v1,v2,v3,v4,v5,v6}
\fmf{dots,width=thick,right=0.4}{v61,v23}
\fmf{dots,width=thick,right=0.4}{v56,v12}
\end{fmfgraph}
\end{gathered}
\;\;+\;\,
\begin{gathered}
\begin{fmfgraph}(40,40)
\fmfsurround{v4,v34,v3,v23,v2,v12,v1,v61,v6,v56,v5,v45}
\fmf{plain,left=0.12}{v1,v12,v2,v23,v3,v34,v4,v45,v5,v56,v6,v61,v1}
\fmfv{decor.shape=circle,decor.filled=full,decor.size=3thick}{v1,v2,v3,v4,v5,v6}
\fmf{dots,width=thick,right=0.4}{v61,v23}
\fmf{dots,width=thick,right=0.4}{v56,v12}
\fmf{dots,width=thick,right=0.75}{v34,v45}
\end{fmfgraph}
\end{gathered}
\end{fmffile}
\;\; \,=\, -i c_s \bigl\{ M^2\,\I[q^4]_i^6 +(-2)\,\I[q^6]_i^6 \bigr\} \,\tr (\gamma^\alpha \Psl \gamma^\beta \Psl \gamma_\alpha \Psl \gamma_\beta \Psl^3) \,,\\[6pt]
&&
\begin{fmffile}{vlf-P6-d}
\begin{gathered}
\begin{fmfgraph}(40,40)
\fmfsurround{v4,v34,v3,v23,v2,v12,v1,v61,v6,v56,v5,v45}
\fmf{plain,left=0.12}{v1,v12,v2,v23,v3,v34,v4,v45,v5,v56,v6,v61,v1}
\fmfv{decor.shape=circle,decor.filled=full,decor.size=3thick}{v1,v2,v3,v4,v5,v6}
\fmf{dots,width=thick}{v23,v56}
\fmf{dots,width=thick,right=0.4}{v12,v34}
\end{fmfgraph}
\end{gathered}
\end{fmffile}
\;\; \,=\, -i c_s M^2\,\I[q^4]_i^6 \,\tr (\gamma^\alpha \Psl \gamma^\beta \Psl \gamma_\alpha \Psl^2 \gamma_\beta \Psl^2) \,,\\[6pt]
&&
\begin{fmffile}{vlf-P6-e}
\begin{gathered}
\begin{fmfgraph}(40,40)
\fmfsurround{v4,v34,v3,v23,v2,v12,v1,v61,v6,v56,v5,v45}
\fmf{plain,left=0.12}{v1,v12,v2,v23,v3,v34,v4,v45,v5,v56,v6,v61,v1}
\fmfv{decor.shape=circle,decor.filled=full,decor.size=3thick}{v1,v2,v3,v4,v5,v6}
\fmf{dots,width=thick}{v12,v45}
\fmf{dots,width=thick}{v61,v34}
\end{fmfgraph}
\end{gathered}
\end{fmffile}
\;\; \,=\, -i c_s \,\frac{1}{2}M^2\,\I[q^4]_i^6 \,\tr (\gamma^\alpha \Psl \gamma^\beta \Psl^2 \gamma_\alpha \Psl \gamma_\beta \Psl^2) \,,\\[6pt]
&&
\begin{fmffile}{vlf-P6-f}
\begin{gathered}
\begin{fmfgraph}(40,40)
\fmfsurround{v4,v34,v3,v23,v2,v12,v1,v61,v6,v56,v5,v45}
\fmf{plain,left=0.12}{v1,v12,v2,v23,v3,v34,v4,v45,v5,v56,v6,v61,v1}
\fmfv{decor.shape=circle,decor.filled=full,decor.size=3thick}{v1,v2,v3,v4,v5,v6}
\fmf{dots,width=thick}{v23,v56}
\fmf{dots,width=thick,right=0.3}{v12,v34}
\fmf{dots,width=thick,right=0.3}{v45,v61}
\end{fmfgraph}
\end{gathered}
\end{fmffile}
\;\; \,=\, -i c_s \,\frac{1}{2}\,\I[q^6]_i^6 \,\tr (\gamma^\alpha \Psl \gamma^\beta \Psl \gamma^\delta \Psl \gamma_\alpha \Psl \gamma_\delta \Psl \gamma_\beta \Psl) \,,\\[6pt]
&&
\begin{fmffile}{vlf-P6-g}
\begin{gathered}
\begin{fmfgraph}(40,40)
\fmfsurround{v4,v34,v3,v23,v2,v12,v1,v61,v6,v56,v5,v45}
\fmf{plain,left=0.12}{v1,v12,v2,v23,v3,v34,v4,v45,v5,v56,v6,v61,v1}
\fmfv{decor.shape=circle,decor.filled=full,decor.size=3thick}{v1,v2,v3,v4,v5,v6}
\fmf{dots,width=thick}{v23,v56}
\fmf{dots,width=thick}{v12,v45}
\fmf{dots,width=thick}{v34,v61}
\end{fmfgraph}
\end{gathered}
\end{fmffile}
\;\; \,=\, -i c_s \,\frac{1}{6}\,\I[q^6]_i^6 \,\tr (\gamma^\alpha \Psl \gamma^\beta \Psl \gamma^\delta \Psl \gamma_\alpha \Psl \gamma_\beta \Psl \gamma_\delta \Psl) \,.
\eeqan
\eseq{trP6fdiag}
All loop integrals appearing in the equations above are finite, so $\O(\epsilon)$ terms can always be dropped when evaluating the traces:
\bseq
\beqa
&& \tr(\Psl^6) \supset -4\,\tr(P^\mu P^\nu P^\rho P_\mu P_\nu P_\rho) +12\,\tr(P^\mu P^\nu P^\rho P_\mu P_\rho P_\nu)\,, \\[2pt]
&& \tr (\gamma^\alpha \Psl^3 \gamma_\alpha \Psl^3) \supset -8\,\tr(P^\mu P^\nu P^\rho P_\mu P_\nu P_\rho) +8\,\tr(P^\mu P^\nu P^\rho P_\mu P_\rho P_\nu)\,, \\[2pt]
&& \tr (\gamma^\alpha \Psl \gamma^\beta \Psl \gamma_\alpha \Psl \gamma_\beta \Psl^3) \supset 32\,\tr(P^\mu P^\nu P^\rho P_\mu P_\rho P_\nu)\,, \\[2pt]
&& \tr (\gamma^\alpha \Psl \gamma^\beta \Psl \gamma_\alpha \Psl^2 \gamma_\beta \Psl^2) \supset 16\,\tr(P^\mu P^\nu P^\rho P_\mu P_\nu P_\rho) -16\,\tr(P^\mu P^\nu P^\rho P_\mu P_\rho P_\nu)\,, \\[2pt]
&& \tr (\gamma^\alpha \Psl \gamma^\beta \Psl^2 \gamma_\alpha \Psl \gamma_\beta \Psl^2) \supset -16\,\tr(P^\mu P^\nu P^\rho P_\mu P_\nu P_\rho) +48\,\tr(P^\mu P^\nu P^\rho P_\mu P_\rho P_\nu)\,, \\[2pt]
&& \tr (\gamma^\alpha \Psl \gamma^\beta \Psl \gamma^\delta \Psl \gamma_\alpha \Psl \gamma_\delta \Psl \gamma_\beta \Psl) \supset -32\,\tr(P^\mu P^\nu P^\rho P_\mu P_\nu P_\rho) -96\,\tr(P^\mu P^\nu P^\rho P_\mu P_\rho P_\nu)\,,\qquad\CR \\
&& \tr (\gamma^\alpha \Psl \gamma^\beta \Psl \gamma^\delta \Psl \gamma_\alpha \Psl \gamma_\beta \Psl \gamma_\delta \Psl )\supset -128\,\tr(P^\mu P^\nu P^\rho P_\mu P_\nu P_\rho)\,,
\eeqan
\eseq{trP6fcalc}
where terms involving $\tr(\dots P^2\dots)$ have been dropped as before. Plugging Eq.~\eqref{trP6fcalc} into \eqref{trP6fdiag}, we can organize the two operator traces into two independent dimension-six pure gauge operators,
\beqa
\L_\text{EFT}^\text{1-loop} \supset && -i c_s \Bigl\{ -\frac{2}{3} M^6\,\I_i^6 +4M^4 \,\I[q^2]_i^6 -\frac{128}{3} \,\I[q^6]_i^6 \Bigr\} \,\tr(P^\mu P^\nu P^\rho P_\mu P_\nu P_\rho) \CR
&& -i c_s \Bigl\{ 2M^6\,\I_i^6 -20M^4 \,\I[q^2]_i^6 +96M^2 \,\I[q^4]_i^6 -128\,\I[q^6]_i^6 \Bigr\}\,\tr(P^\mu P^\nu P^\rho P_\mu P_\rho P_\nu) \CR
= && \pref \frac{1}{M^2} \Bigl\{ -\frac{1}{45} \,\tr(P^\mu P^\nu P^\rho P_\mu P_\nu P_\rho) +\frac{3}{5}\,\tr(P^\mu P^\nu P^\rho P_\mu P_\rho P_\nu) \Bigr\} \CR
\subset && \pref \frac{1}{M^2} \Bigl\{ \frac{2}{15}\,\tr\bigl([P^\mu, G'_{\mu\nu}][P_\rho, G'^{\rho\nu}]\bigr) -\frac{1}{45}\,\tr(G^{\prime\mu}_{\;\;\,\nu} G^{\prime\nu}_{\;\;\,\rho} G^{\prime\rho}_{\;\;\,\mu}) \Bigr\} \CR
= && \frac{1}{16\pi^2}\frac{1}{M^2} \frac{g^2}{60} T(R) \bigl(16\O_{2G} -4\O_{3G}\bigr),
\eeqan
where
\beq
\O_{2G} = -\frac{1}{2}(D^\mu G^a_{\mu\nu})^2 \,,\quad
\O_{3G} = \frac{g}{6} f^{abc} G^{a\mu}_{\;\;\;\;\nu} G^{b\nu}_{\;\;\;\,\rho} G^{c\rho}_{\;\;\;\,\mu} \,.
\eeqn

\subsection{Integrating out a charged scalar singlet: penguin operators}
\label{sec:singlet}

We finally consider an example for one-loop matching involving mixed statistics. The UV theory is the SM extended by a heavy $SU(2)_L$ singlet scalar $h$ with hypercharge $-1$, which couples to the SM Higgs and lepton doublets $\phi$ and $l$. The Lagrangian reads 
\beq
\L_\text{UV} = \L_\text{SM} + |D_\mu h|^2 -M^2 |h|^2 -\alpha|h|^4 -\beta|\phi|^2|h|^2 +\bar l f^\dagger \tilde l h +h^\dagger \bar{\tilde l} f l \,,
\eeqn
where $\tilde l \equiv i\sigma^2 l^\c$, with charge conjugation defined as $l^\c\equiv -i\gamma^2 l^*$. $f$ is a $3\times3$ antisymmetric matrix in generation space; e.g.\ $\bar{\tilde l} f l$ is short for $\bar{\tilde l}_a f_{ab} l_b$ with generation indices $a,b$ summed over. One-loop matching of this model onto the SMEFT is discussed in~\cite{BilenkyS,DKS}, with mixed heavy-light contributions obtained by computing Feynman diagrams. Here, we focus on a subset of dimension-6 operators generated in this model -- the penguin operators -- as an example to demonstrate the use of covariant diagrams involving heavy bosonic and light fermionic loop propagators.

We shall continue to use the four-component notation, treating $l$ as a Dirac fermion field whose right-handed component should be set to zero in the end --- this is legitimate since the unphysical component $l_R$ cannot appear only in the loop. The quadratic terms in $\L_\text{UV}$ needed for our calculation read
\beq
\L_\text{UV, quad} \supset -\frac{1}{2} \Bigl( h'^\dagger \;\; h'^T \;\; \bar l' \;\; \bar{\tilde l}' \; \Bigr) 
\left(
\begin{matrix}
(-P^2+M^2+U_h)_{2\times 2} & (U_{hl})_{2\times2} \\
(U_{lh})_{2\times2} & (-\Psl+U_l)_{2\times2} 
\end{matrix}
\right)
\left(
\begin{matrix}
h' \\ h'^* \\ l' \\ \tilde l' 
\end{matrix}
\right), \qquad
\eeq{Lsingletquad}
where
\beqa
&& U_h = 
\left(
\begin{matrix}
2\alpha \bigl(|h_\c|^2 +h_\c h_\c^\dagger\bigr) +\beta |\phi|^2 & 
2\alpha h_\c h_\c^T \\
2\alpha h_\c^* h_\c^\dagger & 
2\alpha \bigl(|h_\c|^2 +h_\c^* h_\c^T\bigr) +\beta |\phi|^2
\end{matrix}
\right) ,\CR
&& U_l =
\left(
\begin{matrix}
0 & -2f^\dagger h_\c \\
-2 h_\c^\dagger f & 0
\end{matrix}
\right) , \quad
U_{lh} =
\left(
\begin{matrix}
-2 f^\dagger \tilde l & 0 \\
0 & -2 f l
\end{matrix}
\right) ,\quad
U_{hl} =
\left(
\begin{matrix}
-2 \bar{\tilde l}f & 0 \\
0 & -2 \bar l f^\dagger
\end{matrix}
\right) .
\quad
\eeqan
The light fields $\phi$, $l$, $\tilde l$ are understood as background fields $\phi_\b$, $l_\b$, $\tilde l_\b$. Parametrically, $h_\c[\phi, l]\sim \O(f\,l^2)$ at leading order, whose explicit form will not be relevant for our calculation. 
The separations of the complex scalar $h$ into $(h, h^*)$ (with $h^*=h^\dagger$ for a scalar singlet) and the Dirac fermion $l$ into $(l, \tilde l)$ are necessary due to the presence of off-diagonal terms in $U_h$ and $U_l$. As a result, each bosonic (fermionic) field in the field multiplet of Eq.~\eqref{Lsingletquad} effectively has $c_s=\frac{1}{2}$ ($c_s=-\frac{1}{2}$). This is similar to the separation of the SM Higgs field $\phi$ into $(\phi, \tilde\phi)$ in the scalar triplet example in Sections~\ref{sec:triplet-scalar} and~\ref{sec:triplet-gauge}. 

The penguin operators we wish to compute are $\sim\O(P^3 l^2)$. At one-loop level, they can only arise from covariant diagrams with one $U_{hl}$, one $U_{lh}$ and three $P$ insertions. There are nine such diagrams, two of which are hermitian conjugates of each other. They can be easily enumerated by distributing three $P$ insertions on the $h$ and $l$ propagators and contracting the bosonic $P$ insertions and fermionic light propagators (which, unlike the fermionic heavy propagators, cannot be left uncontracted). We will always start reading a covariant diagram from a bosonic propagator, and thus $c_s=\frac{1}{2}$. Dropping $\tr(\dots P^2 \dots)$ terms as before, we have
\bseq
\beqa
\begin{gathered}
\begin{fmffile}{singlet-penguin-a}
\begin{fmfgraph}(40,40)
\fmfsurround{vUlh,v1,fp,v0,vUhl,vP3,vP2,vP1}
\fmf{plain,left=0.2}{vUlh,vP1,vP2,vP3,vUhl}
\fmf{dashes,left=0.2}{vUhl,v0,fp,v1,vUlh}
\fmfv{decor.shape=circle,decor.filled=empty,decor.size=3thick}{vUhl,vUlh}
\fmfv{decor.shape=circle,decor.filled=full,decor.size=3thick}{vP1,vP2,vP3}
\fmf{dots,width=thick,right=0.6}{vP1,vP3}
\fmf{dots,width=thick}{vP2,fp}
\end{fmfgraph}
\end{fmffile}
\end{gathered}
\;\; 
&=&  -i c_s (-2^3) \,\I[q^4]_{i0}^{41} \,\tr(U_{hl} \gamma^\mu U_{lh} P^\nu P_\mu P_\nu)
= \pref \frac{1}{M^2} \frac{1}{9}\,\tr(U_{hl} \gamma^\mu U_{lh} P^\nu P_\mu P_\nu), \CR[-10pt]\\
\begin{gathered}
\begin{fmffile}{singlet-penguin-b}
\begin{fmfgraph}(40,40)
\fmfsurround{vUlh,fp2,vP1,fp1,vUhl,vP3,v0,vP2}
\fmf{plain,left=0.2}{vUlh,vP2,v0,vP3,vUhl}
\fmf{dashes,left=0.2}{vUhl,fp1,vP1,fp2,vUlh}
\fmfv{decor.shape=circle,decor.filled=empty,decor.size=3thick}{vUhl,vUlh}
\fmfv{decor.shape=circle,decor.filled=full,decor.size=3thick}{vP1,vP2,vP3}
\fmf{dots,width=thick}{vP2,fp1}
\fmf{dots,width=thick}{vP3,fp2}
\end{fmfgraph}
\end{fmffile}
\end{gathered}
\;\; 
&=&  -i c_s (-2^2) \,\I[q^4]_{i0}^{32} \,\tr(U_{hl} \gamma^\mu \gamma^\rho \gamma^\nu P_\rho U_{lh} P_\mu P_\nu) \CR
&=& \pref \frac{1}{M^2} \frac{1}{12}\,\tr(U_{hl} \gamma^\mu \gamma^\rho \gamma^\nu P_\rho U_{lh} P_\mu P_\nu) \CR
&\supset& \pref \frac{1}{M^2} \Bigl\{\frac{1}{12} \,\tr\bigl(U_{hl}\gamma^\mu P^\nu U_{lh} (P_\mu P_\nu +P_\nu P_\mu)\bigr) \CR
&&\qquad\qquad\quad +\frac{1}{48}\,\tr\bigl(i\, U_{hl} (\sigma^{\mu\nu} \gamma^\rho +\gamma^\rho \sigma^{\mu\nu}) P_\rho U_{lh} [P_\mu, P_\nu] \bigr) \Bigr\},
\\
\begin{gathered}
\begin{fmffile}{singlet-penguin-c}
\begin{fmfgraph}(40,40)
\fmfsurround{vUlh,fp2,vP1,fp1,vUhl,vP3,v0,vP2}
\fmf{plain,left=0.2}{vUlh,vP2,v0,vP3,vUhl}
\fmf{dashes,left=0.2}{vUhl,fp1,vP1,fp2,vUlh}
\fmfv{decor.shape=circle,decor.filled=empty,decor.size=3thick}{vUhl,vUlh}
\fmfv{decor.shape=circle,decor.filled=full,decor.size=3thick}{vP1,vP2,vP3}
\fmf{dots,width=thick,right=0.6}{vP3,fp1}
\fmf{dots,width=thick,right=0.6}{fp2,vP2}
\end{fmfgraph}
\end{fmffile}
\end{gathered}
\;\; 
&=&  -i c_s (-2^2) \,\I[q^4]_{i0}^{32} \,\tr(U_{hl} \gamma^\mu \gamma^\rho \gamma^\nu P_\rho U_{lh} P_\nu P_\mu) \CR
&=& \pref \frac{1}{M^2} \frac{1}{12}\,\tr(U_{hl} \gamma^\mu \gamma^\rho \gamma^\nu P_\rho U_{lh} P_\nu P_\mu) \CR
&\supset& \pref \frac{1}{M^2} \Bigl\{\frac{1}{12} \,\tr\bigl(U_{hl}\gamma^\mu P^\nu U_{lh} (P_\mu P_\nu +P_\nu P_\mu)\bigr) \CR
&&\qquad\qquad\quad -\frac{1}{48}\,\tr\bigl(i\, U_{hl} (\sigma^{\mu\nu} \gamma^\rho +\gamma^\rho \sigma^{\mu\nu}) P_\rho U_{lh} [P_\mu, P_\nu] \bigr) \Bigr\},
\\
\begin{gathered}
\begin{fmffile}{singlet-penguin-e1}
\begin{fmfgraph}(40,40)
\fmfsurround{vUlh,fp3,vP2,fp2,vP1,fp1,vUhl,v3,v2,vP3,v1,v0}
\fmf{plain,left=0.12}{vUlh,v0,v1,vP3,v2,v3,vUhl}
\fmf{dashes,left=0.12}{vUhl,fp1,vP1,fp2,vP2,fp3,vUlh}
\fmfv{decor.shape=circle,decor.filled=empty,decor.size=3thick}{vUhl,vUlh}
\fmfv{decor.shape=circle,decor.filled=full,decor.size=3thick}{vP1,vP2,vP3}
\fmf{dots,width=thick}{vP3,fp2}
\fmf{dots,width=thick,right=0.4}{fp1,fp3}
\end{fmfgraph}
\end{fmffile}
\end{gathered}
\;\; 
&=& -i c_s (-2) \,\I[q^4]_{i0}^{23} \,\tr (U_{hl} \gamma^\alpha \gamma^\mu \gamma^\rho \gamma^\nu \gamma_\alpha P_\mu P_\nu U_{lh} P_\rho) \CR
&=& \pref \frac{1}{M^2} \Bigl(-\frac{1}{6}\Bigr) \,\tr(U_{hl} \gamma^\nu \gamma^\rho \gamma^\mu P_\mu P_\nu U_{lh} P_\rho) \CR
&\supset& \pref\frac{1}{M^2} \Bigl\{ -\frac{1}{6} \,\tr\bigl(U_{hl} \gamma^\mu (P_\mu P_\nu +P_\nu P_\mu) U_{lh} P^\nu \bigr) \CR
&&\qquad\qquad\quad +\frac{1}{24} \,\tr\bigl(i\, U_{hl} (\sigma^{\mu\nu} \gamma^\rho +\gamma^\rho \sigma^{\mu\nu})[P_\mu, P_\nu] U_{lh} P_\rho \bigr) \Bigr\} ,\\
\begin{gathered}
\begin{fmffile}{singlet-penguin-e}
\begin{fmfgraph}(40,40)
\fmfsurround{vUlh,fp3,vP2,fp2,vP1,fp1,vUhl,v3,v2,vP3,v1,v0}
\fmf{plain,left=0.12}{vUlh,v0,v1,vP3,v2,v3,vUhl}
\fmf{dashes,left=0.12}{vUhl,fp1,vP1,fp2,vP2,fp3,vUlh}
\fmfv{decor.shape=circle,decor.filled=empty,decor.size=3thick}{vUhl,vUlh}
\fmfv{decor.shape=circle,decor.filled=full,decor.size=3thick}{vP1,vP2,vP3}
\fmf{dots,width=thick,left=0.3}{vP3,fp3}
\fmf{dots,width=thick,right=0.75}{fp1,fp2}
\end{fmfgraph}
\end{fmffile}
\end{gathered}
\;\;
&+& \;\,
\begin{gathered}
\begin{fmffile}{singlet-penguin-e2}
\begin{fmfgraph}(40,40)
\fmfsurround{vUlh,fp3,vP2,fp2,vP1,fp1,vUhl,v3,v2,vP3,v1,v0}
\fmf{plain,left=0.12}{vUlh,v0,v1,vP3,v2,v3,vUhl}
\fmf{dashes,left=0.12}{vUhl,fp1,vP1,fp2,vP2,fp3,vUlh}
\fmfv{decor.shape=circle,decor.filled=empty,decor.size=3thick}{vUhl,vUlh}
\fmfv{decor.shape=circle,decor.filled=full,decor.size=3thick}{vP1,vP2,vP3}
\fmf{dots,width=thick,right=0.3}{vP3,fp1}
\fmf{dots,width=thick,right=0.75}{fp2,fp3}
\end{fmfgraph}
\end{fmffile}
\end{gathered}
\CR
&=& -i c_s (-2) \,\I[q^4]_{i0}^{23} \,\tr (U_{hl} \gamma^\alpha \gamma^\mu \gamma_\alpha \gamma^\nu \gamma^\rho P_\mu P_\nu U_{lh} P_\rho +U_{hl} \gamma^\rho \gamma^\mu \gamma^\alpha \gamma^\nu \gamma_\alpha P_\mu P_\nu U_{lh} P_\rho) \CR
&=& \pref \frac{1}{M^2} \Bigl(-\frac{1}{6}\Bigr) \,\tr(U_{hl} \gamma^\mu \gamma^\nu \gamma^\rho P_\mu P_\nu U_{lh} P_\rho +U_{hl} \gamma^\rho \gamma^\mu \gamma^\nu P_\mu P_\nu U_{lh} P_\rho) \CR
&\supset& \pref \frac{1}{M^2} \frac{1}{12}\,\tr\bigl( i\, U_{hl} (\sigma^{\mu\nu} \gamma^\rho +\gamma^\rho \sigma^{\mu\nu})[P_\mu, P_\nu] U_{lh} P_\rho \bigr) ,\\
\begin{gathered}
\begin{fmffile}{singlet-penguin-f}
\begin{fmfgraph}(40,40)
\fmfsurround{fp4,vP3,fp3,vP2,fp2,vP1,fp1,vUhl,v2,v1,v0,vUlh}
\fmf{plain,left=0.12}{vUlh,v0,v1,v2,vUhl}
\fmf{dashes,left=0.12}{vUhl,fp1,vP1,fp2,vP2,fp3,vP3,fp4,vUlh}
\fmfv{decor.shape=circle,decor.filled=empty,decor.size=3thick}{vUhl,vUlh}
\fmfv{decor.shape=circle,decor.filled=full,decor.size=3thick}{vP1,vP2,vP3}
\fmf{dots,width=thick,right=0.75}{fp1,fp2}
\fmf{dots,width=thick,right=0.75}{fp3,fp4}
\end{fmfgraph}
\end{fmffile}
\end{gathered}
\;\;
&=& -i c_s (-1) \,\I[q^4]_{i0}^{14} \,\tr (U_{hl} \gamma^\alpha \gamma^\mu \gamma_\alpha \gamma^\nu \gamma^\beta \gamma^\rho \gamma_\beta P_\mu P_\nu P_\rho U_{lh}) \CR
&=& -i c_s (-4+4\epsilon) \,\I[q^4]_{i0}^{14} \,\tr (U_{hl} \gamma^\mu \gamma^\nu \gamma^\rho P_\mu P_\nu P_\rho U_{lh}) \CR
&=& \pref \Bigl(\frac{1}{36} +\frac{1}{6}\logm{M^2}\Bigr)\,\tr (U_{hl} \gamma^\mu \gamma^\nu \gamma^\rho P_\mu P_\nu P_\rho U_{lh}) \CR
&\supset& \pref \Bigl(-\frac{1}{144} -\frac{1}{24}\logm{M^2}\Bigr)\,\tr\bigl(i\, U_{hl} (\sigma^{\mu\nu} \gamma^\rho [P_\mu, P_\nu] P_\rho +\gamma^\rho \sigma^{\mu\nu}P_\rho [P_\mu, P_\nu] ) U_{lh} \bigr), \CR\\
\begin{gathered}
\begin{fmffile}{singlet-penguin-g}
\begin{fmfgraph}(40,40)
\fmfsurround{fp4,vP3,fp3,vP2,fp2,vP1,fp1,vUhl,v2,v1,v0,vUlh}
\fmf{plain,left=0.12}{vUlh,v0,v1,v2,vUhl}
\fmf{dashes,left=0.12}{vUhl,fp1,vP1,fp2,vP2,fp3,vP3,fp4,vUlh}
\fmfv{decor.shape=circle,decor.filled=empty,decor.size=3thick}{vUhl,vUlh}
\fmfv{decor.shape=circle,decor.filled=full,decor.size=3thick}{vP1,vP2,vP3}
\fmf{dots,width=thick}{fp1,fp4}
\fmf{dots,width=thick,right=0.75}{fp2,fp3}
\end{fmfgraph}
\end{fmffile}
\end{gathered}
\;\;
&=& -i c_s (-1) \,\I[q^4]_{i0}^{14} \,\tr (U_{hl} \gamma^\alpha \gamma^\mu \gamma^\beta \gamma^\nu \gamma_\beta \gamma^\rho \gamma_\alpha P_\mu P_\nu P_\rho U_{lh}) \CR
&=& -i c_s \,\I[q^4]_{i0}^{14} \bigl\{ (-4+2\epsilon)\,\tr (U_{hl} \gamma^\rho \gamma^\nu \gamma^\mu P_\mu P_\nu P_\rho U_{lh}) \CR
&&\qquad\qquad\qquad +2\epsilon\,\tr (U_{hl} \gamma^\mu \gamma^\nu \gamma^\rho P_\mu P_\nu P_\rho U_{lh}) \bigr\} \CR
&=& \pref\frac{1}{M^2} \Bigl\{ \Bigl(-\frac{5}{36}+\frac{1}{6}\logm{M^2}\Bigr)\,\tr (U_{hl} \gamma^\rho \gamma^\nu \gamma^\mu P_\mu P_\nu P_\rho U_{lh}) \CR
&&\qquad\qquad\quad +\frac{1}{6}\,\tr (U_{hl} \gamma^\mu \gamma^\nu \gamma^\rho P_\mu P_\nu P_\rho U_{lh}) \Bigr\} \CR
&\supset& \pref\frac{1}{M^2} \Bigl\{ \Bigl(\frac{5}{18}-\frac{1}{3}\logm{M^2}\Bigr)\,\tr(U_{hl} \gamma^\mu P^\nu P_\mu P_\nu U_{lh} ) \CR
&&\quad +\Bigl(-\frac{11}{144}+\frac{1}{24}\logm{M^2}\Bigr)\,\tr\bigl( i\, U_{hl} (\sigma^{\mu\nu} \gamma^\rho [P_\mu, P_\nu] P_\rho +\gamma^\rho \sigma^{\mu\nu}P_\rho [P_\mu, P_\nu] ) U_{lh} \bigr)\Bigr\}, \CR\\
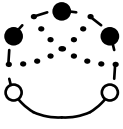
\begin{gathered}
\begin{fmffile}{singlet-penguin-h}
\begin{fmfgraph}(40,40)
\fmfsurround{fp4,vP3,fp3,vP2,fp2,vP1,fp1,vUhl,v2,v1,v0,vUlh}
\fmf{plain,left=0.12}{vUlh,v0,v1,v2,vUhl}
\fmf{dashes,left=0.12}{vUhl,fp1,vP1,fp2,vP2,fp3,vP3,fp4,vUlh}
\fmfv{decor.shape=circle,decor.filled=empty,decor.size=3thick}{vUhl,vUlh}
\fmfv{decor.shape=circle,decor.filled=full,decor.size=3thick}{vP1,vP2,vP3}
\fmf{dots,width=thick,right=0.3}{fp1,fp3}
\fmf{dots,width=thick,right=0.3}{fp2,fp4}
\end{fmfgraph}
\end{fmffile}
\end{gathered}
\;\;
&=& -i c_s (-1) \,\I[q^4]_{i0}^{14} \,\tr (U_{hl} \gamma^\alpha \gamma^\mu \gamma^\beta \gamma^\nu \gamma_\alpha \gamma^\rho \gamma_\beta P_\mu P_\nu P_\rho U_{lh}) \CR
&\supset& -i c_s \,\I[q^4]_{i0}^{14} \Bigl\{ 8\,\tr(U_{hl} \gamma^\mu P^\nu P_\mu P_\nu U_{lh}) \CR
&&\qquad\qquad\qquad -\frac{\epsilon}{2}\,\tr\bigl( i U_{hl} (\sigma^{\mu\nu} \gamma^\rho [P_\mu, P_\nu] P_\rho +\gamma^\rho \sigma^{\mu\nu}P_\rho [P_\mu, P_\nu] ) U_{lh} \bigr) \Bigr\} \CR
&=& \pref\frac{1}{M^2} \Bigl\{ \Bigl(\frac{11}{18} -\frac{1}{3}\logm{M^2}\Bigr) \,\tr(U_{hl} \gamma^\mu P^\nu P_\mu P_\nu U_{lh}) \CR
&&\qquad\qquad\quad -\frac{1}{24}\,\tr\bigl( i\, U_{hl} (\sigma^{\mu\nu} \gamma^\rho [P_\mu, P_\nu] P_\rho +\gamma^\rho \sigma^{\mu\nu}P_\rho [P_\mu, P_\nu] ) U_{lh} \bigr) \Bigr\} ,
\eeqan
\eseq{penguin}
where $\sigma^{\mu\nu}=\frac{1}{2}[\gamma^\mu, \gamma^\nu]$. The $\O(\epsilon)$ terms coming from gamma matrix algebra must be kept when computing the last three diagrams, which involve the divergent master integral $\I[q^4]_{i0}^{14}=\frac{1}{24M_i^2}(\frac{11}{6}-\logm{M_i^2})$, understood as $\frac{1}{24M_i^2}(\frac{2}{\bar\epsilon}+\frac{11}{6}-\logm{M_i^2})$. The following identities, together with the standard gamma matrix formulas, are useful in deriving Eq.~\eqref{penguin},
\bseq
\beqa
\gamma^\mu\gamma^\nu &=& g^{\mu\nu} -i\sigma^{\mu\nu}, \\
\gamma^\mu \gamma^\rho \gamma^\nu 
&=& \frac{1}{2} \bigl( \{\gamma^\mu, \gamma^\rho\} \gamma^\nu +\gamma^\mu \{\gamma^\rho, \gamma^\nu\} -\gamma^\rho \gamma^\mu \gamma^\nu -\gamma^\mu \gamma^\nu \gamma^\rho \bigr) \CR
&=& g^{\mu\rho} \gamma^\nu +g^{\nu\rho} \gamma^\mu -g^{\mu\nu} \gamma^\rho +\frac{i}{2} (\sigma^{\mu\nu} \gamma^\rho +\gamma^\rho \sigma^{\mu\nu}) \\
&=& g^{\mu\rho} \gamma^\nu +g^{\nu\rho} \gamma^\mu -\frac{1}{2} \bigl(\gamma^\rho \{\gamma^\mu, \gamma^\nu\} +\{\gamma^\mu, \gamma^\nu\} \gamma^\rho -\gamma^\rho \gamma^\nu \gamma^\mu -\gamma^\nu \gamma^\mu \gamma^\rho \bigr) \CR
&=& \frac{3}{2} (g^{\mu\rho} \gamma^\nu +g^{\nu\rho} \gamma^\mu) -2g^{\mu\nu} \gamma^\rho -\frac{i}{2} (\sigma^{\rho\nu} \gamma^\mu +\gamma^\nu \sigma^{\mu\rho}), \\
\gamma^\alpha \gamma^\mu \gamma^\beta \gamma^\nu \gamma_\alpha \gamma^\rho \gamma_\beta &=& \frac{1}{2} \bigl\{ -2(\gamma^\nu \gamma^\beta \gamma^\mu \gamma^\rho \gamma_\beta +\gamma^\alpha \gamma^\mu \gamma^\rho \gamma_\alpha \gamma^\nu) +\epsilon(\gamma^\mu \gamma^\beta \gamma^\nu \gamma^\rho \gamma_\beta +\gamma^\alpha \gamma^\mu \gamma^\nu \gamma_\alpha \gamma^\rho) \bigr\} \CR
&=& -8g^{\mu\rho} \gamma^\nu +2\epsilon (g^{\nu\rho} \gamma^\mu +g^{\mu\nu} \gamma^\rho) +\epsilon (\gamma^\nu \gamma^\mu \gamma^\rho +\gamma^\mu \gamma^\rho \gamma^\nu) \CR
&=&-8g^{\mu\rho} \gamma^\nu +3\epsilon (g^{\nu\rho} \gamma^\mu +g^{\mu\nu} \gamma^\rho) +i\epsilon (\sigma^{\mu\nu} \gamma^\rho +\gamma^\mu \sigma^{\nu\rho}) .
\eeqan
\eseq{gammaidentities}
Note that we have been careful to keep all expressions in the intermediate steps of the calculation manifestly hermitian, in order to easily obtain manifestly hermitian final results. This is why we have applied gamma matrix formulas in a symmetric manner in Eq.~\eqref{gammaidentities}. Adding up all diagrams computed in Eq.~\eqref{penguin}, we have
\beqa
\L_\text{EFT}^\text{1-loop} 
&\supset& \pref \frac{1}{M^2} \,\tr\Bigl\{ \,\frac{1}{9} U_{hl}\, \gamma^\mu U_{lh} P^\nu P_\mu P_\nu
+\frac{1}{6} \,U_{hl}\gamma^\mu P^\nu U_{lh} (P_\mu P_\nu +P_\nu P_\mu) \CR
&& -\frac{1}{6} \,U_{hl} \gamma^\mu (P_\mu P_\nu +P_\nu P_\mu) U_{lh} P^\nu
+\Bigl(\frac{8}{9}-\frac{2}{3}\logm{M^2}\Bigr) U_{hl} \gamma^\mu P^\nu P_\mu P_\nu U_{lh} \CR
&& +\frac{1}{8} \,i\, U_{hl} (\sigma^{\mu\nu} \gamma^\rho +\gamma^\rho \sigma^{\mu\nu})[P_\mu, P_\nu] U_{lh} P_\rho \CR
&& -\frac{1}{8} i\, U_{hl} (\sigma^{\mu\nu} \gamma^\rho [P_\mu, P_\nu] P_\rho +\gamma^\rho \sigma^{\mu\nu}P_\rho [P_\mu, P_\nu] ) U_{lh} \Bigr\} \CR
&\subset& \pref \frac{1}{M^2} \,\tr\Bigl\{\, \frac{1}{12} \bigl(U_{hl} \gamma^\mu [P_\nu, [P^\nu, [P_\mu, U_{lh}] ] ] +U_{hl} \gamma^\mu [P_\mu, [P^\nu, [P_\nu, U_{lh}] ] ] \bigr) \CR
&& -\frac{1}{18} U_{hl} \gamma^\nu U_{lh} [P^\mu, [P_\mu, P_\nu]]
+\Bigl(-\frac{4}{9} +\frac{1}{3}\logm{M^2}\Bigr) U_{hl} \gamma^\nu [P^\mu, [P_\mu, P_\nu]] U_{lh} \CR
&& -\frac{1}{8} \bigl(i\,U_{hl} \sigma^{\mu\nu} \gamma^\rho [P_\mu, P_\nu] [P_\rho, U_{lh}] -i[P_\rho, U_{hl}] \sigma^{\mu\nu} \gamma^\rho [P_\mu, P_\nu] U_{lh} \bigr)
\Bigr\} \CR
&=& \pref\frac{1}{M^2} \Bigl\{ -\frac{1}{3} \bigl[ i\,\bar l f^\dagger f (D^2\slashed{D} +\slashed{D}D^2) l +i\,\bar{\tilde l} f f^\dagger (D^2\slashed{D} +\slashed{D}D^2) \tilde l \,\bigr] \CR
&& +\frac{2}{9} \bigl[ (\bar l f^\dagger f \gamma^\nu l) D^\mu (g'B_{\mu\nu}Y_{h^*}) +(\bar{\tilde l} f f^\dagger \gamma^\nu \tilde l) D^\mu (g'B_{\mu\nu}Y_h) \bigr] \CR
&& +\Bigl(\frac{16}{9}-\frac{4}{3}\logm{M^2}\Bigr) \CR
&&\qquad
\Bigl[\, \bar l f^\dagger f \gamma^\nu D^\mu \Bigl(gW^a_{\mu\nu}\frac{\tau^a}{2} +g'B_{\mu\nu}Y_{\tilde l}\Bigr) l
 +\bar{\tilde l} f f^\dagger \gamma^\nu D^\mu \Bigl(gW^a_{\mu\nu}\frac{\tau^a}{2} +g'B_{\mu\nu}Y_l\Bigr) \tilde l \,\Bigr] \CR
&& +\frac{1}{2} \Bigl[\, i\,\bar l f^\dagger f \sigma^{\mu\nu} \Bigl(gW^a_{\mu\nu}\frac{\tau^a}{2} +g'B_{\mu\nu}Y_{\tilde l}\Bigr) \slashed{D} l \CR
&&\qquad 
+i\,\bar{\tilde l} f f^\dagger \sigma^{\mu\nu} \Bigl(gW^a_{\mu\nu}\frac{\tau^a}{2} +g'B_{\mu\nu}Y_l\Bigr) \slashed{D} \tilde l +\text{h.c.} \Bigr] \Bigr\} \CR
&=& \frac{1}{16\pi^2} \frac{1}{M^2} \Bigl\{
-\frac{1}{3} i\,\bar l f^\dagger f (D^2\slashed{D} +\slashed{D}D^2) l
\CR && 
+\frac{2}{3}\Bigl(\frac{5}{3}-\logm{M^2}\Bigr) g' (\bar l f^\dagger f \gamma^\nu l) (D^\mu B_{\mu\nu}) 
+\frac{2}{3}\Bigl(\frac{4}{3}-\logm{M^2}\Bigr) g (\bar l f^\dagger f \gamma^\nu \tau^a l) (D^\mu W^a_{\mu\nu}) 
\CR && 
+\frac{1}{4} \bigl[ ig' (\bar l f^\dagger f \sigma^{\mu\nu} \slashed{D} l) B_{\mu\nu} +\text{h.c.}\bigr] 
+\frac{1}{4} \bigl[ ig' (\bar l f^\dagger f \sigma^{\mu\nu} \tau^a \slashed{D} l) W^a_{\mu\nu} +\text{h.c.}\bigr]
\Bigr\},
\eeqa{penguin-result}
where we have denoted the sigma matrices by $\tau^a$ to avoid clash of notation. Note that the form of $[P_\mu, P_\nu] = igG_{\mu\nu}$ depends on the propagator on which it is inserted, e.g.\ $[P_\mu, P_\nu] = igW^a_{\mu\nu}\frac{\tau^a}{2} +ig'B_{\mu\nu}Y$ and $ig'B_{\mu\nu}Y$ for $SU(2)_L$ doublets and singlets, respectively. Also, we see that terms involving $l$ and $\tilde l$ contribute equally, yielding a factor of 2 which cancels against $c_s=\frac{1}{2}$ in the last line of Eq.~\eqref{penguin-result}. Our results agree with those obtained in~\cite{BilenkyS} by Feynman diagram calculations.

\section{Conclusions}
\label{sec:conclusions}

Matching from an UV theory to a low-energy EFT via gauge-covariant functional methods, as an alternative to Feynman diagrams, will continue to be both theoretically interesting and practically useful. We are now at a stage where one-loop universal master formulas are available~\cite{HLM14,DEQY} and have proven useful in the simplest cases (namely in the absence of mixed heavy-light contributions, open covariant derivatives, etc.), while various proposals exist~\cite{HLM16,EQYZ,FPR} to deal with such additional structures that arise in practical applications. An interesting question to explore at this point is whether ideas from these (or other similar) proposals can be implemented as easily as existing universal master formulas, without the need for additional functional manipulations which might make functional matching methods less accessible.

To this end, we have introduced covariant diagrams as a tool to keep track of functional matching calculations. They are easy to use, and provide physical intuition. Specifically, we carried out a functional matching procedure that builds upon and extends the approach of~\cite{FPR}, from which a set of rules for associating terms in a CDE with one-loop diagrams was derived --- this was done, {\it once and for all,} in Sections~\ref{sec:functionalmatching} through~\ref{sec:cd-ferm}. The rules are reminiscent of conventional Feynman rules, but with a crucial difference that only gauge-covariant quantities are involved. The recipe summarized in Section~\ref{sec:cd-sum} can be easily followed in one-loop matching calculations, including those involving mixed heavy-light contributions, open covariant derivatives and mixed statistics, which are not directly captured by existing universal results. We presented many example calculations in Section~\ref{sec:example}, showing technical details for the sake of pedagogy. They provide nontrivial tests of our covariant diagrams formalism. As a byproduct, some universal results incorporating the additional structures were obtained in the intermediate steps, which are also useful beyond the specific UV models considered in this paper.

Compared with Feynman diagrammatic matching, our formalism inherits some key advantages of functional matching, namely gauge covariance in intermediate steps and the possibility of obtaining universal results as discussed in the Introduction. In addition, compared with recently-proposed functional matching approaches, our covariant diagrammatic formulation has the following highlights:
\begin{itemize}
\item No additional functional manipulations (nor subtraction procedures) are needed. One simply draws diagrams and reads off associated master integrals and operator structures, which is more intuitive conceptually.
\item The step of collecting identical terms in a CDE is automatically achieved by associating a symmetry factor to each covariant diagram, which trivially follows from rotation symmetry of the diagram.
\item Loops with fermions are easily handled. As in the pure bosonic case, vertex insertion rules are directly obtained from the quadratic pieces of the UV Lagrangian without explicitly block-diagonalizing the quadratic operator.
\end{itemize}

An attractive direction to move forward in, as far as functional matching methods are concerned, is trying to fully exploit their universality feature and derive more general universal master formulas for one-loop matching. It is an intriguing possibility that as many interesting UV theories as possible can be matched onto low-energy EFTs simply by applying a few master formulas. In this regard, covariant diagrams provide a useful tool to organize and simplify the calculations involved --- we already saw in Section~\ref{sec:UOLEA} that they allow for a simpler derivation of existing universal results. Meanwhile, even in the absence of complete universal results, one can already take advantage of covariant diagram techniques to facilitate one-loop matching calculations for specific UV models of phenomenological interest, as we did in Sections~\ref{sec:triplet-scalar} through~\ref{sec:singlet}. We also comment that the algorithm of enumerating and computing covariant diagrams can in principle be automated, providing a useful and efficient alternative to automated Feynman diagrammatic matching calculations. Besides, it is interesting to consider the possibility of extending covariant diagram techniques beyond one loop for EFT matching, and more generally for extracting UV information of a quantum field theory (including e.g.\ renormalization group evolution~\cite{HLM16}).

\acknowledgments
I would like to thank Sebastian~Ellis, J\'er\'emie~Quevillon, James~Wells and Tevong~You for useful discussions and comments on the manuscript. I also thank the Kavli Institute for Theoretical Physics China (KITPC) for hospitality where part of this work was completed. This research was supported by the U.S.\ Department of Energy under grant DE-SC0007859, and the Rackham Graduate School of the University of Michigan via the Rackham Summer Award.


\appendix
\section{Master integrals}
\label{app:MI}

In this appendix, we discuss calculations of the master integrals and tabulate some of them that are useful in practical applications of covariant diagrams. The master integrals $\I[q^{2n_c}]_{ij\dots 0}^{n_i n_j\dots n_L}$ are defined by Eq.~\eqref{MIdef}, which we reproduce here for convenience,
\beq
\int\frac{d^dq}{(2\pi)^d} \frac{q^{\mu_1}\cdots q^{\mu_{2n_c}}}{(q^2-M_i^2)^{n_i}(q^2-M_j^2)^{n_j}\cdots (q^2)^{n_L}}
\,\equiv\, g^{\mu_1\dots\mu_{2n_c}} \,\I[q^{2n_c}]_{ij\dots 0}^{n_i n_j\dots n_L} ,
\eeqn
where $g^{\mu_1\dots\mu_{2n_c}}$ is the completely symmetric tensor, e.g.\ $g^{\mu\nu\rho\sigma}=g^{\mu\nu}g^{\rho\sigma} +g^{\mu\rho}g^{\nu\sigma} +g^{\mu\sigma}g^{\nu\rho}$. We first observe that all nondegenerate master integrals, including mixed heavy-light ones (those with $n_L>0$), can be reduced to degenerate ones by recursively applying
\bseq
\beqa
\I[q^{2n_c}]_{ij\dots 0}^{n_i n_j\dots n_L} &=& \frac{1}{\dsq{ij}} \bigl( \I[q^{2n_c}]_{ij\dots 0}^{n_i, n_j-1,\dots n_L} -\,\I[q^{2n_c}]_{ij\dots 0}^{n_i-1, n_j\dots n_L} \bigr) \,,\label{MIreduction-a} \\
\I[q^{2n_c}]_{ij\dots 0}^{n_i n_j\dots n_L} &=& \frac{1}{M_i^2} \bigl( \I[q^{2n_c}]_{ij\dots 0}^{n_i n_j\dots, n_L-1} -\,\I[q^{2n_c}]_{ij\dots 0}^{n_i-1, n_j\dots n_L} \bigr) \,,\label{MIreduction:mixed}
\eeqan
\eseq{MIreduction}
where $\dsq{ij}\equiv M_i^2-M_j^2$, and using the fact that $\,\I[q^{2n_c}]_0^{n_L}=0$. 
As an example, we encountered the master integral $\I[q^2]_{ij}^{22}$ in the calculation of the universal coefficient $f_7^{ij}$ (see Eq.~\eqref{diag:P2U2} and Table~\ref{tab:UOLEA}), which can be reduced as follows,
\beqa
\I[q^2]_{ij}^{22} 
&=& \frac{1}{\dsq{ij}} \bigl(\,\I[q^2]_{ij}^{21}-\I[q^2]_{ij}^{12}\bigr)
= \frac{1}{(\dsq{ij})^2} \bigl(\,\I[q^2]_i^2 -2\,\I[q^2]_{ij}^{11} +\I[q^2]_j^2 \bigr) \CR
&=& \frac{1}{(\dsq{ij})^2} \bigl(\,\I[q^2]_i^2 +\I[q^2]_j^2 \bigr) -\frac{2}{(\dsq{ij})^3} \bigl(\,\I[q^2]_i^1 -\I[q^2]_j^1\bigr) \,.
\eeqan
We also note that a special case of Eq.~\eqref{MIreduction-a} which is useful in reducing the master integrals appearing in Eq.~\eqref{trP0} reads
\beq
\I_{i_1\dots i_N}^{1...1} = \sum_{n=1}^{N} \frac{\I_{i_n}^1}{\prod_{m\ne n} \dsq{i_n i_m}} \,.
\eeqn

With the reduction formulas in Eq.~\eqref{MIreduction} at hand, it is sufficient to tabulate the master integrals of the form $\I[q^{2n_c}]_i^{n_i}$. A general formula for the latter is
\beq
\I[q^{2n_c}]_i^{n_i} = \lf \bigl(-M_i^2\bigr)^{2+n_c-n_i}
\frac{1}{2^{n_c}(n_i-1)!} \frac{\Gamma(\frac{\epsilon}{2}-2-n_c +n_i)}{\Gamma(\frac{\epsilon}{2})} \Bigl(\frac{2}{\bar\epsilon}-\logm{M_i^2}\Bigr) \,,
\eeq{MIformula}
where $\frac{2}{\bar\epsilon} \equiv \frac{2}{\epsilon} -\gamma +\log 4\pi$ with $\epsilon = 4-d$. 
It is understood that with the $\overline{\text{MS}}$ scheme, one replaces $\frac{2}{\bar\epsilon}-\logm{M_i^2}$ by $-\logm{M_i^2}$ in the final result. 
We factor out the common prefactor, $\I = \lf\tilde\I$, and list $\tilde\I[q^{2n_c}]_i^{n_i}$ up to $n_c=3$ and $n_i=6$ in Table~\ref{tab:MIheavy}.

\begin{table}[tbp]
\centering
\begin{tabular}{|c|cccc|}
\hline
$\tilde\I[q^{2n_c}]_i^{n_i}$ & $n_c=0$ & $n_c=1$ & $n_c=2$ & $n_c=3$ \\
\hline
$n_i=1$ 
& $M_i^2 \bigl(1-\logm{M_i^2}\bigr)$ 
& $\frac{M_i^4}{4} \bigl(\frac{3}{2} -\logm{M_i^2}\bigr)$ 
& $\frac{M_i^6}{24} \bigl(\frac{11}{6} -\logm{M_i^2}\bigr)$
& $\frac{M_i^8}{192} \bigl(\frac{25}{12}-\logm{M_i^2}\bigr)$ \\
$n_i=2$ 
& $-\logm{M_i^2}$ 
& $\frac{M_i^2}{2} \bigl(1 -\logm{M_i^2}\bigr)$ 
& $\frac{M_i^4}{8} \bigl(\frac{3}{2} -\logm{M_i^2}\bigr)$
& $\frac{M_i^6}{48} \bigl(\frac{11}{6}-\logm{M_i^2}\bigr)$ \\
$n_i=3$ 
& $-\frac{1}{2M_i^2}$ 
& $-\frac{1}{4}\logm{M_i^2}$ 
& $\frac{M_i^2}{8} \bigl(1 -\logm{M_i^2}\bigr)$
& $\frac{M_i^4}{32} \bigl(\frac{3}{2}-\logm{M_i^2}\bigr)$ \\
$n_i=4$ 
& $\frac{1}{6M_i^4}$ 
& $-\frac{1}{12M_i^2}$
& $-\frac{1}{24}\logm{M_i^2}$
& $\frac{M_i^2}{48} \bigl(1 -\logm{M_i^2}\bigr)$ \\
$n_i=5$ 
& $-\frac{1}{12M_i^6}$ 
& $\frac{1}{48M_i^4}$
& $-\frac{1}{96M_i^2}$
& $-\frac{1}{192}\logm{M_i^2}$ \\
$n_i=6$ 
& $\frac{1}{20M_i^8}$ 
& $-\frac{1}{120M_i^6}$
& $\frac{1}{480M_i^4}$
& $-\frac{1}{960M_i^2}$ \\
\hline
\end{tabular}
\caption{Commonly-used degenerate master integrals $\tilde\I[q^{2n_c}]_i^{n_i}\equiv\I[q^{2n_c}]_i^{n_i}/\lf$, with $\frac{2}{\bar\epsilon} = \frac{2}{\epsilon} -\gamma +\log 4\pi$ dropped. All nondegenerate (including mixed heavy-light) master integrals can be reduced to degenerate master integrals by Eq.~\eqref{MIreduction}.
\label{tab:MIheavy}}
\end{table}

\begin{table}[tbp]
\centering
\begin{tabular}{|c|ccc|}
\hline
$\tilde\I[q^{2n_c}]_{i0}^{n_i n_L}$ & $n_c=0$ & $n_c=1$ & $n_c=2$ \\
\hline
$(n_i, n_L) = (1,1)$
& $1-\logm{M_i^2}$
& $\frac{M_i^2}{4}\bigl(\frac{3}{2}-\logm{M_i^2}\bigr)$
& $\frac{M_i^4}{24}\bigl(\frac{11}{6}-\logm{M_i^2}\bigr)$ \\
\hline
$(n_i, n_L) = (2,1)$
& $-\frac{1}{M_i^2}$
& $\frac{1}{4}\bigl(\frac{1}{2}-\logm{M_i^2}\bigr)$
& $\frac{M_i^2}{12}\bigl(\frac{4}{3}-\logm{M_i^2}\bigr)$ \\
$(n_i, n_L) = (1,2)$
& $\frac{1}{M_i^2} \bigl(1-\logm{M_i^2}\bigr)$
& $\frac{1}{4}\bigl(\frac{3}{2}-\logm{M_i^2}\bigr)$
& $\frac{M_i^2}{24}\bigl(\frac{11}{6}-\logm{M_i^2}\bigr)$ \\
\hline
$(n_i, n_L) = (3,1)$
& $\frac{1}{2M_i^4}$
& $-\frac{1}{8M_i^2}$
& $\frac{1}{24}\bigl(\frac{1}{3}-\logm{M_i^2}\bigr)$ \\
$(n_i, n_L) = (2,2)$
& $-\frac{1}{M_i^4}(2-\logm{M_i^2})$
& $-\frac{1}{4M_i^2}$
& $\frac{1}{24}\bigl(\frac{5}{6}-\logm{M_i^2}\bigr)$ \\
$(n_i, n_L) = (1,3)$
& $\frac{1}{M_i^4} \bigl(1-\logm{M_i^2}\bigr)$
& $\frac{1}{4M_i^2}\bigl(\frac{3}{2}-\logm{M_i^2}\bigr)$
& $\frac{1}{24}\bigl(\frac{11}{6}-\logm{M_i^2}\bigr)$ \\
\hline
$(n_i, n_L) = (4,1)$
& $-\frac{1}{3M_i^6}$
& $\frac{1}{24M_i^4}$
& $-\frac{1}{72M_i^2}$ \\
$(n_i, n_L) = (3,2)$
& $\frac{1}{M_i^6}\bigl(\frac{5}{2}-\logm{M_i^2}\bigr)$
& $\frac{1}{8M_i^4}$
& $-\frac{1}{48M_i^2}$ \\
$(n_i, n_L) = (2,3)$
& $-\frac{2}{M_i^6}\bigl(\frac{3}{2}-\logm{M_i^2}\bigr)$
& $-\frac{1}{4M_i^4}\bigl(\frac{5}{2}-\logm{M_i^2}\bigr)$
& $-\frac{1}{24M_i^2}$ \\
$(n_i, n_L) = (1,4)$
& $\frac{1}{M_i^6}\bigl(1-\logm{M_i^2}\bigr)$
& $\frac{1}{4M_i^4}\bigl(\frac{3}{2}-\logm{M_i^2}\bigr)$
& $\frac{1}{24M_i^2}\bigl(\frac{11}{6}-\logm{M_i^2}\bigr)$ \\
\hline
$(n_i, n_L) = (5,1)$
& $\frac{1}{4M_i^8}$
& $-\frac{1}{48M_i^6}$
& $\frac{1}{288M_i^4}$ \\
$(n_i, n_L) = (4,2)$
& $-\frac{1}{M_i^8}\bigl(\frac{17}{6}-\logm{M_i^2}\bigr)$
& $-\frac{1}{12M_i^6}$
& $\frac{1}{144M_i^4}$ \\
$(n_i, n_L) = (3,3)$
& $\frac{3}{M_i^8}\bigl(\frac{11}{6}-\logm{M_i^2}\bigr)$
& $\frac{1}{4M_i^6}\bigl(3-\logm{M_i^2}\bigr)$
& $\frac{1}{48M_i^4}$ \\
$(n_i, n_L) = (2,4)$
& $-\frac{3}{M_i^8}\bigl(\frac{4}{3}-\logm{M_i^2}\bigr)$
& $-\frac{1}{2M_i^6}\bigl(2-\logm{M_i^2}\bigr)$
& $-\frac{1}{24M_i^4}\bigl(\frac{17}{6}-\logm{M_i^2}\bigr)$ \\
$(n_i, n_L) = (1,5)$
& $\frac{1}{M_i^8}\bigl(1-\logm{M_i^2}\bigr)$
& $\frac{1}{4M_i^6}\bigl(\frac{3}{2}-\logm{M_i^2}\bigr)$
& $\frac{1}{24M_i^4}\bigl(\frac{11}{6}-\logm{M_i^2}\bigr)$ \\
\hline
\end{tabular}
\caption{Commonly-used mixed heavy-light master integrals with degenerate heavy particle masses $\tilde\I[q^{2n_c}]_{i0}^{n_i n_L}\equiv\I[q^{2n_c}]_{i0}^{n_i n_L}/\lf$, with $\frac{2}{\bar\epsilon} = \frac{2}{\epsilon} -\gamma +\log 4\pi$ dropped.
\label{tab:MImixed}}
\end{table}

For convenience let us also tabulate the master integrals of the form $\I[q^{2n_c}]_{i0}^{n_i n_L}$ encountered in Section~\ref{sec:example}. They can be obtained with either Eq.~\eqref{MIreduction:mixed} or a generalization of Eq.~\eqref{MIformula},
\beqa
\I[q^{2n_c}]_{i0}^{n_i n_L} &=& \lf \bigl(-M_i^2\bigr)^{2+n_c-n_i-n_L}
\frac{1}{2^{n_c}(n_i-1)!} \cdot\CR
&&\quad
\frac{\Gamma(\frac{\epsilon}{2}-2-n_c +n_i +n_L)}{\Gamma(\frac{\epsilon}{2})} 
\frac{\Gamma(-\frac{\epsilon}{2}+2+n_c-n_L)}{\Gamma(-\frac{\epsilon}{2} +2+n_c)} 
\Bigl(\frac{2}{\bar\epsilon}-\logm{M_i^2}\Bigr) \,.\quad
\eeqan
Pulling out the loop factor as before, $\I = \lf\tilde\I$, we list $\tilde\I[q^{2n_c}]_{i0}^{n_i n_L}$ up to $n_c=2$ and $n_i+n_L=6$ in Table~\ref{tab:MImixed}.

\section{Explicit expressions of universal coefficients}
\label{app:coefficients}

Here we give explicit expressions of the universal coefficients, namely coefficients of operator traces in the UOLEA master formula Eq.~\eqref{LUOLEA} rederived in Section~\ref{sec:UOLEA} (see Table~\ref{tab:UOLEA}), in terms of heavy particle masses $M_i$, $M_j$, etc. In many cases, our expressions simplify those originally derived in~\cite{DEQY}. We define $f_N = \lf \tilde f_N$ as in~\cite{DEQY}, and list $\tilde f_N$ in the following:
%
\beqa
&& \tilde f_2^i = M_i^2 \left(1-\logm{M_i^2}\right) , \\
&& \tilde f_3^i = -\frac{1}{12} \logm{M_i^2} , \\
&& \tilde f_4^{ij} = \frac{1}{2} \biggl(1-\frac{M_i^2\logm{M_i^2}}{\dsq{ij}}-\frac{M_j^2\logm{M_j^2}}{\dsq{ji}}\biggr) ,\\
&& \tilde f_5^i = -\frac{1}{60M_i^2} , \\
&& \tilde f_6^i = -\frac{1}{90M_i^2} , \\
&& \tilde f_7^{ij} = -\frac{M_i^2 +M_j^2}{4(\dsq{ij})^2} +\frac{M_i^2 M_j^2 \log\frac{M_i^2}{M_j^2}}{2(\dsq{ij})^3} ,\\
&& \tilde f_8^{ijk} = -\frac{1}{3} \biggl( \frac{M_i^2\log M_i^2}{\dsq{ij}\dsq{ik}} + \frac{M_j^2\log M_j^2}{\dsq{ji}\dsq{jk}} + \frac{M_k^2\log M_k^2}{\dsq{ki}\dsq{kj}} \biggr) ,\\
&& \tilde f_9^i = -\frac{1}{12M_i^2} ,\\
&& \tilde f_{10}^{ijkl} = -\frac{1}{4}  \left( \frac{M_i^2\log M_i^2}{\dsq{ij}\dsq{ik}\dsq{il}} + \frac{M_j^2\log M_j^2}{\dsq{ji}\dsq{jk}\dsq{jl}} + \frac{M_k^2\log M_k^2}{\dsq{ki}\dsq{kj}\dsq{kl}} + \frac{M_l^2\log M_l^2}{\dsq{li}\dsq{lj}\dsq{lk}} \right) , \\
&& \tilde f_{11}^{ijk} = \frac{M_i^2 M_j^2 +M_i^2 M_k^2 +M_j^2 M_k^2 -3M_k^4}{2(\dsq{ik})^2(\dsq{jk})^2} +\frac{M_i^2 M_k^2 \log M_i^2}{\dsq{ij}(\dsq{ik})^3} +\frac{M_j^2 M_k^2 \log M_j^2}{\dsq{ji}(\dsq{jk})^3}\CR
&&\qquad\quad  +\frac{\bigl[M_i^2 M_j^2 (M_i^2+M_j^2 -3M_k^2) +M_k^6\bigr] M_k^2 \log M_k^2}{(\dsq{ki})^3(\dsq{kj})^3} ,  \\
&& \tilde f_{12}^{ij} = \frac{M_i^4 +10 M_i^2 M_j^2 +M_j^4}{12(\dsq{ij})^4} - \frac{M_i^2 M_j^2 (M_i^2 +M_j^2) \log\frac{M_i^2}{M_j^2}}{2(\dsq{ij})^5} , \\
&& \tilde f_{13}^{ij} = \frac{2M_i^4 +5M_i^2 M_j^2 -M_j^4}{12M_i^2 (\dsq{ij})^3} -\frac{M_i^2 M_j^2 \log\frac{M_i^2}{M_j^2}}{2(\dsq{ij})^4} \\
&& \tilde f_{14}^{ij} = -\frac{M_i^4 +10 M_i^2 M_j^2 +M_j^4}{6(\dsq{ij})^4} + \frac{M_i^2 M_j^2 (M_i^2 +M_j^2) \log\frac{M_i^2}{M_j^2}}{(\dsq{ij})^5} , \\
&& \tilde f_{15}^{ij} = \frac{2M_i^4 +11M_i^2 M_j^2 -7M_j^4}{18(\dsq{ij})^4} -\frac{M_j^2 (3M_i^4-M_j^4) \log\frac{M_i^2}{M_j^2}}{6(\dsq{ij})^5} \\
&& \tilde f_{16}^{ijklm} = -\frac{1}{5} \biggl( \frac{M_i^2\log M_i^2}{\dsq{ij}\dsq{ik}\dsq{il}\dsq{im}} + \frac{M_j^2\log M_j^2}{\dsq{ji}\dsq{jk}\dsq{jl}\dsq{jm}} + \frac{M_k^2\log M_k^2}{\dsq{ki}\dsq{kj}\dsq{kl}\dsq{km}} \CR
&& \qquad\qquad\qquad + \frac{M_l^2\log M_l^2}{\dsq{li}\dsq{lj}\dsq{lk}\dsq{lm}} + \frac{M_m^2\log M_m^2}{\dsq{mi}\dsq{mj}\dsq{mk}\dsq{ml}} \biggr) , \\
&& \tilde f_{17}^{ijkl} = -\frac{M_i^2 M_j^2 M_k^2 +(M_i^2 M_j^2 +M_i^2 M_k^2 +M_j^2 M_k^2) M_l^2 -3(M_i^2+M_j^2+M_k^2)M_l^4 +5M_l^6}{2(\dsq{il})^2 (\dsq{jl})^2 (\dsq{kl})^2} \CR
&&\qquad\quad +\frac{M_i^2 M_l^2 \log M_i^2}{\dsq{ij}\dsq{ik}(\dsq{il})^3} +\frac{M_j^2 M_l^2 \log M_j^2}{\dsq{ji}\dsq{jk}(\dsq{jl})^3} +\frac{M_k^2 M_l^2 \log M_k^2}{\dsq{ki}\dsq{kj}(\dsq{kl})^3} +\frac{M_l^2\log M_l^2}{(\dsq{li})^3(\dsq{lj})^3(\dsq{lk})^3} \cdot \CR
&&\qquad\quad  \Bigl[ (M_i^2 M_j^2 M_k^2 +M_l^6) \bigl(M_i^2 M_j^2 +M_i^2 M_k^2 + M_j^2 M_k^2 -3 (M_i^2 +M_j^2 +M_k^2) M_l^2 +6M_l^4 \bigr) \CR
&&\qquad\quad +M_l^6 (M_i^4 +M_j^4 +M_k^4 -3M_l^4 ) \Bigr], \CR \\
&& \tilde f_{18}^{ijkl} = 
\frac{1}{4(\dsq{ik})^2(\dsq{jk})^2(\dsq{il})^2(\dsq{jl})^2} \CR
&&\qquad\quad \bigl[-M_i^2 M_j^2 M_k^2 (M_i^4+M_j^4+M_k^4)-M_i^2 M_j^2 M_l^2 (M_i^4+M_j^4+M_l^4) \CR 
&&\qquad\qquad -M_i^2 M_k^2 M_l^2 (M_i^4+M_k^4+M_l^4)-M_j^2 M_k^2 M_l^2 (M_j^4+M_k^4+M_l^4) \CR
&&\qquad\qquad +2 M_i^2 M_j^2 (M_i^2+M_j^2) (M_k^4+M_l^4) +2 M_k^2 M_l^2 (M_i^4+M_j^4) (M_k^2+M_l^2) \CR
&&\qquad\qquad +3 M_i^4 M_j^4 (M_i^2+M_j^2)+3 M_k^4 M_l^4 (M_k^2+M_l^2) \CR
&&\qquad\qquad +3 M_i^2 M_j^2 M_k^2 M_l^2 (M_i^2+M_j^2+M_k^2+M_l^2) \CR
&&\qquad\qquad -7 M_i^4 M_j^4 (M_k^2+M_l^2)-7 M_k^4 M_l^4 (M_i^2+M_j^2) \bigr] \CR
&& \qquad\quad +\frac{\bigl[ M_i^6 +M_k^2 M_l^2 (M_k^2+M_l^2-3M_i^2) \bigr] M_i^2\log M_i^2}{2\dsq{ij}(\dsq{ik})^3(\dsq{il})^3} \CR
&& \qquad\quad +\frac{\bigl[ M_j^6 +M_k^2 M_l^2 (M_k^2+M_l^2-3M_j^2) \bigr] M_j^2\log M_j^2}{2\dsq{ji}(\dsq{jk})^3(\dsq{jl})^3} \CR
&&\qquad\quad +\frac{\bigl[ M_k^6 +M_i^2 M_j^2 (M_i^2+M_j^2-3M_k^2) \bigr] M_k^2\log M_k^2}{2\dsq{kl}(\dsq{ki})^3(\dsq{kj})^3} \CR
&& \qquad\quad +\frac{\bigl[ M_l^6 +M_i^2 M_j^2 (M_i^2+M_j^2-3M_l^2) \bigr] M_l^2\log M_l^2}{2\dsq{lk}(\dsq{li})^3(\dsq{lj})^3} , \CR\\
&& \tilde f_{19}^{ijklmn} = -\frac{1}{6} \biggl( \frac{M_i^2\log M_i^2}{\dsq{ij}\dsq{ik}\dsq{il}\dsq{im}\dsq{in}} + \frac{M_j^2\log M_j^2}{\dsq{ji}\dsq{jk}\dsq{jl}\dsq{jm}\dsq{jn}} + \frac{M_k^2\log M_k^2}{\dsq{ki}\dsq{kj}\dsq{kl}\dsq{km}\dsq{kn}} \CR
&& \qquad\qquad\qquad\quad + \frac{M_l^2\log M_l^2}{\dsq{li}\dsq{lj}\dsq{lk}\dsq{lm}\dsq{ln}} + \frac{M_m^2\log M_m^2}{\dsq{mi}\dsq{mj}\dsq{mk}\dsq{ml}\dsq{mn}} + \frac{M_n^2\log M_n^2}{\dsq{ni}\dsq{nj}\dsq{nk}\dsq{nl}\dsq{nm}} \biggr) .\CR
\eeqan
%
As in the previous appendix, we have used the shorthand notation $\dsq{ij}\equiv M_i^2-M_j^2$, etc.


\end{document}